\documentclass[prd,preprint,tightenlines,floatfix,preprintnumbers,nofootinbib,eqsecnum]{revtex4}

  \usepackage{amssymb}
   \usepackage{amsmath}
    \usepackage{amsfonts}
     \usepackage{epsfig}
      \usepackage{bm} 
      \usepackage[dvipsnames,usenames]{color}

\usepackage[colorlinks,citecolor=blue,urlcolor=blue,linkcolor=blue]{hyperref}

\begin{document}

\title{Odderon and proton substructure from a model-independent L\'evy imaging
of elastic $pp$ and $p\bar p$ collisions}

\author{T. Cs\"{o}rg\H{o}$^{1,2,3}$}
\email{tcsorgo@cern.ch}

\author{R. Pasechnik$^{4,5}$}
\email{Roman.Pasechnik@thep.lu.se}

\author{A. Ster$^{1}$}

\affiliation{
{$^1$\sl 
MTA Wigner FK, H-1525 Budapest 114, POB 49, Hungary
}\\
{$^2$\sl 
EKU KRC, H-3200 Gy\"ongy\"os, M\'atrai \'ut 36, Hungary
}\\
{$^3$\sl 
CERN, CH-1211 Geneva 23, Switzerland
}\\
{$^4$\sl
Department of Astronomy and Theoretical Physics, Lund
University, SE-223 62 Lund, Sweden
}\\
{$^5$\sl Nuclear Physics Institute ASCR, 25068 \v{R}e\v{z}, 
Czech Republic\vspace{1.0cm}
}}

\begin{abstract}
\vspace{0.5cm}
We describe a new and model-independent L\'evy imaging method of quality fits to the published datasets and reconstruct the amplitude of high-energy $pp$ and $p\bar p$ 
elastic scattering processes. This method allows us to determine the excitation function of the shadow (inelasticity) profile $P(b)$, the elastic slope $B(t)$ and the nuclear phase 
$\phi(t)$ functions of $pp$ and $p\bar p$ collisions directly from the data. Surprisingly, notable qualitative differences in $B(t)$ for $pp$ and for $p\bar p$ collisions point towards 
an Odderon effect. As a by-product, we clearly identify the proton substructure with two different sizes at the ISR and LHC energies, that has striking similarity to a dressed quark 
(at the ISR) and a dressed diquark (at the LHC). We present model-independent results for the corresponding sizes and cross-sections for such a substructure for the existing 
data at different energies.
\end{abstract}

\maketitle

\begin{quotation}
{\it ``What we pay attention to is largely determined by our expectations of what should be present.''\\
\null \hfill  Christopher Chabris}
\end{quotation}

\section{Introduction}
\label{s:intro}

The TOTEM Collaboration at the Large Hadron Collider (LHC) at CERN has released recently two new data sets from the first measurements of the total, elastic and differential cross sections, 
as well as the $\rho$-parameter, of elastic $pp$ collisions at the currently highest available energy of $\sqrt{s} = 13$ TeV~\cite{Antchev:2017dia,Antchev:2017yns}. Taken together, 
these papers indicate the influence of an odd-under-crossing (or C-odd) contribution to the elastic scattering amplitude, traditionally called the Odderon~\cite{Lukaszuk:1973nt},
or in more modern language of Quantum Chromo Dynamics (QCD), an odd-gluon (predominantly, three-gluon) bound state, a quarkless excitation sometimes also referred to as a vector glueball.
As far as we know, the properties like the $\sqrt{s}$ and $t$ dependences of the differential cross-section of an Odderon contribution at the LHC energies were determined from $pp$ 
and $p\bar p$ collisions first in Ref.~\cite{Ster:2015esa}.

The new TOTEM preliminary results~\cite{Antchev:2017dia,Antchev:2017yns} generated a burst of high-level and vigorous theoretical debate about the correct interpretation of these data, 
see Refs.~\cite{Samokhin:2017kde}--\cite{Khoze:2018kna}. All possible extreme views were present among these papers, including claims for a maximal Odderon effect~\cite{Martynov:2018nyb} 
and claims of lack of any significant Odderon effects, see Refs.~\cite{Shabelski:2018jfq,Khoze:2018kna}.

In this paper, we investigate earlier, published data and a recently released, new TOTEM preliminary data set~\cite{Frici-la-Biodola:2018vvv}, to look for the Odderon effects in elastic 
$pp$ collisions, and for the interpretation of the data, using a new and model-independent imaging technique, the L\'evy series. Our analysis is based on the TOTEM preliminary data 
as presented by F. Nemes for the TOTEM Collaboration in Ref.~\cite{Frici-la-Biodola:2018vvv}.

We find that clear, but indirect signals of Odderon effects are present in the differential cross-section of elastic $pp$ and $p\bar p$ scattering, as indicated by the difference of 
the four-momentum transfer dependence of the elastic slope $B(t)$ for $pp$ and for $p\bar p$ collisions. A less evident but clear difference is also identified between 
the nuclear phase $\phi(t)$ of $pp$ and $p\bar p$ collisions in the TeV energy range.

Although our analysis was motivated by the search for Odderon effects, our most surprising result is that we find  a clear-cut evidence for a proton substructure having two distinct sizes 
in the GeV and TeV energy ranges, respectively. We estimate these sizes and the corresponding contributions to the total $pp$ cross-section at the ISR and Tevatron/LHC energies.

The structure of the presentation is as follows. In Section~\ref{s:levy-expansion}, we present the model-independent imaging approach of L\'evy expansions, in Section~\ref{s:Data-analysis},
we apply this method in a comprehensive analysis of elastic $pp$ and $p\bar p$ collisions, and in Section~\ref{s:Discussion}, we discuss and interpret the results, 
as well as present the indirect signals of Odderon effects in elastic scattering in the TeV energy range. Finally, we summarize and conclude in Section~\ref{s:conclusions}.

This manuscript is closed by four Appendices. Appendix A details the results of the L\'evy expansion fits to elastic $pp$ scattering data for the whole available region of $t$.
Appendix B shows the individual fits to elastic $p\bar p$ scattering data. Appendix C describes fits to the tails (i.e.~in the large-$|t|$ region just after the dip and the bump structure) 
of the differential cross-section of elastic $pp$ scattering for all the experimentally accessed energies. These fits indicate an evidence for a proton substructure. Appendix D details 
L\'evy fits to the cone (small-$|t|$) regions of elastic $pp$ scattering, indicating that the proton size grows self-similarly in the GeV energy range, but at the LHC energies of 
$\sqrt{s} = $ 7 and 13 TeV, something drastically changes, such that the protons keep on growing but their shape also becomes significantly different from their shape 
in the ISR energy range of $\sqrt{s} = $ 23.5 -- 62.5 GeV. 

\section{Model-independent L\'evy analysis of elastic $pp$ and $p\bar p$ scattering}
\label{s:levy-expansion}

We describe a model-independent L\'evy series, that is a generalization of the Laguerre, Edgeworth~\cite{Csorgo:2000pf} and L\'evy expansion~\cite{DeKock:2012gp} methods proposed 
to analyze nearly L\'evy stable source distributions in the field of particle correlations and femtoscopy. The key point of this method is to have a look at the data, guess their approximate 
shape (for example, Gaussian, exponential or L\'evy-stable shape) and then develop a systematic method to characterize the deviations from the approximate shape with the help of 
a dimensionless variable denoted in this paper by $z$, and a complete orthonormal set of polynomials that are orthogonal with respect to the assumed zeroth order shape or weight function $w(z)$. 
We recommend  Ref.~\cite{Csorgo:2000pf} for detailed discussions and for the convergence criteria of such expansions given in general terms. This way one may find the minimal number of 
necessary expansion coefficients. 

For example, if the data follow the guessed approximate shape precisely, that can be clearly demonstrated by fitting the series to the data and finding that all the expansion coefficients 
that measure deviations from the zeroth-order shape are within errors consistent with zero. Indeed, the PHENIX Collaboration analyzed recently Bose-Einstein correlations of $\sqrt{s_{NN}} = 200$ GeV 
Au+Au collisions~\cite{Adare:2017vig}, and found that they are well described by the L\'evy stable source distributions. The accuracy of the L\'evy description was  tested by a L\'evy expansion 
of the Bose-Einstein correlation functions, as proposed in Ref.~\cite{Novak:2016cyc}, to find that the first-order deviations from the L\'evy stable source distributions within errors are consistent 
with zero~\cite{Adare:2017vig}. 

The data analysis method of Ref.~\cite{Novak:2016cyc} was developed first for the functions that may oscillate between positive and negative values. However, the differential cross-section of 
elastic scattering is measured as an angular-dependent hit distribution, so it is a positive definite function, related to the modulus square of the elastic scattering amplitude. Hence, we describe 
this expansion at the amplitude level, with complex expansion coefficients, and then take the modulus square of such an amplitude to get a positive-definite form.

The differential cross-section of elastic scattering at high energies is measured as a function of the four-momentum transfer squared, the Mandelstam variable $t=(p_1 -p_3)^2<0$. A differential 
cross-section of elastic scattering starts at the optical point at $t=0$, decreases quickly and nearly exponentially in $|t|$, as typically characterized by an exponential slope
parameter $B$, as follows: $\frac{d\sigma}{dt} = A \exp(- B |t|)$. The region, where such a behaviour is approximately valid, is called a diffractive cone, and such a featureless exponential 
decay corresponds to a nearly Gaussian distribution of the centers of elastic scattering~\cite{Block:2006hy}. This region is followed by (one or more) alternating diffractive minima and maxima, 
finally at high four-momentum transfers, a diffractive tail might be seen as well. For more details and for an introduction and review of diffraction before the start of the LHC measurements,
see Ref.~\cite{Block:2006hy}.

We know from the TOTEM analysis of $\sqrt{s}= 8$ TeV elastic $pp$ scattering data, that the differential cross-section in the diffraction cone, at $|t|$ values below the diffraction minimum, 
deviates significantly from an exponential shape~\cite{Antchev:2015zza}. This deviation is a subtle, but a more than  7$\sigma$ effect~\cite{Antchev:2015zza}. Using this knowledge, 
we assume that the elastic scattering amplitude is nearly (but not completely) exponential in $|t|$, i.e.~we assume that it is approximately described by a (Fourier-transformed) L\'evy stable 
source distribution: 
\begin{eqnarray*}
d\sigma/dt \propto \exp\left(-(R^2 |t|)^\alpha\right)\,.
\end{eqnarray*} 
The conventional exponential behaviour corresponds to the $\alpha = 1$ special case. This way the deviation from 
the exponential behaviour can be quantified by a single parameter. If the value of the exponent $\alpha$ is significantly different from unity, it evidences a non-exponential behaviour of 
the differential cross-section of elastic scattering. Later on, we shall see that this is a very fortunate approach, and it connects the imaging of the differential cross-sections of high-energy 
$pp$ and $p\bar p$ collisions with the L\'evy stable source distributions that are ubiquitous in Nature~\cite{Tsallis:1995zz}.

We also know that the differential cross-section of high-energy $pp$ and $p\bar p$ collisions has a diffractive minimum, followed by a second maximum, that is followed by an extended tail.
Thus, the behaviour of the differential cross-sections at large $|t|$ has structures that deviate from a simple L\'evy stable source. In this paper, we attempt to characterize these structures 
with an orthonormalized L\'evy expansion. This way we obtain a new imaging method, that we describe in detail below, and apply it to reconstruct the shadow (inelasticity) profile functions, the $t$-dependent 
slope parameters and the $t$-dependent nuclear phases in high-energy $pp$ as well as $p\bar p$ collisions.

These physical and mathematical assumptions result in the following formulae for the differential cross-section of elastic $pp$ and $p\bar p$ collisions:
\begin{eqnarray}
\label{e:levy-dsigmadt}
  \frac{d\sigma}{dt} & = &   A \, w(z|\alpha) 
   \left| 1 +
   \sum_{j = 1}^\infty c_j l_j (z|\alpha) \right|^2\,, \\
   w(z|\alpha) & = & \exp(-z^\alpha) \,,  \\
   z & = & |t| R^2 \geq 0\,, \\
 c_j & = & a_j + i b_j\,,
\end{eqnarray}
where $w(z|\alpha)$ is the L\'evy weight function, and a dimensionless variable $z$ is introduced as the magnitude of the four-momentum transfer squared $|t|$ multiplied by a L\'evy scale 
parameter $R$ squared, where $R$ is measured in units of fm (natural units $c=\hbar=1$ are adopted here and below). Note that the $w(z|\alpha)$ is also called the stretched exponential
distribution, which, for $\alpha = 1$ limiting case, corresponds to the exponential distribution. This shape actually corresponds to a Fourier-transformed and modulus squared symmetric 
L\'evy-stable source distribution centered on zero. The orthonormal set of  L\'evy polynomials, defined below, are denoted as $l_j(z|\alpha)$. The complex expansion coefficients 
are $c_j$,  with $a_j$ and $b_j$ standing for the real and the imaginary parts of $c_j$, respectively. 

In the forthcoming, we shall refer to the zeroth-order ($c_i = 0$) L\'evy expansion simply as a L\'evy fit. Let us clarify here that these so-called L\'evy fits actually correspond to 
the Fourier-transformed and modulus squared, symmetric L\'evy stable source distributions $t_{el}(b)$, as approximations to the shape of the amplitude of elastic $pp$ scattering. 
This elastic amplitude will be introduced and detailed in Subsection~\ref{ss:cross-sections}. For more details on the application of L\'evy stable source
distributions~\cite{Nolan:2016st,Weisstein:Stable} in particle correlations and femtoscopic measurements, we recommend Refs.~\cite{DeKock:2012gp,Adare:2017vig,Novak:2016cyc,Ster-BGL17,Novak-WPCF18}.

The expansion (\ref{e:levy-dsigmadt}) is expected to converge to nearly L\'evy shaped data, if the order of the series $n$ is chosen to be sufficiently large, i.e.~$n \rightarrow \infty$. 
In practice, however, third-order ($n = 3$) L\'evy series already converged to the data measured at $\sqrt{s} < 1 $ TeV, with confidence levels (i.e.~the probability that the L\'evy series 
or expansion of the elastic scattering amplitude converges to the differential cross-section under investigation) corresponding to a statistically acceptable description. In order to gain 
a statistically marginal or acceptable description of the high precision TOTEM data at 7 TeV and preliminary data at 13 TeV, we had to go to the fourth-order L\'evy series, $n = 4$, 
in these two cases. For reasons of consistency, and in order to eliminate fitting artefacts that may show up if one compares different orders of the L\'evy expansions with one another, 
we decided to re-fit all the $pp$ elastic scattering data with the fourth-order L\'evy expansion and to show only these results. However, when applying a similar procedure to $p\bar p$ 
elastic scattering, it turned out that the range of the data around the dip position was too limited in this case, and the fourth-order L\'evy expansion terms could not be determined 
in a reliable and reasonable manner. So we decided to show only the fit results of the second- and third-order L\'evy expansions for the $p\bar p$ elastic scattering data.

As we explicitly demonstrate in Appendices C and D, in certain limited intervals of $|t|$, L\'evy fits without correction terms (i.e.~for $c_i = 0$, $i \ge 1$)
provided statistically not unacceptable, but in contrast, rather good quality fits and the corresponding confidence levels. These results suggest that the L\'evy
series is a reasonable representation of the scattering amplitude of elastic $pp$ and $p\bar p$ collisions.

The first four  orthogonal (but not yet normalized) L\'evy polynomials denoted as $L_i(z\,|\,\alpha)$ are found as follows
\begin{eqnarray}
 L_0(z\,|\,\alpha) & =&  1 , \\
 L_1(z\,|\,\alpha) & =& 
       \det\left(\begin{array}{c@{\hspace*{8pt}}c}
     \mu_{0,\alpha} & \mu_{1,\alpha}  \\ 
     1 & z \end{array} \right) , \\
 L_2(z\,|\,\alpha) & =& 
       \det\left(\begin{array}{c@{\hspace*{8pt}}c@{\hspace*{8pt}}c}
     \mu_{0,\alpha} & \mu_{1,\alpha} & \mu_{2,\alpha} \\ 
     \mu_{1,\alpha} & \mu_{2,\alpha} & \mu_{3,\alpha}  \\ 
     1 & z & z^2 \end{array} \right),\\
 L_3(z\,|\,\alpha) & =& 
       \det\left(\begin{array}{c@{\hspace*{8pt}}c@{\hspace*{8pt}}c@{\hspace*{8pt}}c}
     \mu_{0,\alpha} & \mu_{1,\alpha} & \mu_{2,\alpha} & \mu_{3,\alpha} \\ 
     \mu_{1,\alpha} & \mu_{2,\alpha} & \mu_{3,\alpha} & \mu_{4,\alpha} \\ 
     \mu_{2,\alpha} & \mu_{3,\alpha} & \mu_{4,\alpha} & \mu_{5,\alpha} \\ 
     1 & z & z^2 & z^3 \end{array} \right)\,, \quad \dots \;
     \mbox{\rm etc} \,,
\end{eqnarray}
where 
$$
\mu_{n,\alpha} = \int_0^\infty dz\;z^{n} \exp( - z^\alpha) = \frac{1}{\alpha}\,\Gamma\left( \frac{n+1}{\alpha}\right) \\
$$
and Euler's gamma function is defined as
\begin{equation}
	\Gamma(x) = \int_0^\infty dz\;z^{x-1}e^{-z} .
    \label{e:Gamma}
\end{equation}

The normalization of these L\'evy polynomials is straightforwardly expressed as follows:
\begin{eqnarray}
	l_j(z\, |\, \alpha) & = & D^{-\frac{1}{2}}_{j} D^{-\frac{1}{2}}_{j+1} 
    	L_j(z\, |\, \alpha), \qquad \mbox{\rm for}\quad j\ge 0 \,,
    \label{e:lj}
\end{eqnarray}
where $D_0 = 1$ and, in general, $D_j \equiv D_j(\alpha)$ stands for the Gram-determinant of order $j$, defined as
\begin{eqnarray}
D_0(\alpha) & =&  1 , \label{e:gram0}\\
D_1(\alpha) & = & \mu_{0,\alpha} , \label{e:gram1}\\
D_2(\alpha) & =& 
       \det\left(\begin{array}{c@{\hspace*{8pt}}c}
     \mu_{0,\alpha} & \mu_{1,\alpha}  \\ 
     \mu_{1,\alpha} & \mu_{2,\alpha} 
       \end{array} \right) , \label{e:gram2}\\
D_3(\alpha) & =& 
       \det\left(\begin{array}{c@{\hspace*{8pt}}c@{\hspace*{8pt}}c}
     \mu_{0,\alpha} & \mu_{1,\alpha} & \mu_{2,\alpha} \\ 
     \mu_{1,\alpha} & \mu_{2,\alpha} & \mu_{3,\alpha} \\ 
     \mu_{2,\alpha} & \mu_{3,\alpha} & \mu_{4,\alpha} 
       \end{array} \right),\label{e:gram3}\\
D_4(\alpha) & =& 
       \det\left(\begin{array}{c@{\hspace*{8pt}}c@{\hspace*{8pt}}c@{\hspace*{8pt}}c}
     \mu_{0,\alpha} & \mu_{1,\alpha} & \mu_{2,\alpha} & \mu_{3,\alpha} \\ 
     \mu_{1,\alpha} & \mu_{2,\alpha} & \mu_{3,\alpha} & \mu_{4,\alpha} \\ 
     \mu_{2,\alpha} & \mu_{3,\alpha} & \mu_{4,\alpha} & \mu_{5,\alpha} \\ 
     \mu_{3,\alpha} & \mu_{4,\alpha} & \mu_{5,\alpha} & \mu_{6,\alpha}
       \end{array} \right), \quad \dots     \;\mbox{\rm etc.} \label{e:gram4}
\end{eqnarray}

These normalized L\'evy polynomials $l_j(z|\alpha)$ are, as far as we know, newly introduced in this work, 
while the unnormalized L\'evy polynomials $L_j(z|\alpha)$ were introduced earlier in Ref.~\cite{DeKock:2012gp}.

\begin{figure}[t]
\begin{center}
\includegraphics[width=0.7\columnwidth]{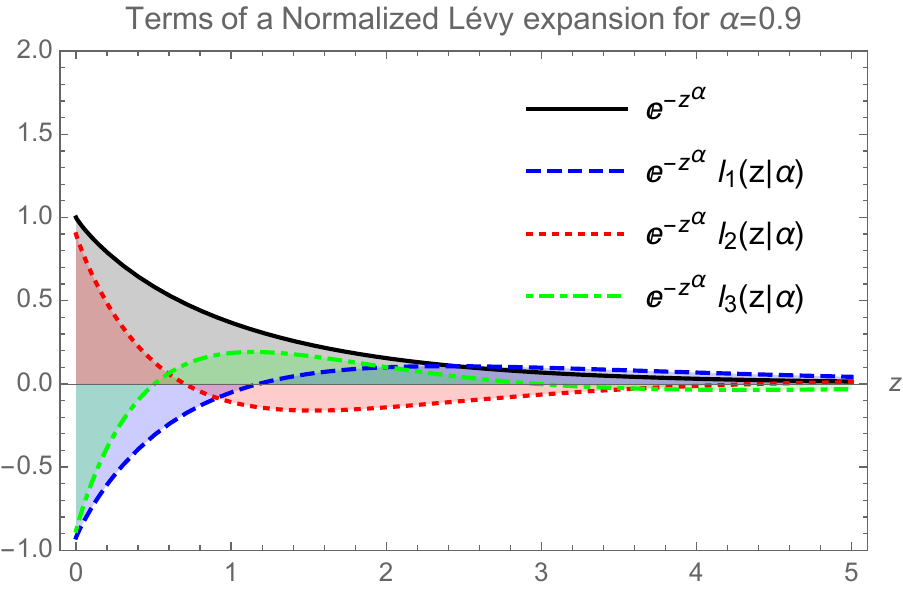}
\caption{Illustration of the four normalised L\'evy polynomials for $\alpha = 0.9$.}
\label{f:levy-expansion}
\end{center}
\end{figure}

The orthonormality of $\left\{l_j(z|\alpha)\right\}_{j=0}^{\infty}$ with respect to a L\'evy or stretched exponential 
weight is expressed by the following relation:
\begin{equation}
	\int_0^{\infty} dz \exp(-z^\alpha) l_n(z\,  |\, \alpha) l_m(z\,  |\, \alpha) = \delta_{n,m} \,.
	\label{e:orthonormal}
\end{equation}
The first few of these orthonormal L\'evy polynomials are illustrated in Fig.~\ref{f:levy-expansion}
for a specific value of $\alpha = 0.9$. More details of them, in particular the $\alpha = 2$ Gaussian, 
the $\alpha = 1$ Laguerre special cases and their explicit forms are being described in 
Refs.~\cite{Ster-BGL17,Novak-WPCF18}.

Once we have a statistically acceptable description of the differential cross-section of elastic $pp$ and $p\bar p$ collisions,
we can build the elastic scattering amplitude with the help of the L\'evy imaging method, and we can then compare the resulting 
shadow profiles of $pp$ and $p\bar p$ collisions without any model-dependent assumptions. Similarly, we can extract the $t$-dependence 
of the nuclear slope $B(t)$ directly from the data, as well as compare its behaviour for $pp$ collisions with that of $p\bar p$ collisions. 
The same is true for the nuclear phase and for the $t$-dependent $\rho$-parameter as well.

In fact, in our analysis we rely only on the convergence of the L\'evy series (or L\'evy expansion), that we have tested by the usual 
$\chi^2$-optimization methods with the CERN Minuit package and by evaluating the confidence level. 

\begin{figure}[!h]
\begin{center}
\includegraphics[width=0.85\columnwidth]{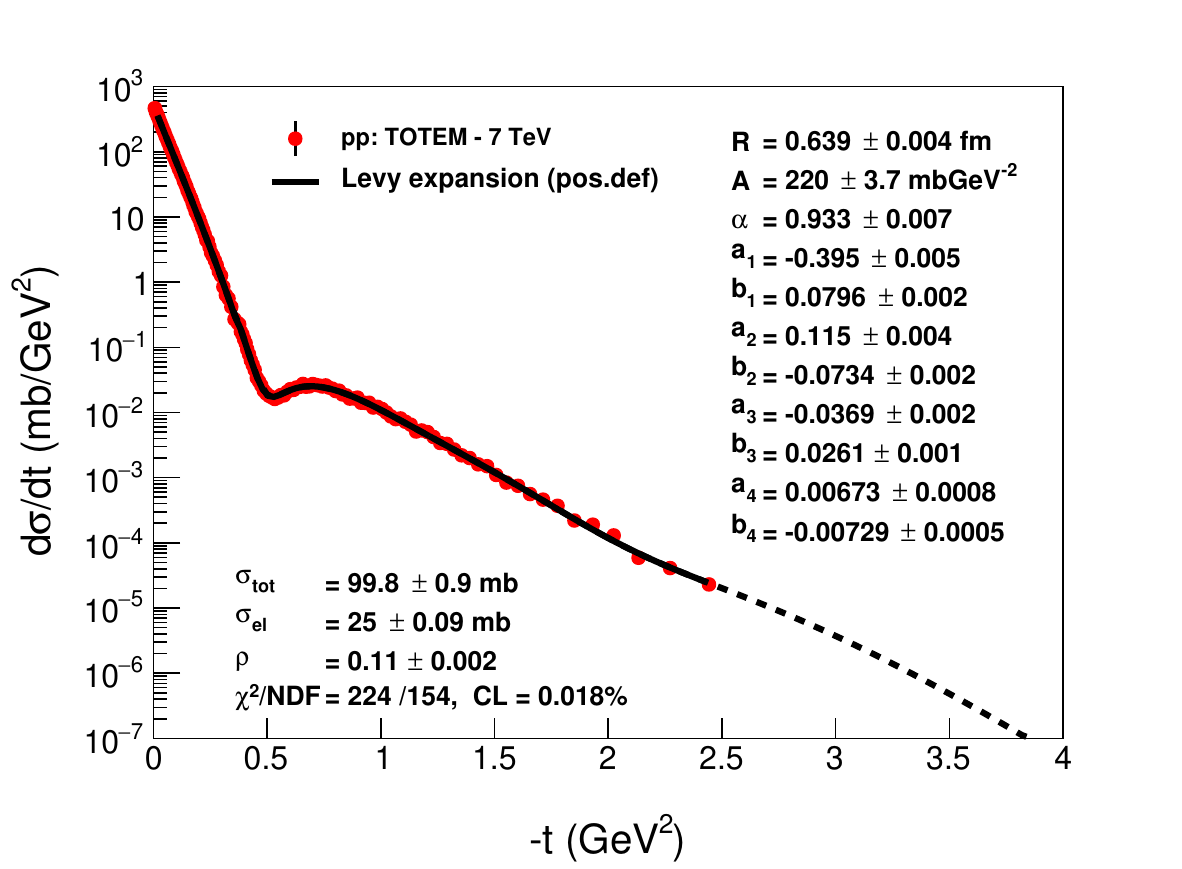}
\caption{Model-independent L\'evy expansion results from fits
        to elastic $pp$ scattering data by the TOTEM Collaboration at 
        the LHC energy of $\sqrt{s} = 7$ TeV. Although the fit quality is marginal,
        CL $\approx 0.02 \%$, the fitted curve follows the data so closely that
        we decided to interpret the fit parameters, noting that the errors 
        on the best values of the parameters are likely underestimated.
}
\label{f:dsigmadt-levyfit-7}
\end{center}
\end{figure}

Based on our experience with extracting $B(t)$, $\rho(t)$ and the shadow
profile $P(b)$ from $pp$ and $p\bar p$ elastic collision data, we can
definitely state that the precise reproduction of the measured data points,
with a statistically acceptable confidence level of CL $> 0.1 \%$ is a
necessary condition for interpreting our fit results. We have achieved such
good quality fits in many cases of the analysis of the published, final data,
except for the 7 TeV $pp$ elastic scattering data, where we reached a marginal
confidence level of CL $\approx 0.02 \%$, as indicated in
Fig.~\ref{f:dsigmadt-levyfit-7}. After scrutinizing this fit, presented in
Fig.~\ref{f:dsigmadt-levyfit-7}, we decided to interpret the parameters of this
fit as well. But in principle, in order to get the final errors of our
parameters, we may need to repeat the analysis by taking into account the full
covariance matrix. 

\begin{figure}[!h]
\begin{center}
\includegraphics[width=0.85\columnwidth]{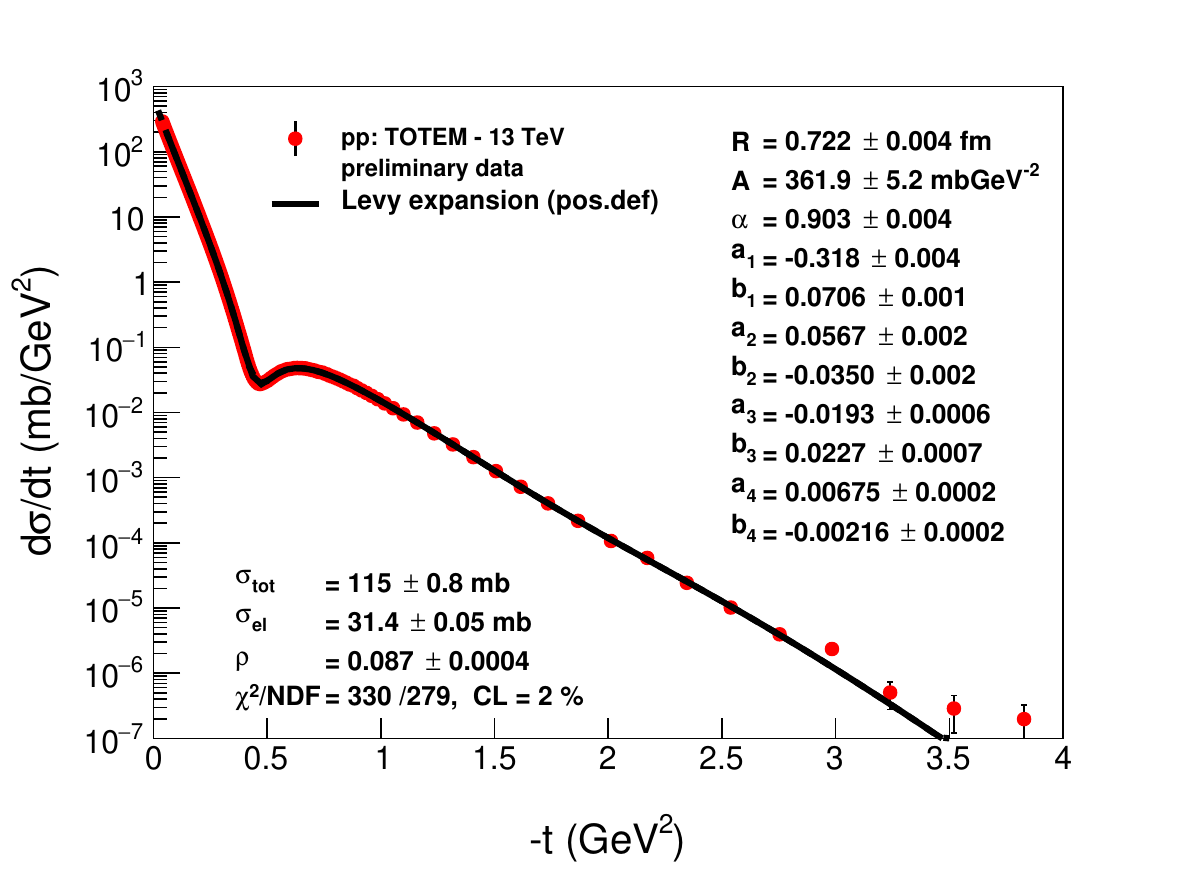}
\caption{Model-independent L\'evy expansion results from fits
        to the TOTEM preliminary elastic $pp$ scattering data at the currently largest 
        LHC energy of $\sqrt{s} = 13$ TeV. The errors on the fit parameters and the fit quality 
        are also preliminary.
}
\label{f:dsigmadt-levyfit-13}
\end{center}
\end{figure}

In case of the TOTEM preliminary 13 TeV data set, we also warn 
the reader that these data points and their errors are still in a preliminary phase, so we have determined the best preliminary values of
the parameters of the L\'evy series from the minimum of the $\chi^2$-distribution. The preliminary value of the confidence level of fits to 
the TOTEM preliminary data at $\sqrt{s} = 13$ TeV CL$ = 2 \%$, as indicated in Fig.~\ref{f:dsigmadt-levyfit-13}, satisfies the criteria 
for good quality fits.

\begin{figure}[!h]
\begin{center}
\includegraphics[width=0.85\columnwidth]{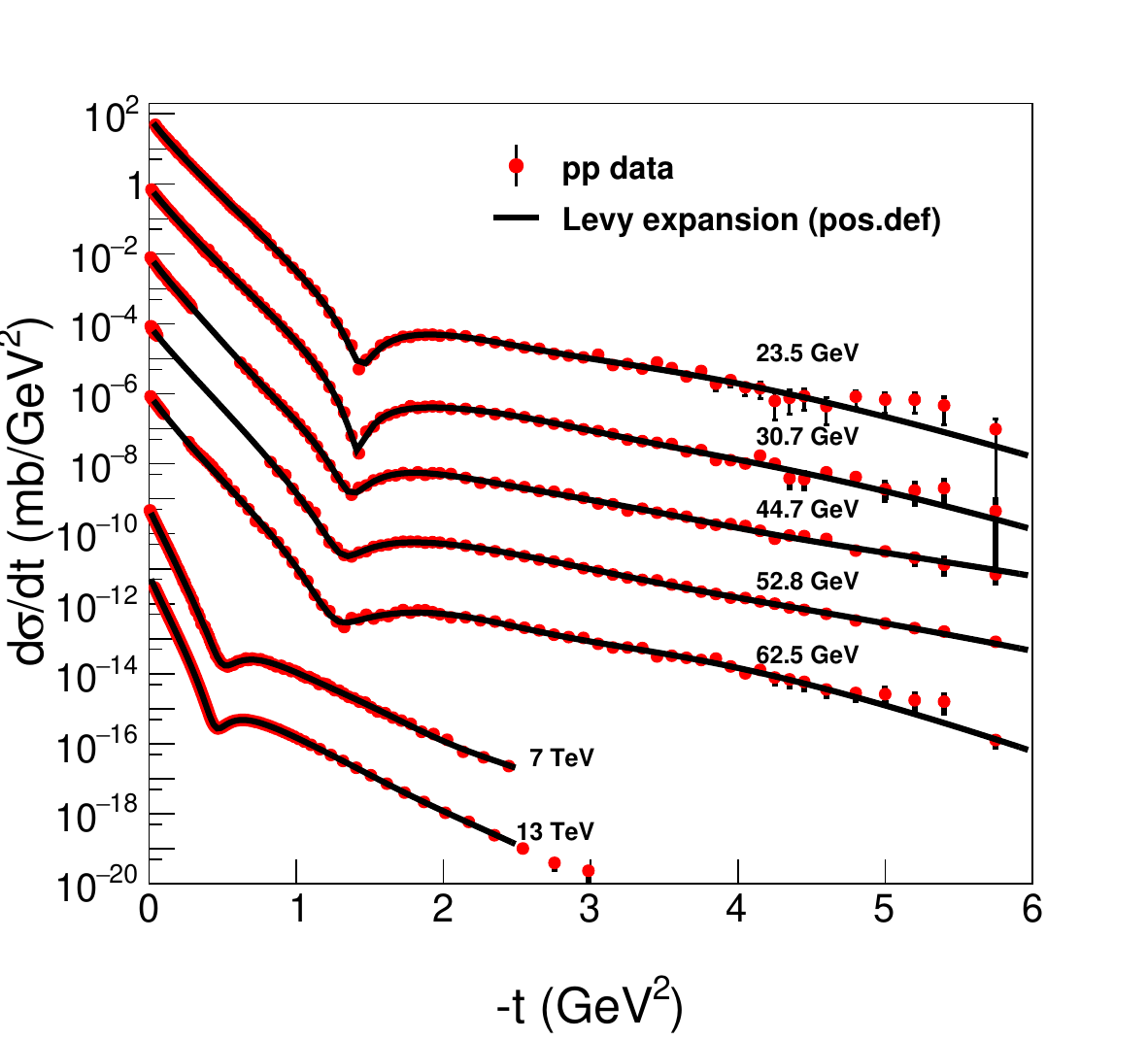}
\caption{Summary plot of the model-independent, fourth-order L\'evy expansion fits 
        to the elastic $pp$ scattering data at ISR and LHC energies ranging from
        from $\sqrt{s} = 23.5$ GeV up to $13$ TeV. 
        These fits are detailed in Appendix A.
}
\label{f:levy-fits-pp-all}
\end{center}
\end{figure}

Fig.~\ref{f:levy-fits-pp-all} summarizes the fits with a fourth-order L\'evy
expansion to all the differential cross-section measurements of elastic
proton-proton collisions from  $\sqrt{s} = 23.5 $ GeV up to 13 TeV.  These fits
are detailed in Appendix A, where the fits to each dataset are shown in detail,
with the fit parameters and confidence levels (or $p$-values) are listed on the
corresponding plots, together with the total and elastic cross-sections as well
as the $\rho(t=0) $ values that are calculated from the fit parameters.

\begin{figure}[!h]
\begin{center}
\includegraphics[width=0.85\columnwidth]{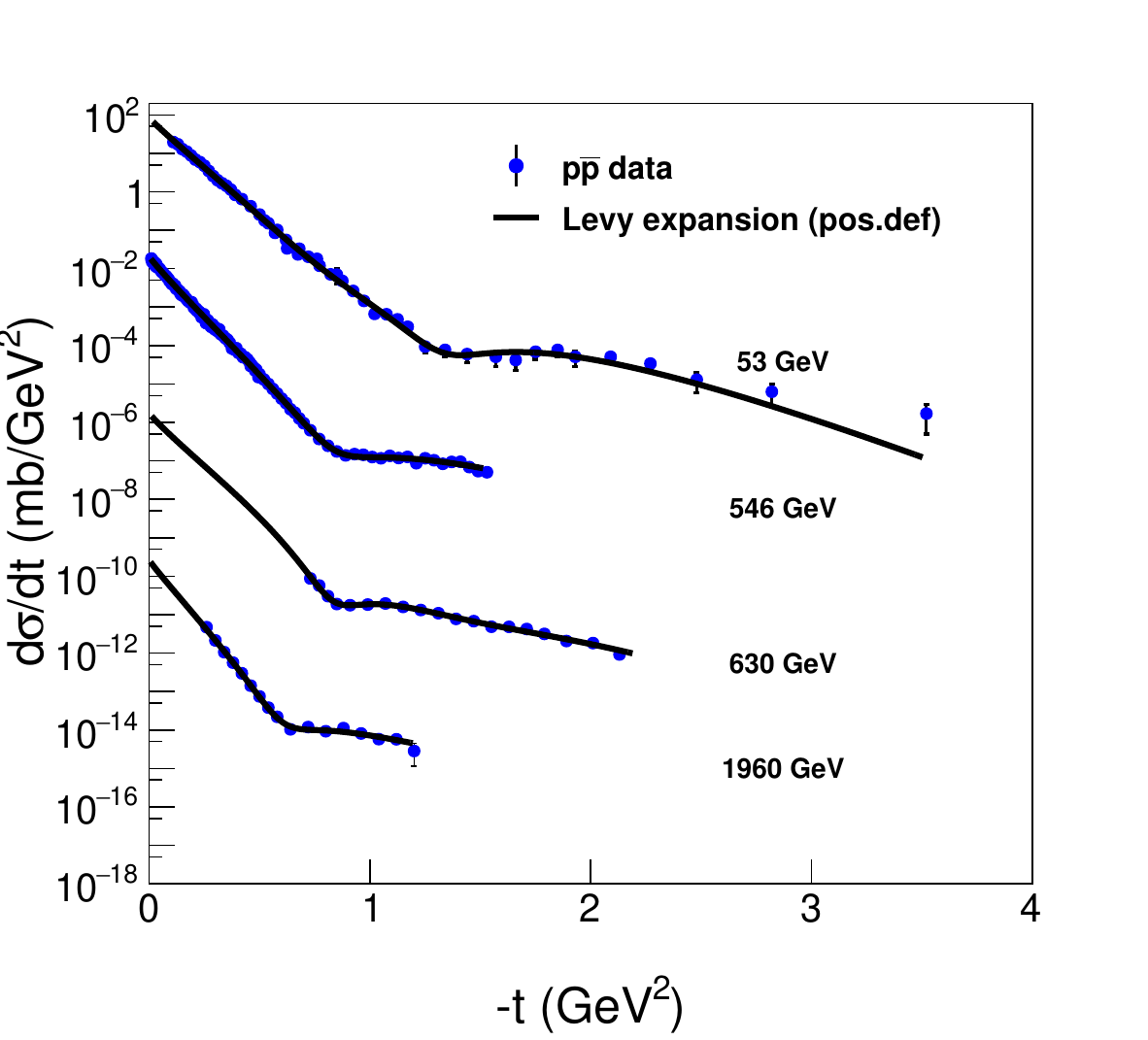}
\caption{Summary plot of the model-independent L\'evy expansion fits to 
the elastic $p\bar p$ scattering data at ISR, S$p\bar{p}$S, and Tevatron energies 
ranging from $\sqrt{s} = 53$ GeV up to $1.96$ TeV. The fits converged, error matrix 
was accurate and CL $\geq 0.1$ $\%$ for these fits, that were obtained with fixed $\alpha = 0.9$. 
These fits are detailed in Appendix B.
}
\label{f:levy-fits-ppbar-all}
\end{center}
\end{figure}

Similarly, Fig.~\ref{f:levy-fits-ppbar-all} summarizes the fits with a
third-order L\'evy expansion to all the differential cross-section measurements
of elastic proton-antiproton collisions from  $\sqrt{s} = 53 $ GeV up to 1.96
TeV.  The fits converged, error matrix was accurate and CL $\geq 0.1$ $\%$ 
for these fits, that were obtained with fixed $\alpha = 0.9$.
These fits are described in greater details in Appendix B,
where the fit parameters and confidence
levels (or $p$-values) are listed on the corresponding plots, together with the
total and elastic cross-sections as well as the $\rho(t=0) $ values that are
calculated from the fit parameters.

\begin{figure}[!h]
\begin{center}
\includegraphics[width=0.85\columnwidth]{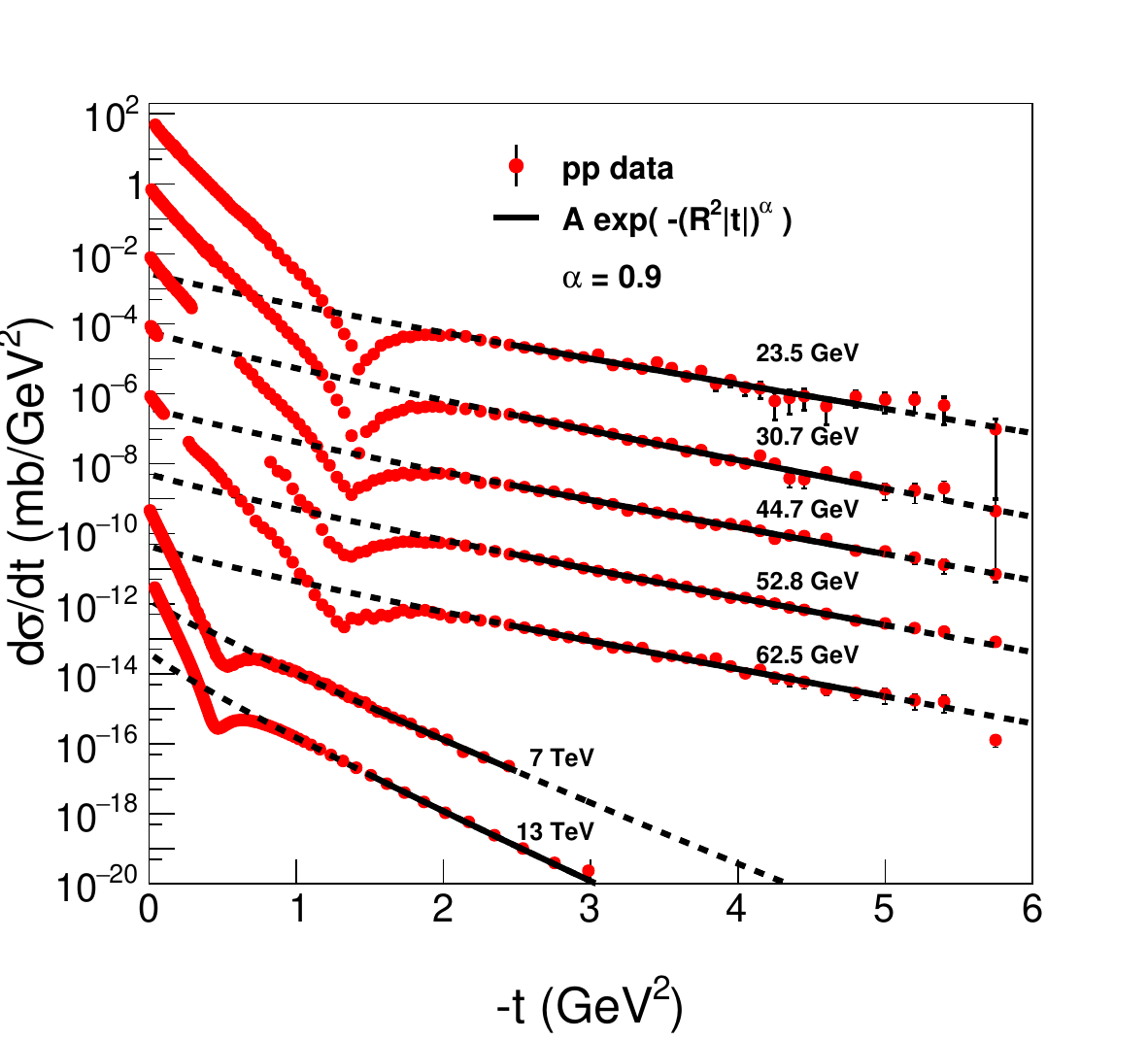}
\caption{Summary plot of the L\'evy fits, $d\sigma/dt = A \exp(-(R^2 |t|)^{\alpha})$
to the tails of the elastic $pp$ scattering data at ISR and LHC energies ranging
from $\sqrt{s} = 23.5$ GeV up to $13$ TeV, with $\alpha = 0.9$ fixed. 
Solid black lines indicate the fitted region, in each case the CL is 
in the acceptable $99.9 \% >$ CL $> 0.1 \%$ range.
Dashed line indicates an extrapolation outside the fitted region. These results 
are detailed in Appendix C and further explained in terms of a proton substructure 
in Subsection~\ref{ss:substructure-results}. The dashed lines continue the fitted, 
solid curves outside the fitted region, to improve the clarity of the presentation.}
\label{f:Summary-tails} 
\end{center}
\end{figure}

Fig.~\ref{f:Summary-tails} represents the summary plot of the L\'evy fits, 
$	d\sigma/dt = A \exp(-(R^2 |t|)^{\alpha})\,, $ that correspond to the
zeroth-order of the L\'evy expansion detailed in this manuscript, to the {\it
tails} of the elastic $pp$ scattering data at ISR and LHC energies from
$\sqrt{s} = 23.5$ GeV up to $13$ TeV, with $\alpha = 0.9$ fixed. These results
are detailed in Appendix C, and are explained in terms of the newly identified
proton substructure in Subsection~\ref{ss:substructure-results}. Namely, rather
elegantly and clearly, a smaller substructure is seen in the ISR energy range, that is
invariant for the (relatively small) change of $\sqrt{s}$, as evidenced by the
dashed lines that (except an overall normalization factor) follow the same
curves. It is apparent from the visualization of Fig.~\ref{f:Summary-tails},
that at $\sqrt{s} =$ 7 and 13 TeV, the slope of these dashed lines changes
dramatically: a proton substructure of a different size is found that is
apparently (within the errors) the same both at 7 and at 13 TeV.

\begin{figure}[!h]
\begin{center}
\includegraphics[width=0.85\columnwidth]{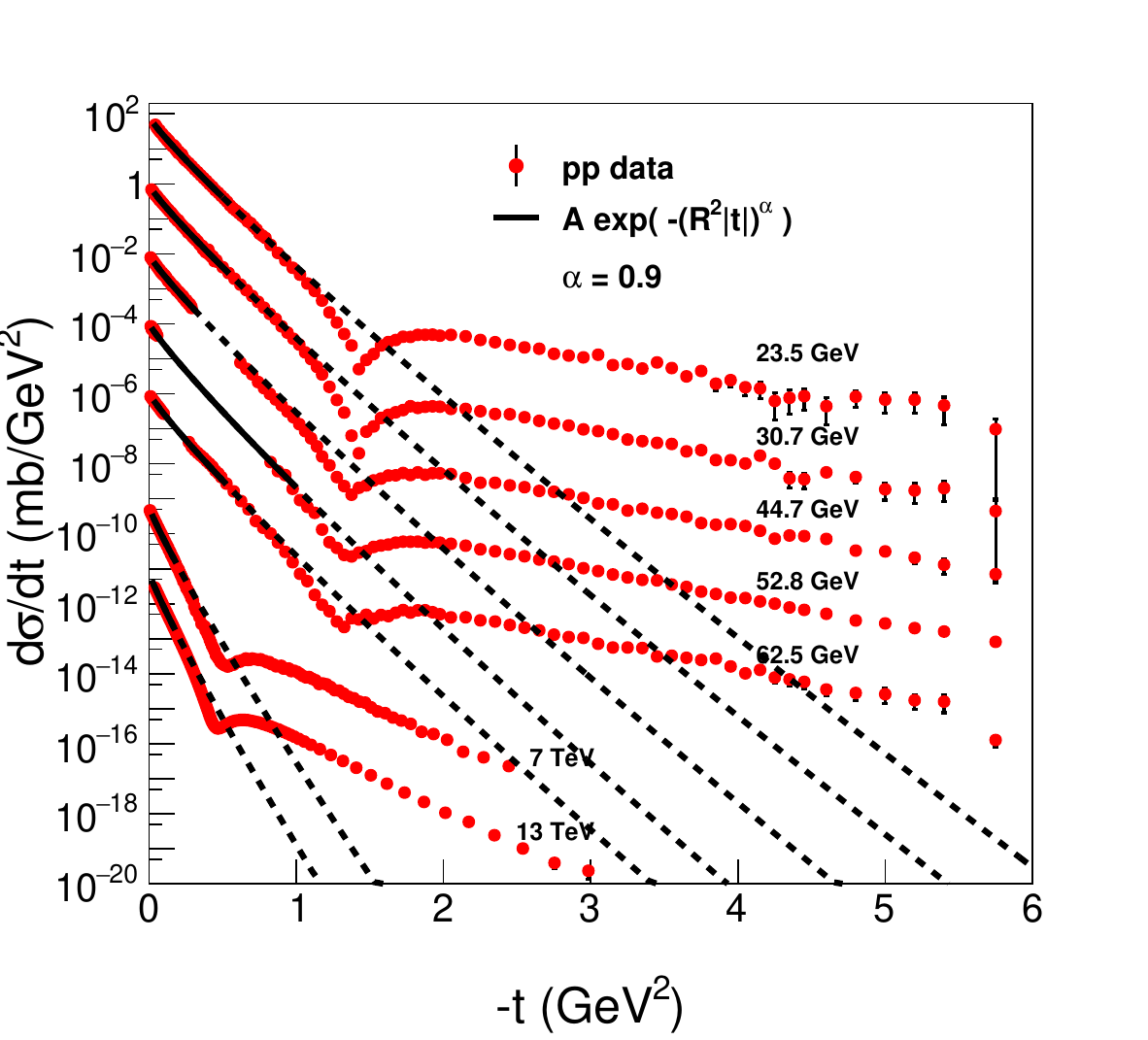}
\caption{Summary plot of the L\'evy fits, $d\sigma/dt = A \exp(-(R^2 |t|)^{\alpha})$ 
to the cone (or low-$|t|$) region of the elastic $pp$ scattering data at ISR and LHC energies
ranging from $\sqrt{s} = 23.5$ GeV to $13$ TeV, with $\alpha = 0.9$ fixed.
The dashed lines continue the fitted, solid curves outside the fitted region,
to improve the clarity of the presentation. These fits are detailed and described 
individually in Appendix D.
}
\label{f:Summary-cone}
\end{center}
\end{figure}

Fig.~\ref{f:Summary-cone} represents a summary plot of the L\'evy fits,
$d\sigma/dt = A \exp(-(R^2 |t|)^{\alpha})$ to the {\it cone} or low-$|t|$ part
of the elastic $pp$ scattering data at ISR and LHC energies from $\sqrt{s} =
23.5$ GeV up to $13$ TeV, with $\alpha = 0.9$ fixed.  All the low-$t$ ISR fits
are successful with the Levy leading order result where there are sufficient
data sets at low-$t$, except those data at $\sqrt{s}=44.7$ and $52.8$ GeV which
exhibit a gap in the measured dataset at low values of $|t|$. These results are
detailed in Appendix D. The gradual steepening of the slope of the fitted
curves indicate, that the L\'evy scale $R$, characterizing the overall size of
the proton, was increasing monotonically with increasing energy $\sqrt{s}$.
This behaviour can be explained in terms of the proton size growing
self-similarly in the ISR energy range, as evident also from the excitation
function of the shadow-profiles at 23.5 $\leq \, \sqrt[]{s} \, \leq\, 62.5$ GeV. This
effect is discussed in greater details in Subsection~\ref{ss:shadow-results}.

However, at $\sqrt{s} = 7$ and 13 TeV, the leading-order Levy fits for fixed
$\alpha = 0.9$ have failed, indicating that not only the size of the proton
increases with an increase of collision energies, but also the shape of the
protons changes. A successful fit in this case is possible only if the first
few data points are taken only into account. 
Note, that  any data in the Coulomb-nuclear interference (CNI) region, i.e. at $|t| < 0.01$
GeV$^2$, were not included in these fits, thus one may  neglect any possible 
CNI effect when interpreting the results of the Levy expansion.

The observation that the impact parameter $b$
dependent shadow profile function $P(b)$ at 7 and 13 TeV deviates from a Levy shape even in the leading order 
confirms that the shape of protons changes at LHC energies in a way which 
is different from that at ISR energies.
These results, detailed in  Subsection~\ref{ss:shadow-results},
can be explained in terms of the evolving shape of the shadow
profiles at 7 and 13 TeV which exhibits plateaux near $b = 0$ (which were not
seen at lower energies). As shown in Subsection~\ref{ss:shadow-results},
one observes a saturation of the shadow profile functions $P(b) \approx 1$ in the $b
\leq 0.4-0.5$ fm region at 7 and 13 TeV, respectively.

\subsection{\it Differential, total and elastic cross-sections}
\label{ss:cross-sections}

The conventional form of the elastic differential cross section
\begin{eqnarray}
\frac{d\sigma}{dt} = \frac{1}{4\pi}|T_{el}(\Delta)|^2 \,, \qquad \Delta=\sqrt{|t|}\,, 
\label{e:dsigmadt-Tel}
\end{eqnarray}
provides us with the key expression for the complex-valued elastic scattering amplitude 
$T_{el}(\Delta)$ in terms of a L\'evy series
\begin{eqnarray}
T_{el}(\Delta) &=& i\sqrt{4\pi A}\, \exp\left({-\frac{1}{2} z^\alpha}\right)\, 
		\left[1+\sum_{i = 1}^\infty c_i l_i (z|\alpha) \right] \, , 
        \label{e:Tel}\\
        z &=& \Delta^2 R^2 \, = \, |t| R^2 \, .
\end{eqnarray}

Then, according to the optical theorem, the total cross section is found as
\begin{equation}
\sigma_{\rm tot} \equiv 2\,{\rm Im}\, T_{el}(\Delta=0)=2\,\sqrt{4\pi A}\,
	\left(1 + \sum_{i = 1}^\infty a_i l_i (0|\alpha) \right)\,,
    \label{e:sigmatot}
\end{equation}
while the ratio of the real to imaginary parts of the elastic amplitude
\begin{equation}
\rho(t)\equiv \frac{{\rm Re}\, T_{el}(\Delta)}{{\rm Im}\, T_{el}(\Delta)} =
	- \left. \frac{\sum_{i = 1}^\infty b_i l_i (z|\alpha)}{1+\sum_{i = 1}^\infty a_i l_i (z|\alpha)}\right|_{z= t R^2}
    \label{e:rho}
\end{equation}
is known as the $\rho$-parameter, in consistency with the traditional form of the forward elastic differential cross section
\begin{eqnarray}
\frac{d\sigma}{dt}\Big|_{t\to 0}=\frac{(1+\rho_0^2)\sigma_{\rm tot}^2}{16\pi}\,, \qquad \rho_0=\rho(t=0)\,.
\label{e:optical-point}
\end{eqnarray}

\subsection{\it Four-momentum transfer dependent elastic slope $B(t)$}
\label{ss:Bt}

The $t$-dependent elastic slope $B(t)$ is traditionally defined as
\begin{eqnarray}
	B(t) \equiv \frac{d}{dt}\left(\ln \frac{d\sigma}{dt} \right)\,.
	\label{e:Bslope}
\end{eqnarray}
There is a great current interest in the value of this function at $t= 0$ at
LHC energies. Traditionally, the elastic slope is determined as $B = B(t=0)$.
Let us note that this requires an extrapolation of the measured differential
cross-sections to the $t=0$ optical point. Frequently, an exponential
approximation is applied, however, at the very low-$|t|$ region the
CNI complicates such an extrapolation as
well~\cite{Antchev:2015zza}.

At this point, it is important to emphasize, that the L\'evy series utilized in
this paper to represent the elastic scattering amplitude is not an analytic
function at $t = 0$ if $\alpha < 1$, hence formally our expressions for $B$ may
not exist, as $B(t)$ is well-defined only for $|t| >0$ in this case. However,
if $\alpha = 1$, the cone region decreases exponentially and the elastic
scattering amplitude becomes an analytic function at $t = 0$. Hence, it is very
important to determine the value of $\alpha$ precisely from the analysis of the
elastic differential cross-section data.

Note that $B$ is related to the root-mean-square (RMS) radius of the
impact-parameter $b$-dependent elastic amplitude $t_{el}(b)$.  It is well known
that for L\'evy-stable source distributions, that are our zeroth-order choices
for the impact parameter dependent elastic amplitudes, the RMS of the source is
divergent, if the L\'evy index of stability $\alpha_L < 2$
\cite{Weisstein:Stable}, with the exception of the Gaussian case, corresponding
to $\alpha_L = 2$, when the RMS of the source is finite.
Due to the importance
of this point, we have dedicated Subsection~\ref{ss:Levy-example}
to compare the differential cross-section of L\'evy-stable sources
with Gaussian sources and dedicated 
Appendices C and  D to investigate if such a
non-analytic model at $t=0$ as the L\'evy-stable source distribution
$t_{el}(b)$ can describe reasonably well the $pp$ elastic differential
cross-section data in limited kinematic regions.
Actually, we find that this is indeed a very good
approximation to the  data at low values of $|t|$ in the ISR energy range of 23.5 $\leq \, \sqrt{s} \,
\leq$ 62.5 GeV, as explicitly demonstrated with $\alpha = 0.9$ fixed fits in
Appendix D. Let us note here that the cone region of TOTEM data at $\sqrt{s} =$
7 and 13 TeV can also be described in the same cone (or low-$|t|$) region only approximately,
if the parameter $\alpha$ is released, corresponding to a change of the proton
shape in the TeV energy range.

For more detailed examples and for the introduction of L\'evy stable source
distributions to femtoscopy in high-energy particle and nuclear physics, with
an emphasis at their non-analytic nature of their Fourier-transform, we
recommend Ref.~\cite{Csorgo:2003uv}.

\subsection{\it Four-momentum transfer dependent nuclear phase $\phi(t)$}
\label{ss:phi-t}

The nuclear phase $\phi(t)$ can also be introduced at this point. 
A conventional definition of this phase $\phi(t)$ is referring to it as
the phase of the elastic scattering amplitude in the complex plane, as follows:
\begin{equation}
T_{el}(t) = |T_{el}(t)| \exp( i \phi(t) ) \,.
\label{e:phi-complex}
\end{equation}
An alternative definition was used recently by the TOTEM Collaboration, 
that related $\phi(t)$ to the $t$-dependent $\rho(t)$ parameter as
\begin{equation}
\phi_2(t) = \frac{\pi}{2} - \arctan\rho(t) \,.
\label{e:phiprincipal}
\end{equation}
These definitions are equivalent if the nuclear phase satisfies $0 \le \phi(t)\le \pi$. The mathematical definition
of the inverse tangent function $\arctan(x)$  is as follows.
For a real number $x$, $\theta = \arctan(x)$ represents the radian angle measure $\theta$ with  
$- \frac{\pi}{2} < \theta < \frac{\pi}{2}$ such that $\tan(\theta) = x$. Thus by definition,
$\phi_2(t) $ of eq.~(\ref{e:phiprincipal}) 
satisfies $ 0 \le \phi_2(t) < \pi$ : it stands for the principal value of the nuclear phase $\phi(t)$.
Given that for complex arguments, $\arctan(z)$ has branch cut discontinuities on the complex plane
the principal value and the continuous definitions of the nuclear phase with the help of Eqs.~(\ref{e:phiprincipal}) and (\ref{e:phi-complex})
are, in general,  inequivalent.

It is also interesting to evaluate the total elastic cross-section, $\sigma_{el}$. In this calculation, we utilize
the orthonormality of the L\'evy polynomials $l_n(z|\alpha)$ to obtain
\begin{equation}
     \sigma_{el} = \int_{0}^\infty d|t| \frac{d\sigma}{dt} =
     \frac{A}{R^2}\left[\frac{1}{\alpha} \Gamma\left(\frac{1}{\alpha}\right) + \sum_{i= 1}^\infty (a_i^2 + b_i^2) \right]\,.
     \label{e:sigmael}
\end{equation}

One may argue that the overall phase $\chi(z)$ of the elastic amplitude (\ref{e:Tel}) is not constrained by the fits of the elastic 
cross section data. Indeed, an overall phase factor may in principle cause a redefinition of the complex expansion coefficients 
in the amplitude as
\begin{eqnarray*}
\tilde{T}_{el}(\Delta) &=& i\sqrt{4\pi A}\, \exp\left({-\frac{1}{2} z^\alpha}\right)\, 
		\left[e^{i\chi(z)}+\sum_{i = 1}^\infty \tilde{c}_i(z) l_i (z|\alpha) \right]  \,,
        \label{e:Tel-phase}
\end{eqnarray*}
such that the new coefficients $\tilde{c}_i(z)=\exp(i\chi(z))c_i$ could not be uniquely determined by the fits with the L\'evy expansion. 
However, under certain additional assumptions, the L\'evy expansion and the reconstructed phase is unique, as outlined below.

By means of the optical theorem, the measurable total cross section uniquely determines the imaginary part of the elastic 
amplitude at $t= 0$, as indicated by Eq.~(\ref{e:sigmatot}). Measurements of the nuclear phase at $t=0$ use the CNI region to determine $\rho(t=0)$ with typical values at LHC of the order of 0.1.
This implies that not only the imaginary but also the real part of the forward scattering amplitude can also be uniquely determined
at the optical point, $t=0$,
not allowing for an arbitrary phase at $z=0$: the value of $\phi(t=0)$ is  fixed by measurements  unambiguously. 
One can also determine the nuclear phase at the dip position $t_{\rm dip}$, where the differential cross-section of elastic proton-proton collisions has a diffractive minimum. At this point, the imaginary part of the elastic scattering amplitude approximately vanishes, hence the differential cross-section measures the real part of the forward scattering amplitude and the nuclear phase is an integer multiple of $\pi$.
Under the additional assumption that the nuclear phase $\phi(t)$ is an analytic function of $t$, it can be continued with the help of our expansion method from its known value to arbitrary values of $0 \le -t < \infty$.
In the subsequent applications, the nuclear phase $\phi(t)$ can thus be uniquely determined, if a well-defined diffractive minimum is included
to the measured region, due to the following reasons: The arbitrary phase $\chi(z)$, based on the construction of 
Eq.~(\ref{e:Tel}), has to have a vanishing initial value both at small $z$, i.e. $\chi(z)\to 0$ for $z\to 0$, as well as at the value of $z_{\rm dip}$ that corresponds to $t_{\rm dip}$. 

If a non-vanishing $\chi(z)$ emerges at $z>0$, that would apparently destroy the orthonormality of the Levy terms (\ref{e:orthonormal}) 
in the expansion for the amplitude, given that  
\begin{equation}
	\int_0^{\infty} dz \cos\chi(z) \exp(-z^\alpha) l_n(z\,  |\, \alpha) l_m(z\,  |\, \alpha) \not= \delta_{n,m} \,.
	\label{e:orthonormal-phase}
\end{equation}
which holds, by construction, for $\chi(z) = 0$.  Actually, the condition of the
orthonormality may only be restored for functions that satisfy two conditions:
$\cos(\chi(z))=1$, and $\chi(z=0) = 0$. There are infinitely many piecewise
continuous functions that satisfy both requirements, by jumping between 0 and $2\pi$ at various values of $z$,
however, the requirement that the arbitrary phase $\chi(z)$ is a continuous function of its argument $z$,
uniquely determines that $\chi(z) \equiv 0$.
One may also observe that the overall normalization coefficient $A$ and the zeroth order expansion coefficient $a_0+ i b_0$
would appear in the fits with Eq.~(\ref{e:levy-dsigmadt}) in a product form only.
By absorbing these possible zeroth order  constants into the overall normalization constant $A$,
the zeroth order expansion simplifies as $A \exp(-z^\alpha)$,
that can be uniquely fitted to the data points. Furthermore all the fit parameters of Eq.~(\ref{e:levy-dsigmadt}) can be uniquely determined under these above listed conditions, that imply that $\chi(z=0) \equiv 0$.

This proof indicates, that with the help of the L\'evy expansion method, the
nuclear phase of the elastic scattering amplitude can be unambiguously
determined, if this nuclear phase is measurable at  $t= 0 $ (as evident from
total cross-section and $\rho_0$ measurements at $t=0$) and if it is assumed to
be a continuous function of its argument $t$ (apart from branch cuts, where
uniqueness requires that we specify on which branch we define the value of the
nuclear phase). 

A posteriori, our method is validated by the excellent reproduction of the $\rho_0$ value measured in elastic $pp$
collisions at $\sqrt{s} = 13$ TeV by the TOTEM Collaboration~\cite{Antchev:2017yns} using the data in the CNI region.
Fits detailed in Appendices A and B indicate that indeed the 4th
and 3rd order L\'evy expansions reproduce well measured values of $\rho_0$
not only at $\sqrt{s} = 13 $ TeV but at lower energies as well, 
if the fit quality is satisfactory. Results summarized in Appendices  C and D
also indicate that the zeroth order L\'evy fits are not suitable to determine
the nuclear phase and $\rho_0$: one has to measure an interference to extract
information about the phase, which interference is natural to find in the
dip-bump region of elastic proton-proton (or, proton-antiproton) reactions.
Note that this proof however does not extend to the investigation of the convergence properties
of the $\phi(t) $ measurements. We recommend to investigate the stability of the 
reconstructed $\phi(t)$ by fitting the data with higher and higher order L\'evy expansions
and to look for the numerical convergence of the reconstructed $\phi(t)$ functions 
and to determine the domain of convergence also numerically.

\subsection{\it Shadow profile functions}
\label{ss:shadow-intro}

Turning to the impact parameter space, we get
\begin{eqnarray}\nonumber
	t_{el}(b)&=&\int \frac{d^2\Delta}{(2\pi)^2}\, e^{-i{\bm \Delta}{\bm b}}\,
	T_{el}(\Delta)\\
	&=&\frac{1}{2\pi}\int J_0(\Delta\,b)\,T_{el}(\Delta)\,\Delta\, d\Delta \,,	\label{tel-b}\\ 
	\Delta&\equiv&|{\bm \Delta}|\,, \quad b\equiv|{\bm b}|\,, \nonumber
\end{eqnarray}
This Fourier-transformed elastic amplitude $t_{el}(b)$ can be represented in the eikonal form
\begin{eqnarray}
	t_{el}(b)=i\left[ 1 - e^{-\Omega(b)} \right] \,,
	\label{e:tel-eikonal}
\end{eqnarray}
where $\Omega(b) $ is the so-called opacity function, which is in general complex. Thus, a statistically acceptable description of the elastic scattering data
provides us with a direct access to the opacity $\Omega(b)$ (known also as the eikonal function) and, in particular, to the shadow profile function defined as
\begin{eqnarray}
	P(b) = 1-\left|e^{-\Omega(b)}\right|^2 \,.
    \label{e:shadow}
\end{eqnarray}

\subsection{\it A simple example -- Gaussian versus L\'evy stable source distributions}
\label{ss:Levy-example}

To gain intuition about the meaning of the characteristic Levy scale parameter of the proton $R$, before we go deeper to the data analysis, let us consider first the usual 
$\alpha = 1$ case, neglecting all but the leading order (unity) term in the series that defines the L\'evy expansion of the differential cross-section in Eq.~(\ref{e:levy-dsigmadt}).

In this Gaussian case, the differential cross section is apparently a structureless exponential in $t$,
\begin{equation}
	\frac{d\sigma}{dt} = A \exp( - R^2 |t|)\,,
    \label{e:Gaussian}
\end{equation}
that corresponds to a Gaussian parametrization of the elastic scattering amplitude, based on Eq.~(\ref{e:dsigmadt-Tel}):
\begin{equation}
t_{el}(b) \propto \exp\left(-\frac{b^2}{2 R^2}\right)\,.
\end{equation}

The Gaussian distributions correspond to central limit theorems, when several
random elementary processes are convoluted to yield the final distribution. The
Gaussian appears as a limiting distribution, if the elementary processes have
finite means and variances, regardless of further details of the elementary
probability distributions. Generalized central limit theorems describe limiting
distributions for a large number of elementary processes, when the resulting
elementary distributions have infinite means or variances. In these cases a
limiting distribution exists, that remains stable for adding one more random
elementary process. Due to this reason, such distributions are called stable,
or, L\'evy-stable distributions. They are denoted by $S_n (x | \alpha_L, \beta,
\gamma, \delta)$ where $x$ is the variable of the distribution, $\alpha_L$
stands for the L\'evy index of stability, $\beta$ is the asymmetry parameter,
$\gamma$ is the so called scale parameter, and $\delta$ is the location
parameter of this distribution, while $n$ determines the
convention~\cite{Nolan:2016st,Weisstein:Stable,Csorgo:2003uv}.  In this paper, we follow the
conventions defined in Ref.~\cite{Csorgo:2003uv}, that correspond to $n= 2$. In
this case, for $\alpha \ne 1$, the Fourier-transformed L\'evy stable source
distribution reads as

\begin{eqnarray} \nonumber
\tilde S_2(q |\alpha_L, \beta, \gamma, \delta) &=& \exp\Big(i q \delta - \gamma^{\alpha_L} q^{\alpha_L} \\ 
&\times&
\left[ 1 - i \beta \, \mbox{\rm sgn}(u) \tan(\frac{1}{2} \pi \alpha_L)\right]\Big) \,.
\end{eqnarray}

By now, the L\'evy stable distributions are implemented into commercially
available software packages, for example, Mathematica~\cite{Weisstein:Stable}.
In practice, we utilize the parameterization that is continuous in the L\'evy
index of stability $\alpha_L$~\cite{Weisstein:Stable}.  Given that the
Gaussians correspond to L\'evy-stable source distributions with $\alpha_L = 2$
(the value of the exponent in the Fourier-transformed Gaussians) and taking
into account, that in our analysis the Gaussian elastic amplitude $t_{el}(b) $
has the exponent $\alpha = 1$, we conclude that the L\'evy index of stability
$\alpha_L$ is simply twice the exponent of our L\'evy series, i.e.

\begin{equation}
 \alpha_{L} = 2 \alpha \ .
\end{equation}

Recently, the TOTEM Collaboration published a high precision measurement of the
low-$|t|$ region of the differential cross-section of elastic $pp$ scattering
at $\sqrt{s} = 8$ TeV~\cite{Antchev:2015zza}. This demonstrated a significant,
more than 7$\sigma$ deviation from a simple exponential cone behaviour,
corresponding to a Gaussian representation of the elastic scattering amplitude.
In our language, this means that $\alpha < 1$, or, using the standard form of
the L\'evy index of stability, $\alpha_{L} = 2 \alpha < 2$ for this data set.

Subsequently, let us present the results of our data analysis and indicate the
best L\'evy expanison fits to elastic $pp$ and $p\bar p$ differential
cross-sections. Let us proceed to evaluate the shadow profiles $P(b,s)$ and the
slope parameters $B(s,t)$ as well as the nuclear phases $\phi(s,t)$ for the
available values of $s$ and for both proton-proton and proton-antiproton
elastic scattering reactions, to find their excitation functions, and to
compare proton-proton and proton-antiproton results, as described in the next
section.

\section{Data analysis}
\label{s:Data-analysis}

Let us test the power of our L\'evy expansion method first on the already
published differential cross-section data of elastic $pp$ collisions at
$\sqrt{s} = 7 $ TeV, utilizing the published TOTEM data set of
Ref.~\cite{Antchev:2013gaa}. Fig.~\ref{f:dsigmadt-levyfit-7} indicates that the
7 TeV TOTEM data set can be represented by a fourth-order L\'evy expansion with
a reasonable $\chi^2 / {\rm NDF} = 224/154$, that corresponds to a marginal
confidence level of CL $\approx 0.02 \%$.  Inspecting
Fig.~\ref{f:dsigmadt-levyfit-7} by eye suggests also that the parameters of the
L\'evy expansion in Eq.~(\ref{e:levy-dsigmadt}) can be interpreted as they
closely represent the data. These parameters are printed on the right-hand side
of the top panel of
Figs.~\ref{f:dsigmadt-levyfit-7}--\ref{f:dsigmadt-levyfit-13}.

Let us also investigate in detail the recently released, new 13 TeV TOTEM
preliminary data set~\cite{Frici-la-Biodola:2018vvv}, to look for crossing (C)
odd effects in the comparison of elastic $pp$ and $p\bar p$ collisions. In what
follows, we consider four different aspects of the TOTEM data in comparison
with elastic scattering data at lower energies, both for $pp$ and $p\bar p$
collisions. Namely, we compare the shadow profile functions, the $t$-dependence
of the elastic slope parameter $B$, the same for the $\rho$-parameter and the
so-called nuclear phase $\phi(t)$, that measures the argument of the elastic
scattering amplitude. Finally, we show in a simple and straightforward analysis
of a large-$|t|$ region beyond the diffractive minimum and maximum, that the
differential cross-section of elastic $pp$ scattering evidences a proton
substructure of two distinct sizes for GeV and TeV energy ranges, respectively.

\subsection{\it Looking for Odderon effects}

As noted in Refs.~\cite{Jenkovszky:2011hu,Ster:2015esa}, the only direct way to see the Odderon is by comparing the particle and antiparticle scattering at 
sufficiently high energies provided that the high-energy $pp$ and $p\bar p$ elastic scattering amplitude is a difference or a sum of even and odd C-parity contributions.
The even-under-crossing part consists of the Pomeron and the $f$ Reggeon trajectory, while the odd-under-crossing part contains the Odderon and a contribution 
from the $\omega$ Reggeon, i.e.
\begin{eqnarray}
T_{el}^{pp}(s,t) & = & T_{el}^{+}(s,t) + T_{el}^{-}(s,t), \\
T_{el}^{p\overline{p}}(s,t) & = & T_{el}^{+}(s,t) - T_{el}^{-}(s,t) , \\
 T_{el}^{+}(s,t) & = & T_{el}^{P}(s,t) +T_{el}^{f}(s,t),\\
 T_{el}^{-}(s,t) & = & T_{el}^{O}(s,t) +T_{el}^{\omega}(s,t) \,.
\end{eqnarray}
It is clear from the above formulae that the odd component of the amplitude can be extracted from the difference of the $p\bar p$ and the $pp$ scattering amplitudes. 
At sufficiently high energies, the relative contributions from secondary Regge trajectories is suppressed, as they decay as negative powers of the colliding energy $\sqrt{s}$. 
The vanishing nature of these Reggeon contributions offers a direct way of extracting the Odderon as well as the Pomeron contributions, $T_{el}^{O}(s,t)$ and $T_{el}^{P}(s,t)$,
respectively, from the elastic scattering data at sufficiently high colliding energies.

In Refs.~\cite{Ster:2015esa}, the authors argued that the LHC energy scale is already sufficiently large to suppress the Reggeon contributions,
and they presented the $(s,t)$-dependent contributions of an Odderon exchange to the differential and total cross-sections at LHC energies. That analysis, however, 
relied on a model-dependent, phenomenological extension of the Phillips-Barger model~\cite{Phillips:1974vt} and focussed on fitting the dip region
of elastic $pp$ scattering, but it did not analyse in detail the tail and cone regions. In fact, that analysis relied heavily on the extrapolation of fitted model 
parameters of $pp$ and $p\bar p$ reactions to exactly the same energies. Similarly, Ref.~\cite{Lebiedowicz:2018eui} also argued that the currently highest LHC energy of 
$\sqrt{s} = $ 13 TeV is sufficiently high to see the Odderon contribution, given that the Pomeron and the Odderon contributions can be extracted from the elastic scattering 
amplitudes at sufficiently high energies as
\begin{eqnarray}
 T_{el}^{P}(s,t) & \simeq & \frac{1}{2} \left(T_{el}^{pp}(s,t) +T_{el}^{p\overline{p}}(s,t)\right),\\
  T_{el}^{O}(s,t) & \simeq & \frac{1}{2} \left(T_{el}^{pp}(s,t) - T_{el}^{p\overline{p}}(s,t)\right) \,.
\end{eqnarray}

One of the problems is that the elastic $pp$ and $p\bar p$ scattering data have not been measured at the same energies in the TeV region so far. So, we strongly emphasize the need 
to run the LHC accelerator at the highest Tevatron energies of 1.96 TeV, in order to make such direct comparisons possible. Another problem is a lack of precision data at the low- and 
high-$|t|$, primarily, in $p\bar p$ collisions. Nevertheless, we show that robust features of the already performed measurements provide not only an Odderon signal, but they also 
indicate the existence of a proton substructure.

In this paper, we take the data as given and do not attempt to extrapolate the model parameters for their unmeasured values. Instead, we look for even-under-crossing and odd-under-crossing 
contributions by comparing $pp$ and $p\bar p$ collisions at different energies, looking for robust features that can be extracted in a model-independent manner. In addition, we build upon 
Ref.~\cite{Ster:2015esa} by assuming, as justified by that analysis, that the Reggeon contributions to the elastic scattering amplitudes are negligible if $\sqrt{s} \geq$ 1.96 TeV.

Let us first of all compare the behaviour of the shadow profile functions $P(b)$ before investigating the four-momentum transfer dependent $B(t)$ functions 
for both $pp$ and $p\bar p$ reactions. 

\subsection{\it Excitation function of the shadow profiles}
\label{ss:shadow-results}

\begin{figure}[!h]
\begin{center}
\includegraphics[width=0.7\columnwidth]{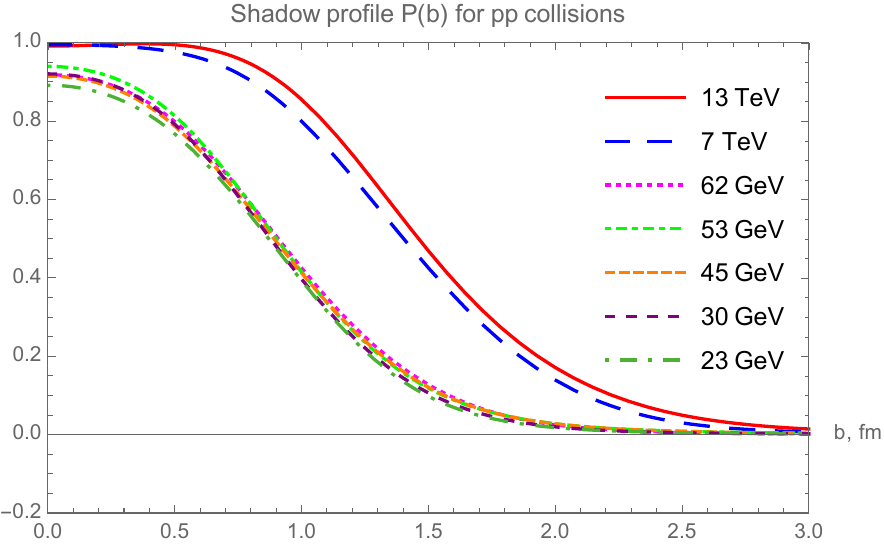} 
\caption{
Shadow profile functions for $pp$ collisions from $\sqrt{s} = 23.5$ GeV to 13 TeV.}
\label{f:Shadow-pp}
\end{center}
\end{figure}

\begin{figure}[!h]%
\begin{center}
\includegraphics[width=0.7\columnwidth]{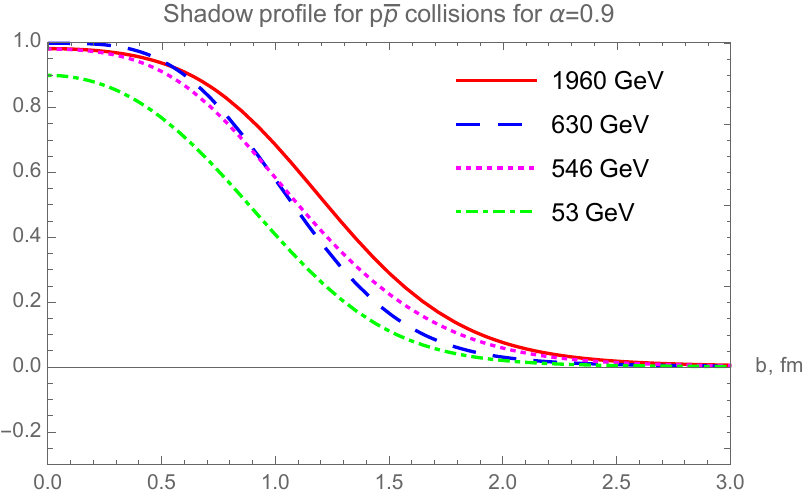}
\caption{
Shadow profiles for  $p\bar p$ collisions, $\sqrt{s} = $ 53 GeV to 1.96 TeV.}
\label{f:Shadow-ppbar}
\end{center}
\end{figure}

The excitation of the shadow profiles is obtained from the elastic scattering
amplitude obtained by fits to $pp$ elastic scattering cross-section data from
$\sqrt{s} = 23.5$ GeV to $13$ TeV, as illustrated in Fig.~\ref{f:Shadow-pp}.
The excitation function of the shadow profile functions for $p\bar p$ reactions
is indicated in Fig.~\ref{f:Shadow-ppbar}.

In $pp$ collisions at the lower ISR energies, the shadow profile functions look
nearly Gaussian, and their values at zero impact parameter are below unity,
$P(b=0) < 1$. The picture changes at the LHC energies of 7 and 13 TeV, where
the shadow profile functions seem to saturate, with $P(b= 0) > 99.9 \%$ in an
extended range, for $b < 0.4$ fm at 7 TeV and $b < 0.5 $ fm at 13 TeV.  This
indicates that the black disc limit is reached in the center of these
collisions, corresponding to $P(b) \approx 1$. However, outside the 0.4 or 0.5
fm saturated regions, the $P(b)$ decreases nearly in the same manner, as at
lower energies. One may conclude that a new, black region opens up in the TeV
energy region, which increases with growing colliding energies, and it is
surrounded by a gray hair or skin region, that has a ``skin-width" that is
approximately independent of the energy of the colliding protons. Thus, with an
increase of colliding energies, the protons become blacker, they do not become
edgier but become larger. This is the so called BnEL
effect~\cite{Nemes:2015iia}, which can be contrasted to the earlier
expectations, the so-called BEL effect suggesting that with increasing energy
of the collisions, the protons might become blacker, edgier and larger.

A new trend opens up with the 7 TeV TOTEM data, that indicates a black region,
$P(b) \simeq 1$ up to a radius of about 0.4 fm and the size of this black
region is increasing with an increase of colliding energies. Note also that at
the ISR energy range, $\sqrt{s} $ $\le $ $ 62.5 $ GeV, the shadow profiles are
very similar, however, at the TeV energy range, $pp$ and $p\bar p$ collisions
evolve somewhat differently. For example, in the shadow-profile of the elastic
$p\bar p$ collisions at $\,\sqrt[]{s} = 1.96 $ TeV, the nearly flat region with
$P(b) \approx 1$ is not yet present, while this region is present and it is
rather extended in the shadow profiles of elastic $pp$ collisions 
at $\,\sqrt[]{s}  = $ 7 and 13 TeV.

\begin{figure}[!h]%
\begin{center}
\includegraphics[width=0.7\columnwidth]{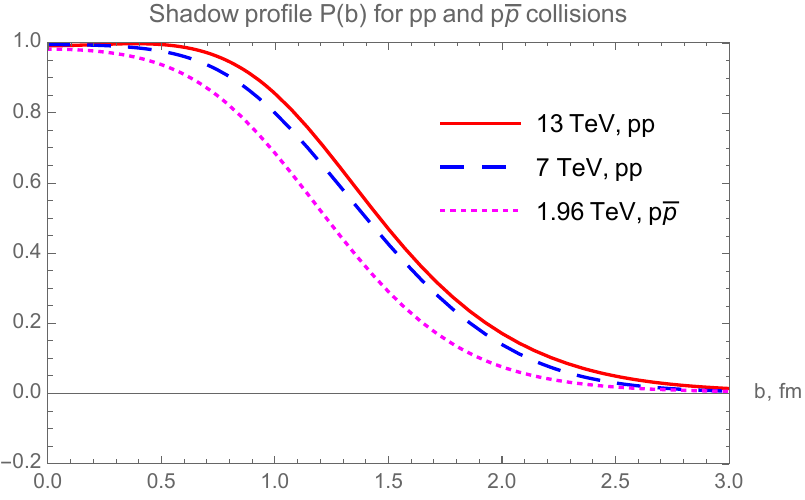}
\caption{
Shadow profile functions of $pp$ collisions at $\sqrt{s}=13$ and 7 TeV, as compared to the shadow 
profile function for $\sqrt{s} = 1.96$ TeV $p\bar p$ collisions.}
\label{f:Shadow-pp-vs-ppbar}
\end{center}
\end{figure}

In both Figs.~\ref{f:Shadow-pp} and \ref{f:Shadow-ppbar}, one can observe that
the proton becomes blacker and larger with increasing energies, however, its
edge is apparently nearly constant, looks like a nuclear skin that has the same
skin-width regardless of the energy of the collision. These results are similar
to earlier observations, published in Refs.~\cite{Nemes:2015iia,Kohara:2014waa,
Lipari:2013kta,Dremin:2017vtf}. To highlight this point, we have plotted the
shadow profile functions $P(b|pp, \, 13\, \mbox{\rm TeV})$, $P(b|pp, \, 7\,
\mbox{\rm TeV})$ and $P(b|p\overline{p}, \, 1.96\, \mbox{\rm TeV})$ together in
Fig.~\ref{f:Shadow-pp-vs-ppbar} containing the effects that come from evolution
of the structure of the proton with increasing $\sqrt{s}$.  We find that the
energy evolution of the shadow profiles is similar for the C-even $pp$
collisions and for the C-odd $p\bar p$ collisions.

We summarize that from the shadow profile functions it is very difficult to
draw strong conclusions given that the model-independent method does not allow
estimation of the collision energy dependence of the model parameters yet, so
it is very hard to tell if the obvious difference between the $pp$ and the
$p\bar p$ collisions is due to the difference in the energy of the collisions
or not. 

This observation underlines the importance of data-taking at the LHC, a $pp$
collider, with $\sqrt{s}$ decreased to as close as reasonably possible to 1.96
TeV or 1.8 TeV, the energy range of the $p\bar p$ collisions measured at
Tevatron.

\subsection{\it Results for the nuclear slope parameter $B(t)$}

Let us move on to the analysis of another important characteristics, namely, 
the $t$-dependent nuclear slope $B(t)$. 

\begin{figure}[!h]%
\begin{center}
\includegraphics[width=0.9\columnwidth]{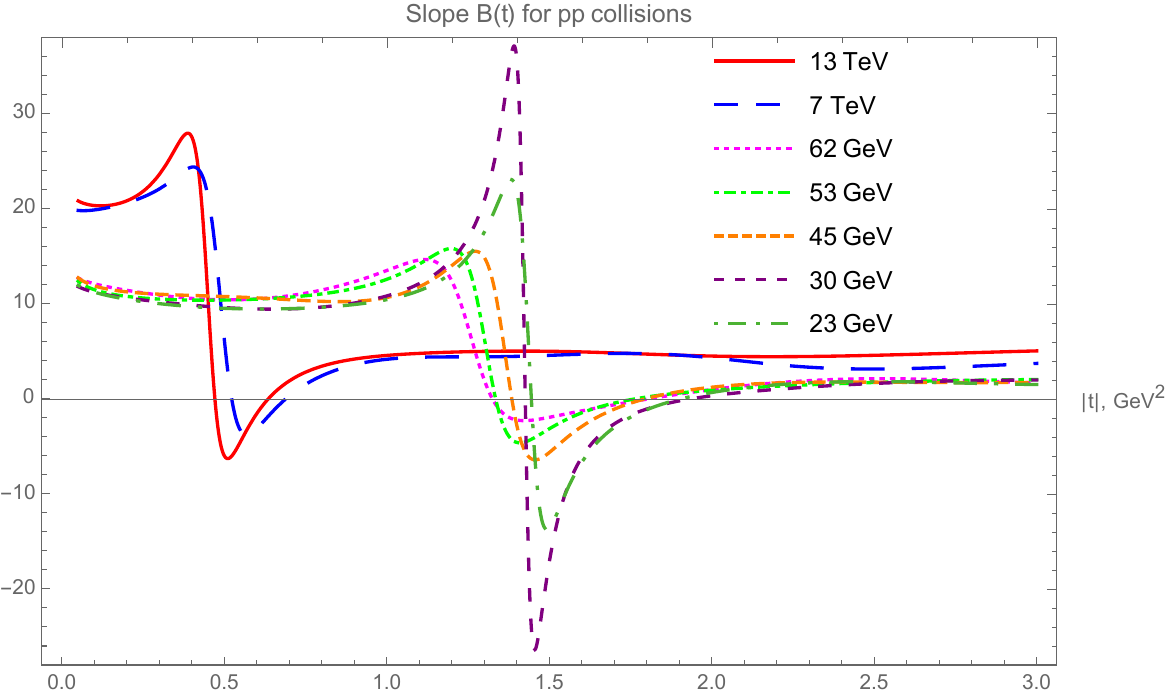}
\caption{
Slope parameter $B(t)$ for elastic $pp$ collisions. The diffractive minimum followed by a diffractive maximum is present in each case, 
as evidenced by $B(t) $ crossing the $B(t) = 0$ line twice in each case.}
\label{f:Slope-pp}
\end{center}
\end{figure}

In $pp$ collisions, the analysis of the four-momentum transfer and
center-of-mass energy dependent nuclear slope, $B(t,s)$ is summarized in
Fig.~\ref{f:Slope-pp}.  Surprisingly, in the low-$|t|$ region, where a
diffraction cone is expected, we find that $B(t)$ is actually not exactly
constant, but a $t$-dependent function, so the exponential behaviour can only
be considered as an approximation, as clearly shown in Fig.~\ref{f:Slope-pp}.
In the ISR energy range 23.5 GeV $\leq \sqrt{s} \leq 62.5$ GeV, the nuclear
slope can be considered as roughly constant both in the $|t|$ $ \leq$ 1.0
GeV$^2$ (diffractive cone) and in the 2.0 $\leq |t| \leq$ 3.0 GeV$^2$ (tail)
region, with nearly ISR energy independent $B_{\rm cone}(pp|{\rm ISR}) \approx
10$ GeV$^{-2}$ and $B_{\rm tail}(pp|{\rm ISR}) \approx 2 $ GeV$^{-2}$, rather
surprisingly. At the LHC energy scales $\sqrt{s} = $7  and 13 TeV, the cone
region shrinks, as expected, down to $|t|$ $ \leq$ 0.3 GeV$^2$, however, rather
unexpectedly and surprisingly, the tail region opens up with a featureless,
nearly flat $B(t)$ function that extends the tail region to 1.0 $\leq$ $|t|$ $
\leq$ 3.0 GeV$^2$. An obvious and numerically very stable observation is that
the low-$|t|$ approximate value, $B_{\rm cone}(pp|{\rm LHC}) \approx 20 $
GeV$^{-2}$ is nearly a factor of two larger than the corresponding values at
ISR, but also it is clear that these values are significantly $t$-dependent. On
the other hand, in the large-$|t|$ region, $B_{\rm tail}(pp|{\rm LHC}) \approx
5 $ GeV$^{-2}$, valid in a broad, 2 GeV$^{-2}$ wide range of $|t|$, with not
larger than 20 \% level variations over this range. This suggests the existence 
of some substructures inside the protons that we detail in the next subsubsection.

\subsubsection{Substructures of protons from $B(t)$ at large $t$}

It is important to realize, that the asymptotic value $B_{\rm tail}(pp|{\rm
LHC}) \approx 5 $ GeV$^2$ is nearly independent of the 7 or 13 TeV colliding
energies and it is apparently significantly larger than the same asymptotic
values at the ISR energies. It is rather clear that these values of $B(t)$,
nearly constant over such rather extended 1 or 2 GeV$^2$ ranges (plateaux) of
$|t|$, indicate two different and nearly Gaussian shaped proton substructures,
a smaller  substructure at ISR energies, that  corresponds to $B_{\rm
tail}(pp|{\rm ISR}) \approx 2 $ GeV$^{-2}$,  and a larger substructure at LHC
energies, that corresponds to $B_{\rm tail}(pp|{\rm LHC})$ $ \approx$ $ 5 $
GeV$^{-2}$. It is remarkable that the size of these substructures is not
changing when the center of mass energy of the collision is varied in the
$\sqrt{s} = 23.5 $ to $62.5$ GeV ISR range, or in the LHC energy range of
$\sqrt{s} = 7 $ to $13$ TeV. It is desirable to take more data and to
investigate what happens in between these energy ranges. In particular, we suggest that the
colliding energy of the LHC accelerator be varied in the broadest possible
range, from $\sqrt{s} = 900 $ GeV to the designed top colliding energy of 14
TeV, to see  particularly if a smooth or sudden transition is seen in $B_{\rm tail}(pp)$ 
between 2.76 TeV and 7 TeV, or not.

\begin{figure}[!h] 
\begin{center}
	\includegraphics[width=0.7\columnwidth]{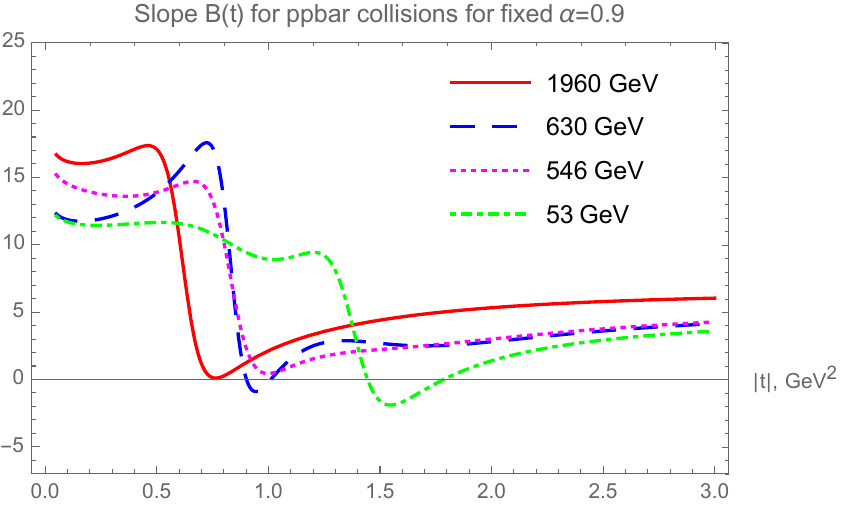}
	\caption{ Slope parameter $B(t)$ for elastic $p\bar p$ collisions.}
	\label{f:Slope-ppbar} 
\end{center} 
\end{figure}

In $p\bar p$ elastic collisions, a similar analysis of the four-momentum
transfer squared and center-of-mass energy dependent nuclear slope, $B(t,s)$ is
summarized in Fig.~\ref{f:Slope-ppbar}. In this analysis, the data are less
detailed and only one data set is analyzed in the ISR energy range,
corresponding to $\sqrt{s} = 53 $ GeV.  This data set has points both in the
cone and in the tail regions, while the dataset at $\sqrt{s} = 546$ GeV is
detailed in the low-$|t|$ region but lacks data in the tail.  The data at
$\sqrt{s} = 630$ GeV lacks data in the cone region, but extends more to the
tail range, finally the data that we analyze at the Tevatron energy scale, at
$\,\sqrt[]{s} = 1.96$ TeV, have limited $|t|$-range that only partially covers
the cone and the tail regions.

For a kind of uniformity of the comparisons of $p\bar p$ elastic scattering
data at various energies, we thus rely on fits and extrapolations with fixed
$\alpha = 0.9 $, as detailed in Appendix B, and also summarized in
Fig.~\ref{f:levy-fits-ppbar-all}. In the low-$|t|$ region, where a diffractive
cone is expected, we find that $B(t)$ is actually not exactly constant, but a
$t$-dependent function, so the exponential behaviour can only be considered as
an approximation at all the considered energies, as clearly seen in
Fig.~\ref{f:Slope-ppbar}.  In $p\bar p$ collisions, the nuclear slope is
approximately a constant both in the $|t| \leq$ 0.5 GeV$^2$ (cone), as well as
in the 2.0 $\leq |t| \leq$ 3.0 GeV$^2$ (tail) region.  These data sets cover a
broad energy range, and Fig.~\ref{f:Slope-ppbar} clearly indicates, that the
approximate average values of $B_{\rm cone}(p\overline{p})$ increase
monotonically with an increase of $\sqrt{s}$. It is remarkable, that the slope
parameter $B(t)$ is, rather surprisingly, tending to an energy-scale
independent asymptotic value of  $B_{\rm tail}(p\overline{p}) \approx 5 $
GeV$^{-2}$. 
The range, over which the asymptotic exponential
region prevails, is apparently extending at least up to 3 GeV$^2$ or more. The
larger the colliding energy, the broader this region, which starts to open at
$|t| \simeq 1.5$ GeV$^2$ at $\sqrt{s} = 1.96$ TeV. The asymptotic value
$B_{\rm tail}(p\overline{p}|1.96\, {\rm TeV}) \approx 5 $ GeV$^2$ is nearly the same as
the value of $B_{\rm tail}(pp|{\rm LHC})$ 
at $\sqrt{s} = $ 7 or 13 TeV colliding energies and is thus almost independent of the type of
the collisions as well.

It is clear that these two different, but nearly constant asymptotic values of
$B(t)$, corresponding to $B_{\rm tail}(p\overline{p}|1.96\, {\rm TeV}) \approx B_{\rm
tail}(pp|{\rm LHC}) \approx 5 $ GeV$^{-2}$ and $B_{\rm tail}(pp|{\rm ISR})
\approx 2 $ GeV$^{-2}$ over extended, 1 or 2 GeV$^2$ wide four-momentum
transfer squared ranges exhibit a domain with a nearly $\exp(-B|t|) $
behaviour. Thus, this domain reveals the existence of a proton substructure
with a nearly Gaussian elastic scattering amplitude distribution, $ t_{el}(b)
\propto \exp (-b^2/(2 R^2))$. As is well known, an approximate value of the
slope parameter $B_{\rm tail}$ is proportional to the squared Gaussian radius
of such a substructure. The larger radius of this substructure is observed in
the TeV energy range, both in $pp$ and in $p\bar p$ collisions, while a
smaller-size substructure is seen in $pp$ collisions in the $\sqrt{s} = 23.5 -
62.5 $ GeV ISR energy range. From the relation $R^2 = 4 B$ (in natural
units)~\cite{Block:2006hy,Csanad:2016add}, the Gaussian radius of the
substructure is about $R_{\rm LHC} \approx 0.9  $ fm (TeV energies) and $R_{\rm
ISR}\approx 0.6$ fm (few 10 GeV energies).  The analysis of the proton substructure
and the determination of its contribution to the total cross-section is
detailed in the next subsection, as well as in Appendix C.

The values $R_{\rm LHC}$ and  $R_{\rm ISR}$ are strikingly similar to the radii
of an effective diquark ($R_d$) and quark ($R_q$), respectively, that were
independently obtained to characterize the substructures inside the protons in
Ref.~\cite{Nemes:2015iia}. It turned out that the unitarized quark-diquark
model of elastic $pp$ scattering (called the Real extended Bialas-Bzdak or ReBB
model) predicted the $\sqrt{s}$ dependence of the total cross-section, the dip
position and even certain scaling properties of the differential cross-section
of elastic $pp$ scattering at $\sqrt{s}= 13$ TeV with a reasonably good
accuracy, based on its tuning at ISR energies and the TOTEM data set at
$\sqrt{s} = 7 $ TeV. So, the hypothesis about a proton substructure gains a
larger weight and evidence in our analysis and, thus, definitely deserves more
detailed investigations -- that, however, go beyond the scope of the
model-independent approach elaborated in this work. In particular, an important
question is whether the observed two distinct scales $R_{\rm LHC}$ and  $R_{\rm
ISR}$ correspond to the dressed diquark and the dressed quark, respectively, or
simply represent a single substructure whose size grows with energy, remains
open.

A dynamical model for the elastic amplitude based upon a two-scale structure of
the proton was previously proposed also in
Refs.~\cite{Kopeliovich:2000ef,Kopeliovich:2000pc,Kopeliovich:2012yy}.  In this
model, while the first scale was associated with the confinement radius
$R_c\simeq 1$ fm and can be attributed to the proton ``shell'', the second
semi-hard scale $r_0\approx 0.3$ fm originates due to non-perturbative
interactions of gluons and characterizes an effective gluonic ``spot'', or a
cloud, around each of the valence quarks. Despite somewhat different values of
the physical scales adopted in this model, it has appeared to predict the
energy dependence of the total and elastic cross sections quite accurately, at
least, in a parameter-dependent way.  In our current work, however, instead of
reviewing various possible model interpretations, we study the
model-independent properties of the elastic scattering data and search for
C-odd (or Odderon) effects. We employ our model-independent imaging method to
sharpen the picture of the proton as can be ``seen'' by elastic scattering
measurements at different energies.
\begin{figure}[!h]
	\includegraphics[width=0.7\columnwidth]{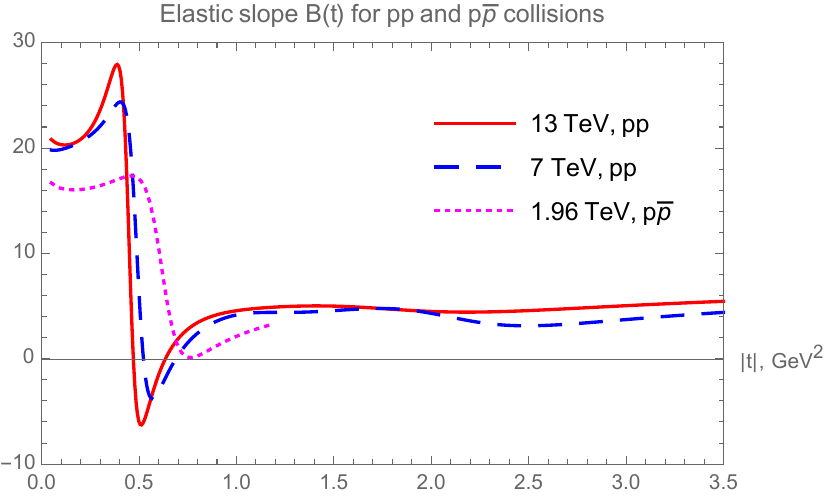}
	\caption{ Slope parameter $B(t)$ for elastic $pp$ collisions at
	$\sqrt{s} = $ 13 and 7 TeV, compared to the slope parameter of 
	$p\bar p$ collisions at $\sqrt{s} = $ 1.96 TeV.} 
	\label{f:Slope-pp-ppbar}
\end{figure}

\subsubsection{Odderon effect and the difference between $B(t|pp)$ and $B(t|p\bar p)$ }

We can observe in Fig.~\ref{f:Slope-pp}, that in $pp$ collisions $B(t)$ starts
in the cone region with nearly constant values. However, after the cone region,
$B(t)$  increasea shortly then   falls sharply, to cut through the $B(t) = 0$ line and to reach
a deep minimum with $B(t) \ll 0$ values. Then this function $B(t)$ starts to increase,
it cuts through the $B(t) = 0$ line second time, from below, to approach slowly its
asymptotic value, $B_{\rm tail}$.

Remarkably, for $p\bar p$ collisions, the $B(t)$ functions behave apparently in
a qualitatively different way. In this case, after the cone region, $B(t)$
approaches zero and it may marginally cross zero, but not so deeply and sharply
as in the case of $pp$ collisions. Taking into account that the published error
on $B$ in $p\bar p$ collisions is about 0.5 GeV$^{-2}$~\cite{Amos:1991bp}, the
error on the extrapolated $B(t)$ function can be estimated also. The latter
appears to be similar or even larger as compared to the error of $B$ at the
optical point $t = 0$. So it seems to us, that the crossing of the $B(t)$
function below zero is within errors and thus it is likely not a significant
effect in any of the $p\bar p$ collision data. 

To clarify this point more, we compare the $B(t)$ functions for $pp$ collisions
at $\sqrt{s} =$ 7 and 13 TeV with that of the $p\bar p$ collisions at $\sqrt{s}
=$ 1.96 TeV, see Fig.~\ref{f:Slope-pp-ppbar}. The nuclear slope in $pp$
collisions becomes clearly and significantly negative in an extended $|t|$
region, starting from the diffractive minimum (dip) and lasting to the
subsequent diffractive maximum (bump). These dip and bump structures are
clearly visible in the corresponding data sets, as visualized in
Figs.~\ref{f:dsigmadt-levyfit-7} and~\ref{f:dsigmadt-levyfit-13} as well. In
contrast, for $p\bar p$ collisions in the Tevatron energy range, we do not find
any dip and bump structure, that would correspond to a $|t|$-region where
$B(t)$ were negative. 

We have performed further tests to cross-check if, within the errors of the
analysis, the dip and bump structure is indeed absent in $p\bar p$ collisions
at $\,\sqrt[]{s} = 1.96 $ TeV, or not.  First of all, one can directly inspect
Appendix B to see that any reasonably smooth (e.g. spline) extrapolation of the
$p\bar p$ data points would lack a diffractive minimum structure at $\sqrt{s} =
$ 1.96 TeV.  One may argue that we do not see the minimum because the values of
$\alpha$ and $R$ were fixed. However, these numbers specify the approximate
L\'evy shape only, $d\sigma/dt \propto \exp\left(-(R^2 |t|)^\alpha\right)$,
that decreases monotonically, so their variation cannot cause diffractive
minima or maxima, as also apparent on the dashed lines of partial fits
described in Appendices C and D. In any case, we have also tested numerically
that changing $\alpha$ in the region of fixed 0.8 -- 1.0 does not qualitatively
change the behaviour of $B(t)$. Within the allowed range of variation of the
essential Levy expansion parameters $c_i$, we find that the diffractive minimum
is lacking in elastic $p\bar p$ collisions at 1.96 TeV.

This lack of diffractive minimum as well as the lack of the subsequent
diffractive maximum in elastic $p\bar p$ collisions is contrasted to the strong
diffractive minimum and maximum (the dip and bump structure) in elastic $pp$
collisions at all investigated energies, hence it indicates a rather evident
C-odd contribution to the elastic scattering amplitude, the so called Odderon
effect, shown also in Fig.~\ref{f:Slope-pp-ppbar}.

\subsection{\it Evidence for proton substructure from L\'evy fits at large $|t|$}
\label{ss:substructure-results}

L\'evy fits to the tails of the differential cross-section of $pp$ elastic
scattering data from $\sqrt{s} = $ 23.5 GeV to 13 TeV are shown on a summary
plot in Fig.~\ref{f:Summary-tails} and detailed in Appendix C.  Note, that all
the tails of the differential cross-sections are nearly linear on a log-linear
plot, indicating a nearly Gaussian substructure of the proton. As was discussed
earlier, two distinct sizes of such a substructure are seen, given the two
different values of the slope in the ISR energy range of $\sqrt{s} = $ 23.5 --
62.5 GeV, and in the LHC energy range of $\sqrt{s} = $ 7 -- 13 TeV.  After the
dip-bump structure, the differential cross-section of elastic $pp$ collisions
can be described by a simple $A \exp\left(-(|t| R^2)^\alpha\right) $ form, with
$\alpha = 0.9 \pm 0.1$ value.  Thus, for illustration this plot was done for
fixed value of $\alpha = 0.9$.

We found that in the 23.5 $\,\leq\, \sqrt[]{s}\, \leq\,$ 62.5 GeV range, a proton
substructure of nearly constant size (within errors) was present, with a
characteristic L\'evy length scale of $R_{\rm ISR} = 0.3 \pm 0.1 $ fm, and with
the corresponding contribution to the total cross-section $\sigma_{\rm ISR} =
0.3 ^{+0.3}_{-0.1}$ mb, where the quoted errors take into account also the
errors coming from the variation of the value of $\alpha$ between 0.8 and 1.0,
see also the ISR plots of Appendix C for details.

Fig.~\ref{f:Appendix-C-13TeV-log-log} indicates the TOTEM preliminary elastic
scattering data at $\sqrt{s} = 13$ TeV with their fourth-order L\'evy expansion
fits. A power-law tail would show up as a straight line on this plot, but
apparently it does not yet show up on the currently available $|t|$-range that
extends up to about $t_{\rm max} = $ 4 GeV$^{2}$. Although a straight line fit
to the tail of this distribution is perhaps possible starting from $|t| \geq 2$
GeV$^2$, these points are getting close to the end of the TOTEM acceptance for
this data set, and the error bars are getting large.  To clarify the existence
of such a possible power-law tail, more data at larger values of $|t|$ would be
desirable. In contrast, the data in the well measurable $|t|$-range are
sufficient to demonstrate the nearly exponential behaviour of the differential
cross-section in the tail region that follows the dip and bump structure. The
existence of the nearly Gaussian substructure is thus an obvious and rather
robust feature of the data (for more details, see Appendix C). This claim is
also supported by the nearly energy- and $t$-independent values of the
slope-parameter $B(t)$ in the tail regions, see Figs.~\ref{f:Slope-pp} and
\ref{f:Slope-ppbar}.
\begin{figure}[!h]
\begin{center}
\includegraphics[width=0.85\columnwidth]{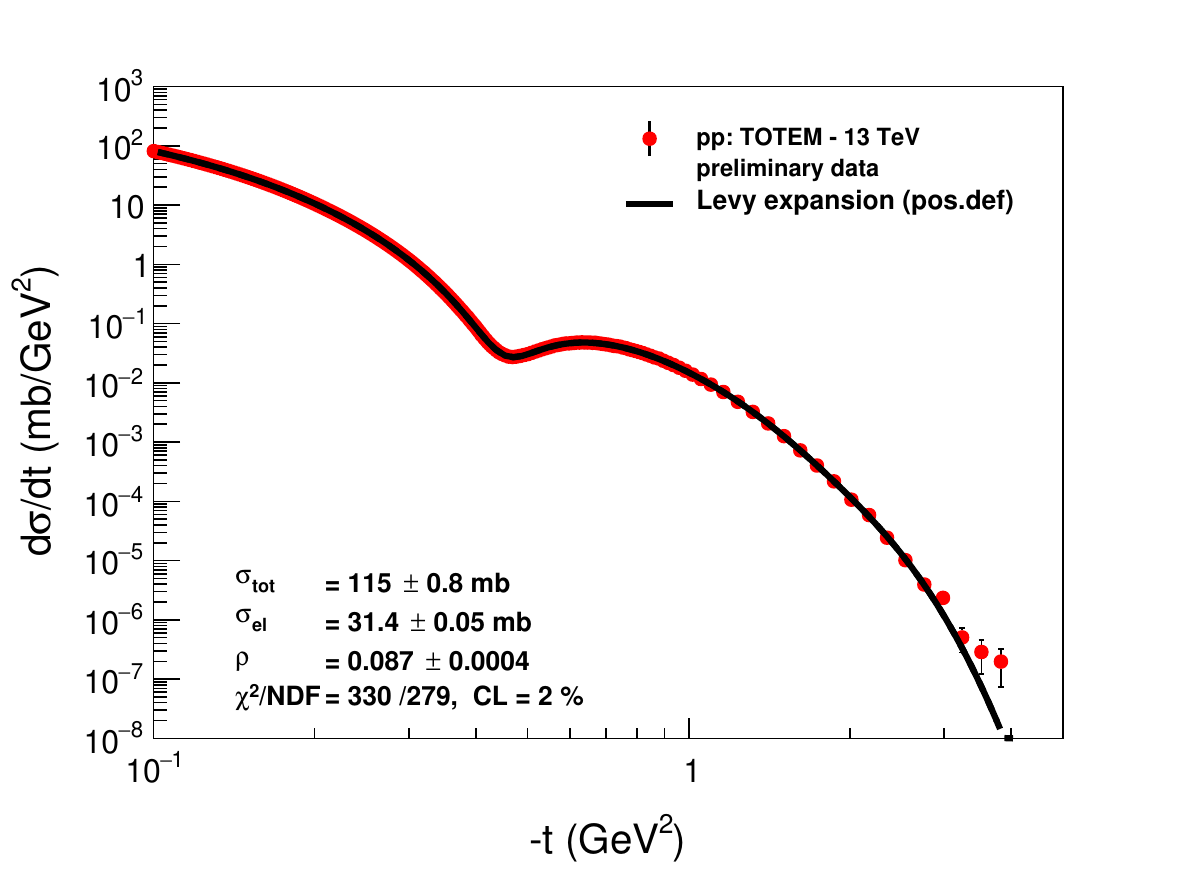}
\caption{
TOTEM preliminary data at $\sqrt{s} = 13$ TeV with their fourth-order L\'evy expansion fits on a log-log plot. 
A power-law tail would show up as a straight line on this plot.
}
\label{f:Appendix-C-13TeV-log-log}
\end{center}
\end{figure}

\subsection{\it Results for $\rho(t)$}

\begin{figure}[!h]
\begin{center}
\includegraphics[width=0.7\columnwidth]{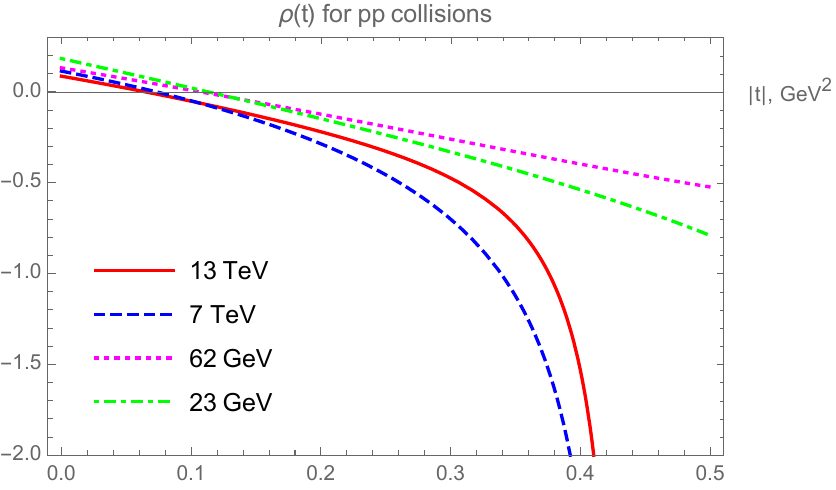}
\caption{
$\rho(t)$-parameter for $pp$ elastic scattering collisions.}
\label{f:Rho}
\end{center}
\end{figure}

By reconstructing the elastic scattering amplitude from the data, we have also
found the $t$-dependent ratio of its real to the imaginary parts in $pp$
collisions, the $\rho$-parameter. Such a result is illustrated in
Fig.~\ref{f:Rho} and indicates that the $\rho$-parameter is significantly
$t$-dependent. This dependence is initially nearly linear in $t$. 

\begin{figure}[!h]
\begin{center}
\includegraphics[width=0.7\columnwidth]{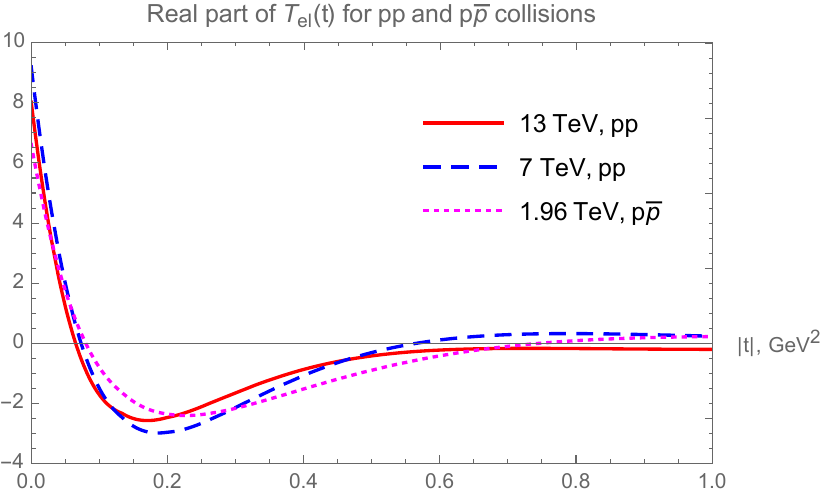}
\caption{
Real part of the elastic scattering amplitude as a function of $t$ 
for three distinct energies of $pp$ and $p\bar p$ collisions.}
\label{f:Re}
\end{center}
\end{figure}

\begin{figure}[!h]
\begin{center}
\includegraphics[width=0.7\columnwidth]{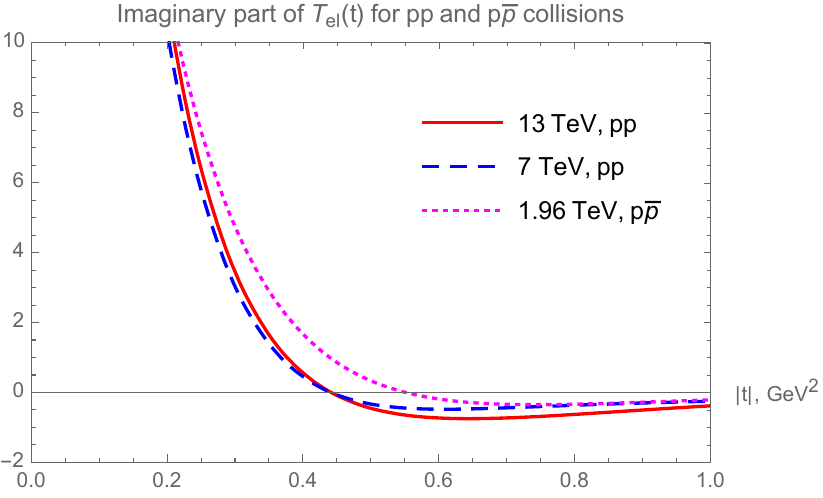}
\caption{
Imaginary part of the elastic scattering amplitude as a function of $t$ 
for three distinct energies of $pp$ and $p\bar p$ collisions.}
\label{f:Im}
\end{center}
\end{figure}

However, at the 7 and 13 TeV LHC energies, $\rho(t)$ starts to diverge to minus
infinity which corresponds to a zero point or node of the imaginary part of the
elastic scattering amplitude. In order to illustrate this point, we show the
real and imaginary parts of the elastic amplitude in Figs.~\ref{f:Re} and
\ref{f:Im}. 

Let us now analyze in detail the nuclear phase $\phi(t)$, the argument of the
complex elastic scattering amplitude $T_{el}$, that is traditionally measured
in units of $\pi$, as described in the next subsection.

\subsection{\it Results for the nuclear phase $\phi(t)$}

In this subsection, let us investigate in detail if we can identify the Odderon
effects in the $t$-dependence of the nuclear phase $\phi(t)$. As we clarify in
Appendices A and B, such a phase can be reconstructed mostly from the
differential cross-section at low-momentum transfers squared. In Appendix B we
have demonstrated that $\rho(t=0)$ cannot be reliably extracted from the
studied elastic $p\bar p$ collision data, in particular, due to a significant
lack of the low-$|t|$ data points. However, in Appendix A we demonstrated that
$\rho(t)$ can be extracted from L\'evy fits to elastic $pp$ scattering, with
the exception of $\sqrt{s} = 44.7 $ GeV and 52.8 GeV data, where the confidence
level of our fits is not in the statistically acceptable domain.

We have evaluated the $t$-dependent nuclear phase for $pp$ collisions in the
both ISR and LHC energy ranges. Unfortunately, the ISR range is a cumbersome
one, where the Reggeon as well as the Pomeron and Odderon contributions mix
with each other. In this energy range, for some of the $pp$ collision data
sets, we found that $\phi(t)$ reaches $\pi$ value at the same (or close) $|t|$
values above 1 GeV$^2$ while our sensitivity studies indicate that in this
region our method to reconstruct the nuclear phase has increasing systematic
difficulties. So, we cannot at present reliably test if the points where
$\phi(t)=\pi$ coincide or not at low energies in $pp$ and $p\bar p$ collisions.
We can however make a statement that the $pp$ collisions at the ISR energy
range of 23.5 $\leq\sqrt{s}\leq$ 62.5 GeV, the nuclear phase does not reach
$\pi$ for $|t| \leq $ 1 GeV$^2$.

\begin{figure}[!h] \begin{center}
	\includegraphics[width=0.7\columnwidth]{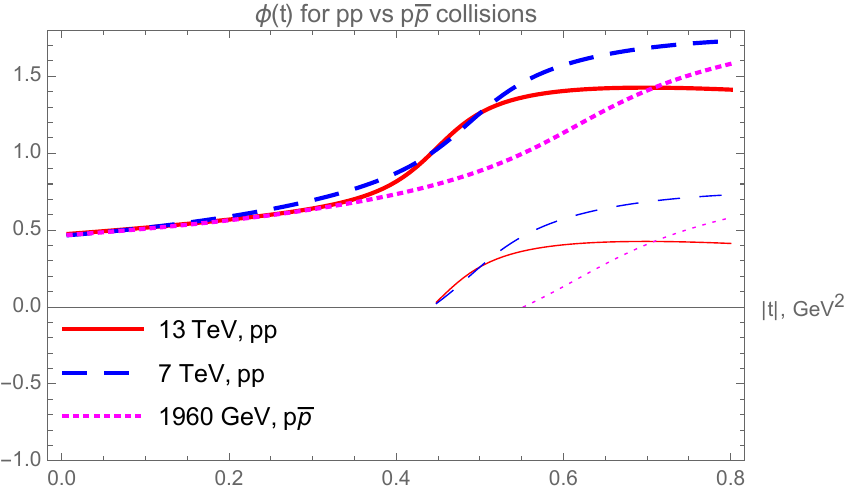}
	\caption{ The nuclear phase $\phi(t)$, shown in units of $\pi$ as a
	function of the four-momentum transfer squared $|t|$, for $pp$
	collisions at $\sqrt{s} =$ 13 TeV and 7 TeV, as compared to the nuclear
	phase for $p\bar p$ collisions at $\sqrt{s} = 1.96 $ TeV. Here, for
	illustration we also kept the corresponding principal values for the
	nuclear phase satisfying $0<\phi_2(t)< \pi$, also shown in units of $\pi$. The latter are indicated by thinner and discontinuous
	curves explicitly seen at large $|t| > |t_{\rm dip}|$.} \label{f:phi-full-range}%
\end{center} \end{figure}

Fortunately, a similar analysis in the TeV energy range gave interesting
results. Fig.~\ref{f:phi-full-range} indicates that the phase $\phi(t)$ reaches
the $\pi$ value at 7 and 13 TeV simultaneously at $|t_0|_{(pp)}\\ \approx 0.45 \pm
0.05 $ GeV$^2$ in $pp$ collisions, while $\phi=\pi$ is reached at a rather
different value of $|t_0|_{(p\overline{p})} \approx 0.70 \pm 0.05$ for 1.96 TeV
$p\bar p$ collisions.

This is an important qualitative feature of $\phi(t)$, that indicates a significant  Odderon contribution, that apparently cannot be attributed to an $s$-dependent effect. 
This subtle Odderon signature is clearly illustrated 
in Fig.~\ref{f:phi-full-range} and cannot be directly seen in the differential cross sections. Thus, the $\sqrt{s}$ independence of the crossing point of the nuclear phase $\phi(t)=\pi$ 
in the TeV energy range for $pp$ collisions and its strong shift in $p\bar p$ collisions in the TeV energy range, where the Reggeon contributions are apparently negligible, 
can be considered as a second reliable signature of the Odderon.

\section{Discussion}
\label{s:Discussion}

Our experience is that many of our results, for example, for $\rho(t)$ and also for the nuclear phase $\phi(t)$ in the $|t| > 1$ GeV$^2$ region are very sensitive to the precise 
details of the fits, so interpretation of the TOTEM results at $\sqrt{s} = 13$ TeV depends critically and sensitively on the currently preliminary TOTEM data points and 
their error bars. Certain features of our analysis, for example, the behaviour of $B(t)$ in the range where the slopes could directly be evaluated from the data are more robust and stable.

Thus, we feel strongly motivated to warn the astute reader against the over-interpretation of model results that indicate certain features of the elastic scattering data correctly only 
on the qualitative level, but fail miserably on a confidence level test. Actually, the Odderon effects that we discuss in detail in this work are due to some robust and model independent 
features of the data, but we have investigated other more subtle effects too that we do not emphasize in this work.

Our key point is that the significance of the new Odderon effects can be revealed only if the data sets are final, with published statistical and systematic errors, and if they can be correctly 
and faithfully represented by theoretical calculations. So we recommend to carefully evaluate the confidence levels of all subsequent theoretical analysis of final TOTEM data in future analyses 
and to determine the significance of the presence of novel effects like the Odderon contributions. Our method, presented in this work, allows for such sensitivity and significance analysis, 
however, given that the TOTEM data at $\sqrt{s} = 13 $ TeV are preliminary, the evaluation of the systematic errors of the our fit results would most likely also be premature at present.

Nevertheless, we may warn the careful readers that descriptions of possible Odderon effects or the lack of them, based on data analysis with zero confidence levels might have apparently 
been over-interpreted recently: the significance of the interpretation of fits that do not describe the data in a statistically acceptable manner is not particularly well defined. 
Based on our experience, we recommend against the over-interpretation of the data in terms of models that do not have a confidence level of at least CL $\geq 10^{-3}\%$, but in a final analysis, 
we strongly recommend to rely on descriptions that have a confidence level of at least  CL $> 0.1 \% $ before the model results can be interpreted. The authors should also check if their optimization 
procedure has converged or not and test if the error matrix is accurate and the estimated distance to the real minimum is sufficiently small. 

In our studies, we have found that small variations in the fit range or in the values of the fit parameters do not change the following robust features of the data, that we highlight below.

\subsection{\it Robust qualitative features}
\label{ss:Robust}

While searching for the differences between the differential cross-sections of elastic $pp$ and $p\bar p$ collisions, that could exhibit a C-odd contribution, or an Odderon effect in 13 TeV $pp$ 
elastic scattering, we have evaluated the shadow profile functions at 7 and 13 TeV $pp$ collisions using a novel imaging method, the model-independent L\'evy expansion.

We have compared the shadow profile functions of $pp$ and $p\bar p$ collisions at various energies. We have found that the shadow profiles saturate at the LHC energies of 7--13 TeV:
for small values of the impact parameter, a $P(b)\simeq 1$ region opens up. With increasing the collision energies, the protons become blacker, but not edgier, and larger, confirming 
the BnEL effect, that was reported in Ref.~\cite{Nemes:2015iia}. 

We see a significant difference between the shadow profile functions $P(b)$ of protons and anti-protons in the TeV region, but from the current analysis we cannot 
determine uniquely, if this difference is an Odderon effect, or, an effect of saturation that is apparent also in $pp$ collisions with an increase of collision energy. We would need 
$pp$ and $p\bar p$ elastic scattering data at exactly the same collision energies, that can be realized these days only by running the LHC accelerator at energies close to the Tevatron energy scales of $\sqrt{s} = $ 1.8 -- 1.96 TeV.

In Subsection~\ref{ss:Bt},  we have analyzed the dependence of the $B$-slope parameter on the four-momentum transfer squared $t$ in $pp$ as well as in $p\bar p$ reactions. We have found, that the $B(t)$ functions indicate an Odderon effect very clearly.

Surprisingly, we have identified a $|t|$ region after the dip and the bump
structure, where a clear-cut evidence is seen in our analysis for a proton
substructure of two distinct sizes in two experimentally probed GeV and TeV
energy ranges. In every case, such a substructure is characterized by a L\'evy
exponent of $\alpha = 0.9 \pm 0.1$. Possible evidence for a new substructure
inside the proton at LHC energies was, as far as we know, first pointed out by
Dremin in Ref.~\cite{Dremin:2018uwt}. In this paper, we explored this
possibility in detail and identified the characteristic L\'evy exponent as
$\alpha = 0.9 \pm 0.1$, characterized the substructure with an approximate
Gaussian and L\'evy scales at the ISR and LHC energies, respectively, and
determined the corresponding contribution to the total $pp$ cross-section as
well, as described in Subsection~\ref{ss:Bt} and in Appendix C. Based on the
Gaussian sizes found in Subsection~\ref{ss:Bt}, it is tempting to note that
they are strikingly similar to a dressed quark and a dressed diquark, found to
describe elastic $pp$ scattering in an earlier, model dependent
analysis~\cite{Nemes:2015iia}. Their presence seems to provide a
phenomenological support for the quark-diquark picture of the proton, which is
deeply related to the solution of the confinement problem in QCD proposed
recently by Brodsky and collaborators in Ref.~\cite{Brodsky:2017qno}. This,
however, does not exclude the possibility for a single substructure growing
with energy in such a way that the substructure grows only in between the ISR and the LHC
energies, but remains constant between $\sqrt{s} = $ 23.5 and 62.5 GeV, then it grows
but again remains of constant size between $\sqrt{s} = $ 7 and 13 TeV. 
More measurements 
would be strongly desirable to justify the quark-diquark picture or to investigate 
the growth of a single substructure (a dressed quark) with increased colliding  energy.
Presently this is feasible only by running the LHC accelerator at decreased energies, 
varying the collision energies between $\sqrt{s} = $ 900 GeV and 7 TeV, to identify the 
transition energy. Another measurement at the desinged top LHC energy of 14 TeV is 
already planned and approved, as far as we know, by the TOTEM Collaboration at CERN LHC.

\subsection{\it Highlighted results}
\label{ss:Highlights}

Let us highlight some of the important points of our study: 
\begin{enumerate}
\item 
	We have found a solid, stable and clear-cut evidence for a proton
		substructure, with two different sizes extracted for two
		distinct (GeV and TeV) energy ranges that are similar to the
		sizes of a dressed quark and a dressed diquark, respectively,
		as discussed in Ref.~\cite{Nemes:2015iia}, and as also derived
		from QCD in Ref.~\cite{Brodsky:2017qno}.
		Fig.~\ref{f:Slope-pp-ppbar}, even without the quantitative
		results, demonstrates the existence of such a proton
		substructure, corresponding to the second extended plateau in
		the large-$|t|$ region (besides the usual elastic cone region
		at low $|t|$) with a nearly exponential contribution to the
		differential cross-section of elastic $pp$ scattering.  We
		noticed that such a plateau corresponds to a dressed
		quark-scale substructure in the lower $\sqrt{s} = $ 23.5 --
		62.5 GeV energy range, while it resembles a larger, dressed
		diquark-scale substructure in the 7 -- 13 TeV energy range, see
		Fig.~\ref{f:Summary-tails}.
		This, however, does not exclude the possibility for a single substructure growing
		with energy in such a way that the substructure grows only stepwise: 
		within the resolution errors, it remains of a constant size between 
		the $\sqrt{s} = $ 23.5 and 62.5 GeV ISR energies, then it grows 
		but its size remains constant within the experimental resolution error
		of about 0.1 fm in the energy region between $\sqrt{s} = $ 7 and 13 TeV.
		In short we observed two significantly different in size substructures
		inside the protons at ISR and LHC energies but the physical interpretation
		of this substructures is an open question, and requires more experimental
		and theoretical investigations.
\item 
	At each energy and for each investigated data set, the scattering
		amplitude of elastic $pp$ and $p\bar p$ collisions was
		described by our new L\'evy series expansion method.  With the
		help of the elastic scattering amplitude, we have reconstructed
		values for the total cross-section and for the differential
		cross-section of elastic scattering. The published values of
		the total cross-sections were reproduced within errors and the
		fits to the differential cross-section looked fine. In case of
		several data sets they have also passed the more stringent
		tests of mathematical statistics, namely the confidence level
		of most of our fits was not unacceptable from the point of view
		of mathematical statistics, either, with confidence levels CL
		$> 0.1$ \%.
\item
	For all those data-sets, where the confidence level of the fit was not unacceptable from the point of view of mathematical statistics, we found that the exponent $\alpha$ was significantly 
	less than unity in $pp$ collisions, the deviation being a more than 5$\sigma$ effect. Given that exponential behaviour corresponds to the case of $\alpha = 1$, and fits with CL $ > 0.1$ \% represent 
	the data undoubtedly, we find that the differential cross-section at low $|t|$ is apparently non-exponential in 23.5, 30.7, 62.5 GeV and in 13 TeV $pp$ elastic scattering data. It is quite remarkable, 
	that the corresponding values of the non-exponentiality are $\alpha = 0.88 \pm 0.01$, $0.89 \pm 0.02$, $0.90 \pm 0.01$, $0.90 \pm 0.01$, when rounded up to two decimal digits. This implies 
	that an average  value of $\alpha = 0.89 \pm 0.02$ is consistent with all the measurements in a very broad energy range from 23.5 GeV to 13 TeV. The energy independence of this $\alpha = 0.89 \pm 0.02$ 
	value calls for a physics interpretation, and for further studies.
\item
	Signals of non-exponentiality in the cone region are also indicated in $p\bar p$ elastic scattering data at all energies, where we have been able to describe all the analyzed datasets with 
	an $\alpha = 0.9$ fixed value, from $\sqrt{s} = 53 $ GeV to 1.96 TeV. At a first sight, difference of $\alpha$ from unity may imply non-linearity of the Regge trajectories. However, it is known 
	that the extrapolation of the Regge trajectories from masses to negative values of momentum transfer $t$ depends on the assumed analyticity of the elastic scattering amplitude. It is worth 
	to mention here that our amplitudes are singular at $t = 0$ for $\alpha < 1$, and this behaviour does not allow for an easy analytic continuation and, hence, an interpretation of non-exponentiality
	is not straightforward in terms of Regge theory.
\item
	This non-exponential nature of the differential cross-\-section in the low-$|t|$ region implied that the slope parameter $B=B(t)$ is a function that is strongly dependent on the
	$|t|$ range, where it is determined. We have evaluated $B(t)$ both numerically and analytically in the whole $|t|$ region, where it is defined. The extrapolation of $B(t)$ to the optical point 
	of $t=0$ turned out to be model-dependent, not only due to the fact that there is a Coulomb effect that induces CNI terms and modifies the slope at very 
	small $|t|$, but also due to the analytic result that $\lim_{t\rightarrow 0} B(t) = \infty$, i.e.~our L\'evy expansion method is non-analytic at the optical point. This closely corresponds 
	to the physical picture that we allow for the underlying source contributions that may have infinite root mean square, which is typical for a L\'evy stable source distribution. 
	(For a typical example, one may consider a Lorentzian distribution, that is a symmetric, L\'evy stable distribution with L\'evy index of stability $\alpha_L = 1$, corresponding
	to our non-exponential parameter $\alpha = 0.5$.
\item
     The failure of the leading-order Levy fits for fixed $\alpha = 0.9$ at $\sqrt{s} = 7$ and 13 TeV, while their success at any lower energy, have indicated that not only the size of the proton 
     increases with an increase of collision energies, but also the shape of the protons changes. These results can be explained due to an emergence of the saturated plateaux $P(b) \approx 1$ 
     in the small $b \leq 0.4-0.5$ fm region at $\sqrt{s} = 7$ and 13 TeV which were not seen at lower energies. In our analysis, the data points from the CNI 
     region, i.e. at $|t| < 0.01$ GeV$^2$, were not included neglecting any possible Coulomb effect in the Levy expansion.
\item
	The lack of detailed data in the very low- or very large-$|t|$ regions in $p\bar p$ collisions prevented us to determine precisely the $\rho$ and the $B$ parameters of this case.
\item 
    The analysis of the four-momentum transfer squared and the center of mass energy dependent nuclear slope, $B(t,s)$ in Fig.~\ref{f:Slope-pp} not only confirms the existence 
	of a proton substructure (corresponding to the existence of large regions in $t$ where $B(t)$ is approximately but not exactly constant), but also indicates a sharp difference
	between the $B(t) $ functions of $pp$ and $p\bar p$ collisions, when comparing Figs.~\ref{f:Slope-pp} and \ref{f:Slope-ppbar}. This is a clear-cut and significant 
	Odderon effect. For $pp$ collisions, $B(t)$ starts with positive values, then it cuts sharply through the $B(t) = 0$ line and returns above it shortly but very significantly, corresponding to the dip 
	and bump structure in the differential cross-section for elastic scattering in each of the investigated data sets. In contrast, in $p\bar p$ collisions, $B(t)$ approaches zero but 
	within the errors of the analysis it does not cross it at $\,\sqrt[]{s} = 1.96 $ TeV. At two of the lower energy scales  of $\sqrt{s} =$ 630 and 53 GeV, $B(t)$ apparently crosses zero 
	and develops minima whose values are $2$ and $1$ GeV$^{-2}$, respectively. However, taking into account that the published error on $B$ in $p\bar p$ collisions is about 0.5 GeV$^{-2}$~\cite{Amos:1991bp}, 
	the error on the extrapolated $B(t)$ function can be estimated to be similar or larger as compared to the error of $B$ at the optical point of $t = 0$. So it seems to us, that the crossing 
	of the $B(t)$ function below zero is within errors likely not a significant effect in any of the $p\bar p$ collision data. In the only data set that we could access in the TeV energy range, 
	where the complicated Reggeon contributions are already negligible, $B(t)$ does not cross zero, so the diffractive minimum and maximum, the dip-bump structure is lacking in these 
	$\sqrt{s} = 1.96$ TeV $p\bar p$ elastic collisions. Such a behaviour is in sharp contrast to the $pp$ differential cross-sections at all energies. Apparently, this is a clear-cut Odderon effect, 
	as illustrated in Fig.~\ref{f:Slope-pp-ppbar}.
\item
	In addition, we have also found a surprisingly clear Odderon effects in the $t$-dependent nuclear phase $\phi(t)$. Our analysis indicates that this phase reaches $\pi$ value at the very different, 
	high LHC energies of 7 and 13 TeV in elastic $pp$ collisions at the same value of $|t_0| \approx 0.45 $ GeV$^2$, that suggests that this $|t_0|$ value has a negligibly small dependence 
	on the collision energy, $\sqrt{s}$. However, in elastic $pp$ collisions at $\sqrt{s} = 1.96 $ TeV, such a crossing point in the nuclear phase $\phi(t)=\pi$ is at 
	a very different location from $pp$ collisions. 
\item
	We have also found a weak difference between the shadow profile functions $P(b)$ belonging to $p\bar p$ collisions at $\sqrt{s} = 1.96$ TeV as compared to that of $pp$ collisions 
	at 7 and 13 TeV. However, we also found a significant evolution of the shadow profile functions from 23.5 to 62.5 GeV and from 7 TeV to 13 TeV in elastic $pp$ collisions. 
\item
	In order to clarify if the difference between the shadow profiles of
	$pp$ and $p\bar p$ collisions occurs due to the change of the colliding
	system type or due to the change of t he center of mass energy of the
	collisions, as well as to clarify differences of the other observables
	like $B(t)$ and $\phi(t)$, we strongly recommend to run the LHC
	measurements at lower energies, preferably as close to $\sqrt{s}
	=  1.96 $ TeV, as reasonably achievable, to measure the difference between $pp$  and $p\bar p$
	collisions  exactly and to clarify the Odderon contribution without any
	possible energy evolution and extrapolation effects.
\item 
	We recommend extreme care before drawing big conclusions, given that we
	see the sensitivity of some of the details like $\phi(t)$ at large
	$|t|$ for tiny details in the data and in changing some of the higher
	order coefficients of the fits. Possibly these tiny details differ in
	some papers that may apparently draw big, but contradicting, and not
	particularly well founded conclusions about the existence or
	non-existence of the Odderon effects. When looking for a robust
	conclusion about the Odderon contribution, we recommend to look at the
	summary plot of the L\'evy fits to the slope of the differential
	cross-sections of elastic scattering, as indicated
	in Fig.~\ref{f:Slope-pp-ppbar}.	
\end{enumerate}

\section{Summary and conclusions}
\label{s:conclusions}

In summary, we conclude that we have found clear-cut and post-factum rather obvious differences between the differential cross-sections of elastic $pp$ and $p\bar p$
collisions, indicating a C-odd contribution: the Odderon effect. This corresponds to a small difference of the $t$-dependent slope parameters between $pp$ collisions at 13 TeV and 7 TeV
collision energies at the LHC as compared to a large change of the $t$-dependence of the slope parameter $B(t)$ in 1.96 TeV $p\bar p$ collisions. Another characteristic Odderon signal is 
the difference is between the existence of a diffractive minimum and maximum in both 13 and 7 TeV elastic $pp$ scattering, corresponding to two distinct crossing-points of the $B(t)$ functions 
with the $B(t) = 0 $ line, as contrasted to the monotonically decreasing differential cross-section of elastic $p\bar p$ collisions, with a $t$-dependent elastic slope that is $B(t) > 0$ significantly, 
a function that never crosses the $B(t) = 0 $ line.

\begin{figure}[!h]
\begin{center}
\includegraphics[width=0.85\columnwidth]{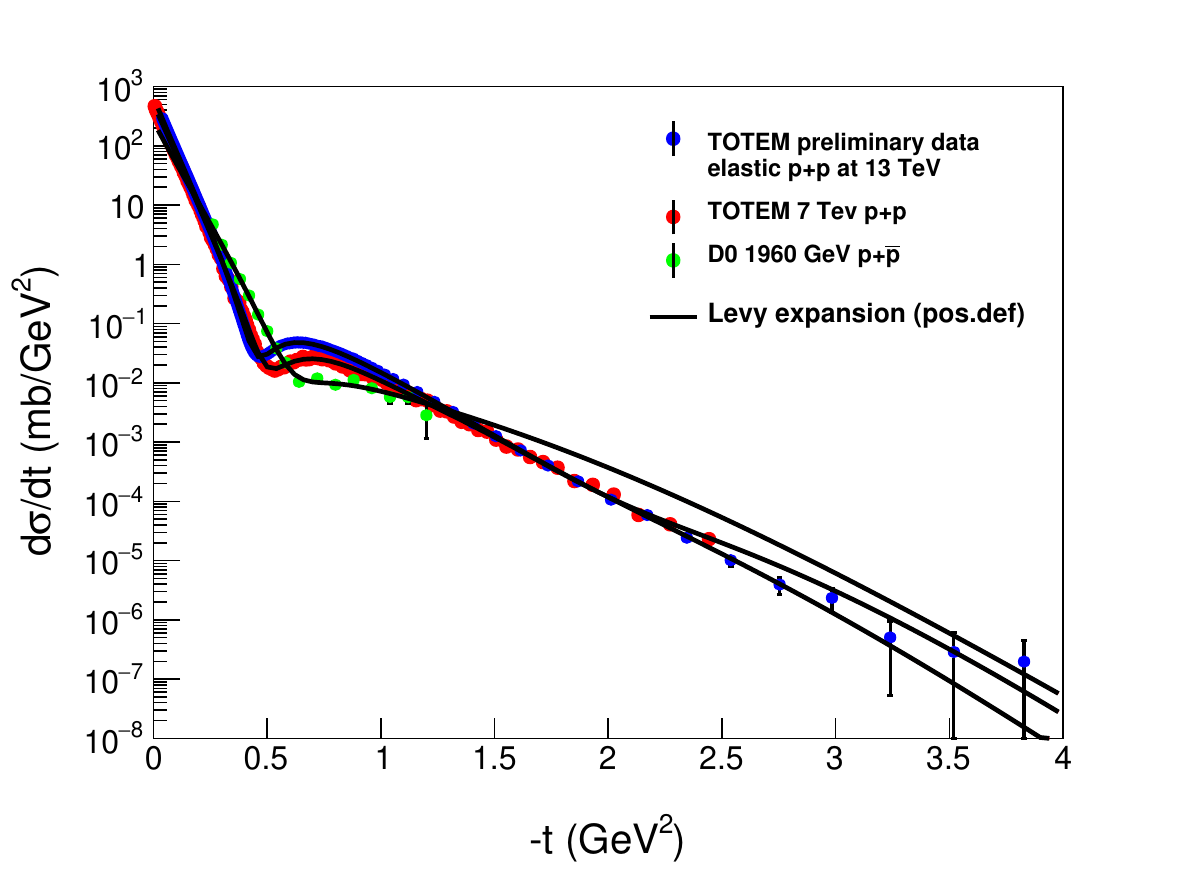}
\caption{
A direct comparison of the differential cross-sections of elastic $pp$ scattering at the LHC energies of 7 and 13 TeV,
with $p\bar p$ elastic scattering at the Tevatron energy of 1.96 TeV.}
\label{f:punchline}
\end{center}
\end{figure}

These Odderon signals, the change in $B(t)$ and the disappearance of the diffractive minimum as well as the diffractive maximum, when changing from $pp$ reactions to $p\bar p$ reactions, 
are rather obvious, stable and clear-cut effects. Once they are identified, they can be directly seen on the data sets, when one plots the 13 TeV and 7 TeV differential cross-section of elastic $pp$ 
scattering on the same plot with the differential cross-section of elastic $p\bar p$ scattering, as illustrated in Fig.~\ref{f:punchline}.

We have confirmed such Odderon effects with a more refined and subtle analysis, that indicated the lack of energy dependence of the crossing point $|t_0|$ of the nuclear phase $\phi(t)$ both 
in the TeV region $\phi(t) = \pi$ at the same value of $|t_0|(pp) \approx 0.45 \pm 0.05$ GeV$^2$ both at 7 and 13 TeV. When evaluating the crossing point of the nuclear phase for $p\bar p$ collisions 
at $\sqrt{s} = 1.96 $ TeV, a significantly different value of $|t_0|(p\overline{p}) = 0.70\pm 0.05$ GeV$^2$ was obtained for the position of this point. The difference between
$|t_0|(pp)$ and $|t_0|(p\overline{p})$ is apparently a clear but subtle Odderon effect, that cannot be obviously obtained by directly inspecting Fig.~\ref{f:punchline}, 
but it supports the same conclusion about the presence of Odderon effects in the few TeV elastic scattering data. 

As a by-product, but perhaps even more importantly, we have also found a
clear-cut evidence for a proton substructure, as shown by the presence of the
second, nearly exponential region in the differential cross-sections of elastic
$pp$ collisions at large $|t|$. At the ISR region, $23.5 \le \sqrt{s} \le 62.5$
GeV, from the asymptotic value of the $t$-dependent slope parameter of $B_{\rm
ISR}\approx 2 $ GeV$^{-2}$ a substructure with a Gaussian radius of $R_{{\rm
ISR},G} \approx 0.6$ fm, while at the LHC energies of 7 $\leq \sqrt{s} \leq$
13 TeV, a substructure with a different Gaussian radius of $R_{{\rm LHC},G}
\approx 0.9 $ fm is identified. Apparently, the size of this structure found at
the ISR and LHC matches reasonably well the size of the dressed quarks and
diquarks, respectively, as found recently in a unitarized Bialas-Bzak model
analysis of 7 TeV elastic $pp$ scattering~\cite{Nemes:2015iia}. These results
may provide a phenomenological support for the quark-diquark picture of hadron
confinement as obtained recently by Brodsky using AdS/QCD
techniques~\cite{Brodsky:2017qno}, although they may not exclude 
other reasonable interpretations.

Our analysis indicates that the proton substructure contributes to the total
$pp$ cross-sections with $\sigma_{\rm ISR} = 0.3 ^{+0.3}_{-0.1}$ mb at ISR, and
$\sigma_{\rm LHC} \approx 8.2^{+7.9}_{-4.7} $ mb at the LHC energies.

We have also found that this substructure can be better characterized by a
L\'evy source with $\alpha = 0.9$ as compared to a Gaussian source
(corresponding to $\alpha = 1$).  Using the characteristic L\'evy length scale
$R_L$, we find that these substructures of the protons are characterized of
of $R_{{\rm ISR},L} = 0.3 \pm 0.1 $ fm at ISR, and $R_{{\rm LHC},L} = 0.5 \pm
0.1$ fm at the LHC energies.

From the analysis of the cone region, we clearly demonstrated that the shape of
the protons actually changes in the 7 -- 13 TeV energy range, corresponding to
an opening of a new channel, as clearly demonstrated by the appearance of a
saturated $P(b) \approx 1$ region in the shadow profile functions in the TeV
energy range.

Finally, based on our experience with precision description of the differential
cross-sections of elastic $pp$ and $p\bar p$ collisions let us warn the careful
readers against over-inter\-preting reasonably looking fit results in cases when
the fitted function does not represent the data with a statistically not
unacceptable confidence level.

We hope that this data analysis method of L\'evy series expansion, detailed for
the first time in this manuscript for a positive definite function, may find
several important applications in the future, in a broad range of quantitative
sciences. Essentially this method is able to characterize the deviations from
Fourier-transformed and symmetric L\'evy stable source distributions.  Given
the ubiquity of L\'evy distributions in Nature, we hope that our new method
will be relevant in several areas of human knowledge, that extend far beyond
the science of physics.

\begin{acknowledgements}

We gratefully acknowledge inspiring physics discussions with professors W.
	Guryn, G. Gustaf\-son, L. L\"onnblad, and M. \v{S}umbera. T.Cs. and A.S.
	thank the Lund university for kind hospitality and support during their
	visit. T.Cs. would like to thank to Prof. R.J. Glauber for inspiring
	discussions and for proposing a series expansion to describe the
	elastic scattering amplitude. We are thankful to the TOTEM
	Collaboration at CERN LHC, in particular, to S. Giani, the
	Spokesperson, and K. \"Osterberg, the Physics Coordinator, for making the
	TOTEM preliminary 13 TeV data set available for us.  R.P. is partially
	supported by the Swedish Research Council, contract numbers
	621-2013-4287 and 2016-05996, by CONICYT grants PIA ACT1406 and
	MEC80170112, as well as by the European Research Council (ERC) under
	the European Union's Horizon 2020 research and innovation programme
	(grant agreement No 668679). This work was supported in part by the
	Ministry of Education, Youth and Sports of the Czech Republic, project
	LTC17018. The work has been performed in the framework of COST Action
	CA15213 ``Theory of hot matter and relativistic heavy-ion collisions''
	(THOR). Our research was partially supported by the Hungarian NKIFH
	grants No. FK-123842 and FK-123959, the Hungarian EFOP
	3.6.1-16-2016-00001 project and the exchange programme of the Hungarian
	and the Ukrainian Academies of Sciences, grant NKM-92/2017.

\end{acknowledgements}

\appendix

\section{L\'evy expansion fits to elastic $pp$ collisions: full acceptance region}
\label{s:AppendixA}

\begin{figure}[!h]
\begin{minipage}{0.7\textwidth}
 \centerline{\includegraphics[width=1.0\textwidth]{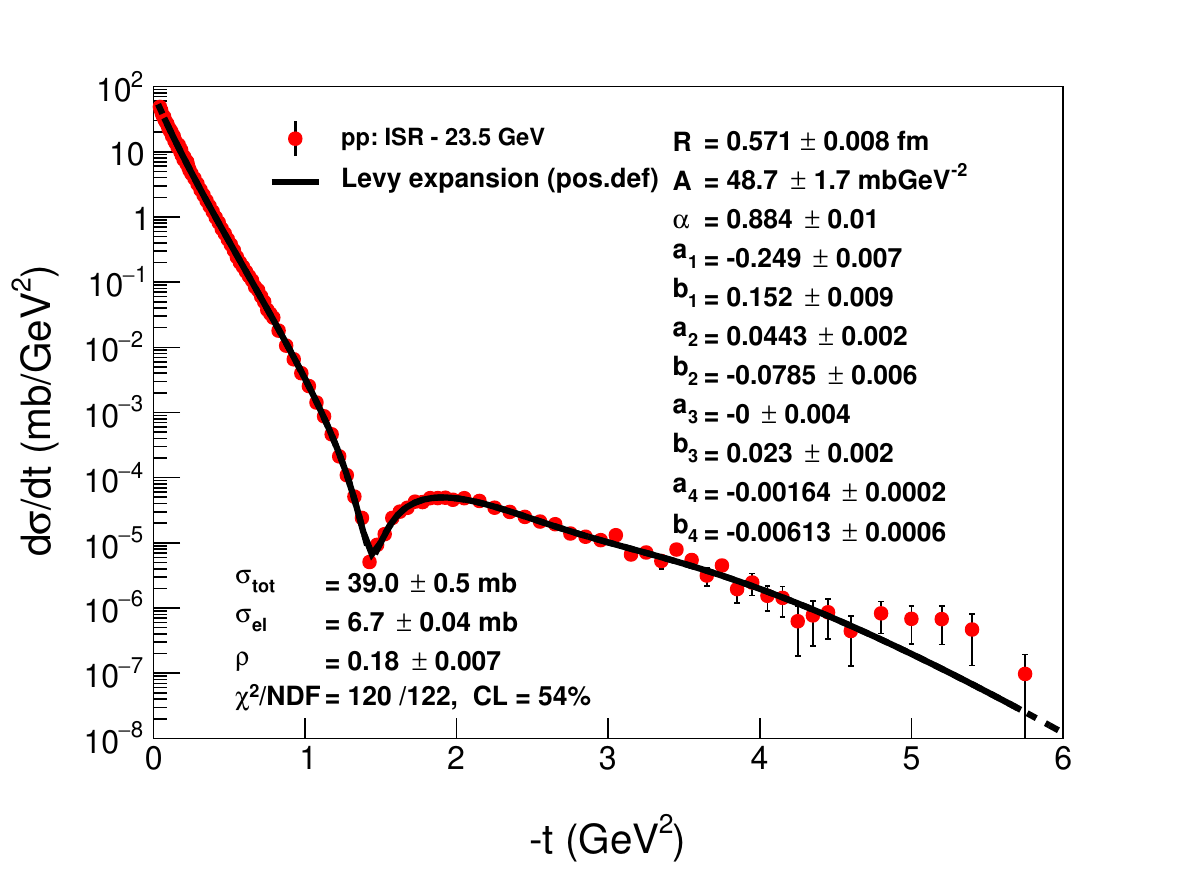}}
\end{minipage}
\caption{Fourth-order L\'evy expansion fits of the differential cross-section data from Ref.~\cite{Amaldi:1979kd} 
on $pp$ elastic scattering at $\sqrt{s} = 23.5 $ GeV.
}
\label{f:pp1}
\end{figure}

\begin{figure}[!h]
\begin{minipage}{0.7\textwidth}
 \centerline{\includegraphics[width=1.0\textwidth]{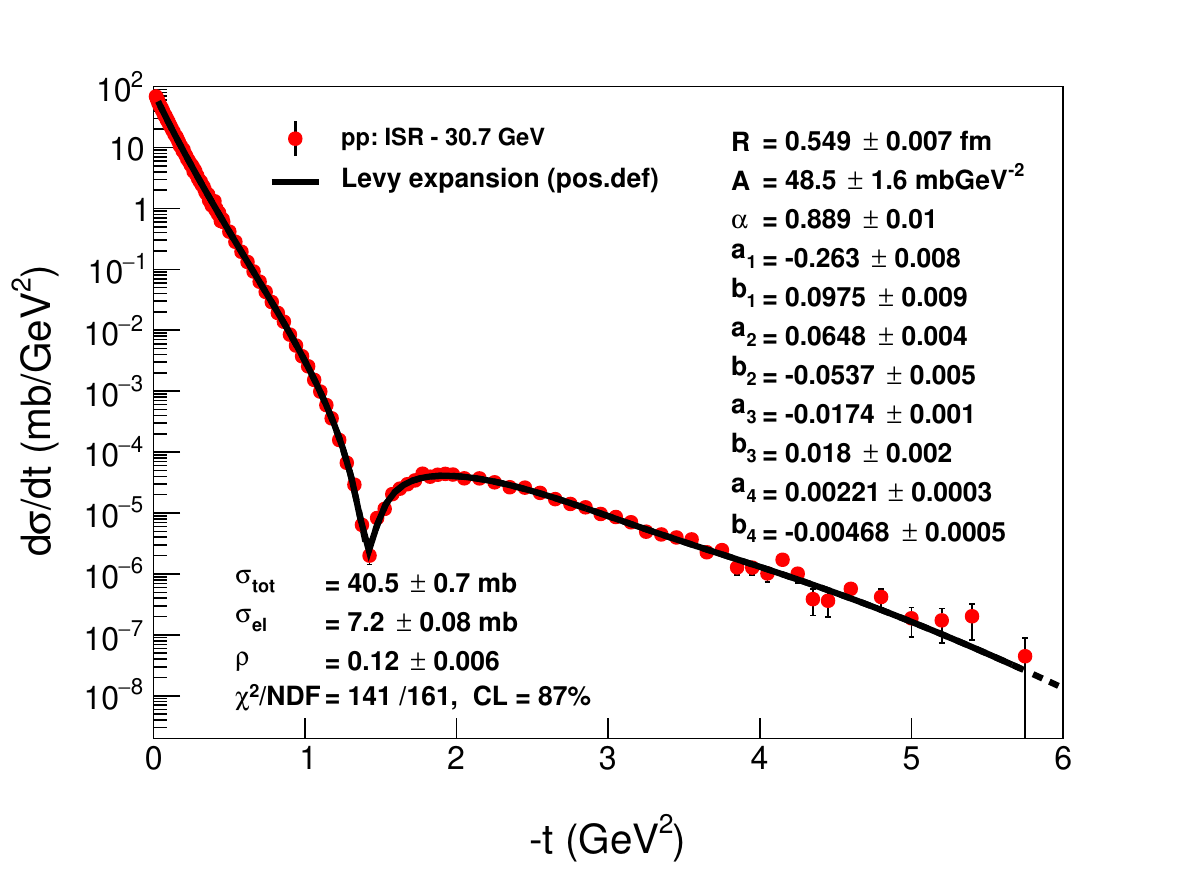}}
\end{minipage}
\caption{Fourth-order L\'evy expansion fits of the differential cross-section data from Ref.~\cite{Amaldi:1979kd} 
on $pp$ elastic scattering at $\sqrt{s} = 30.7 $ GeV.
}
\label{f:pp2}
\end{figure}

\begin{figure}[!h]
\begin{minipage}{0.7\textwidth}
 \centerline{\includegraphics[width=1.0\textwidth]{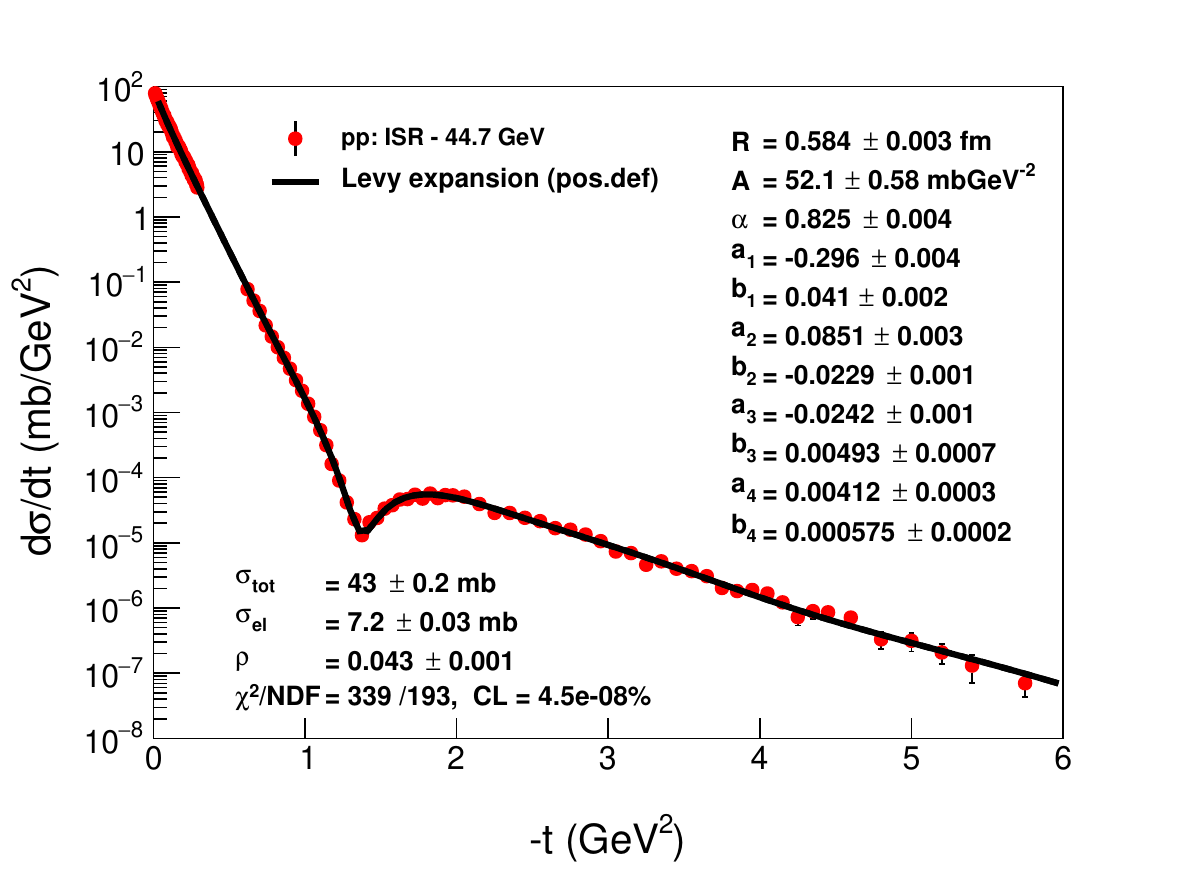}}
\end{minipage}
\caption{Fourth-order L\'evy expansion fits of the differential cross-section data from Ref.~\cite{Amaldi:1979kd} 
on $pp$ elastic scattering at
$\sqrt{s} = 44.7 $ GeV.
}
\label{f:pp3}
\end{figure}

\begin{figure}[!h]
\begin{minipage}{0.7\textwidth}
 \centerline{\includegraphics[width=1.0\textwidth]{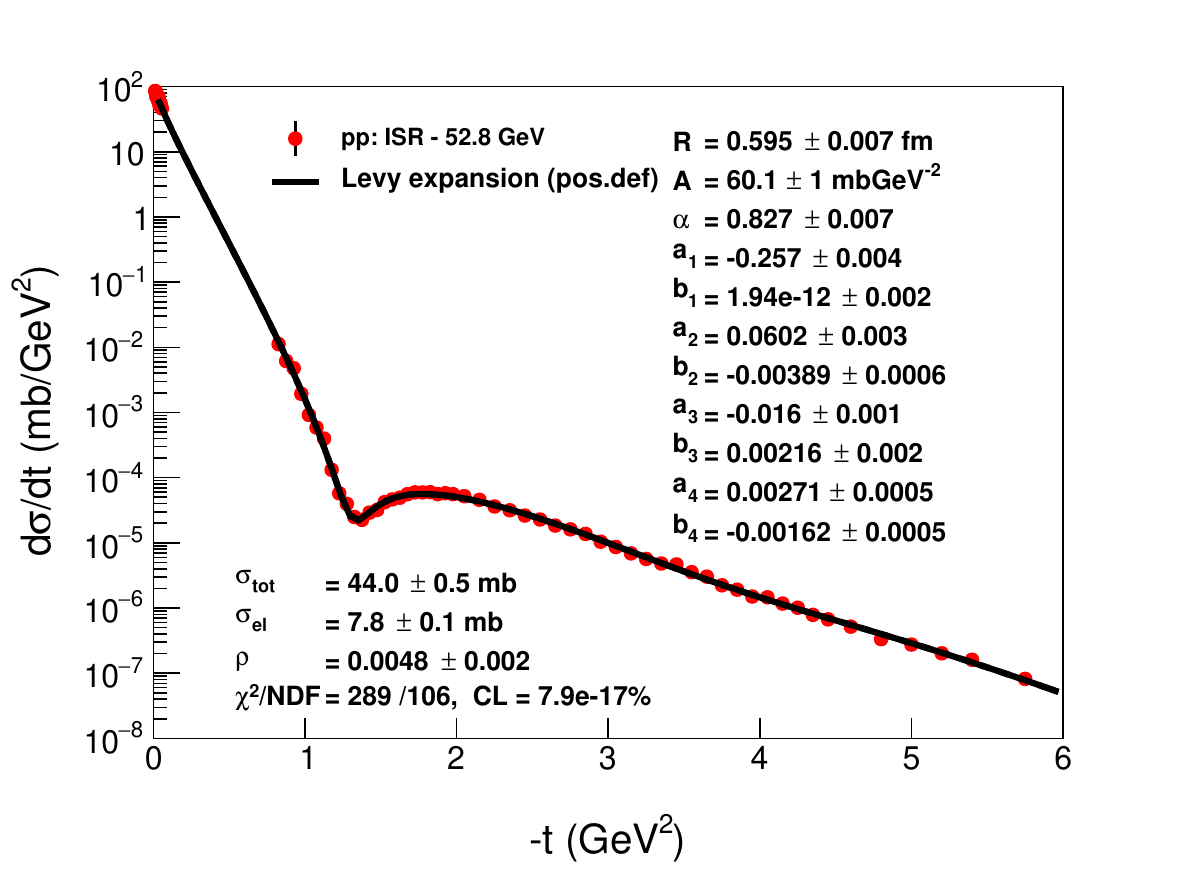}}
\end{minipage}
\caption{Fourth-order L\'evy expansion fits of the differential cross-section data from Ref.~\cite{Amaldi:1979kd} 
on $pp$ elastic scattering at $\sqrt{s} = 52.8 $ GeV.
}
\label{f:pp4}
\end{figure}

\begin{figure}[!h]
\begin{center}
\begin{minipage}{0.7\textwidth}
 \centerline{\includegraphics[width=1.0\textwidth]{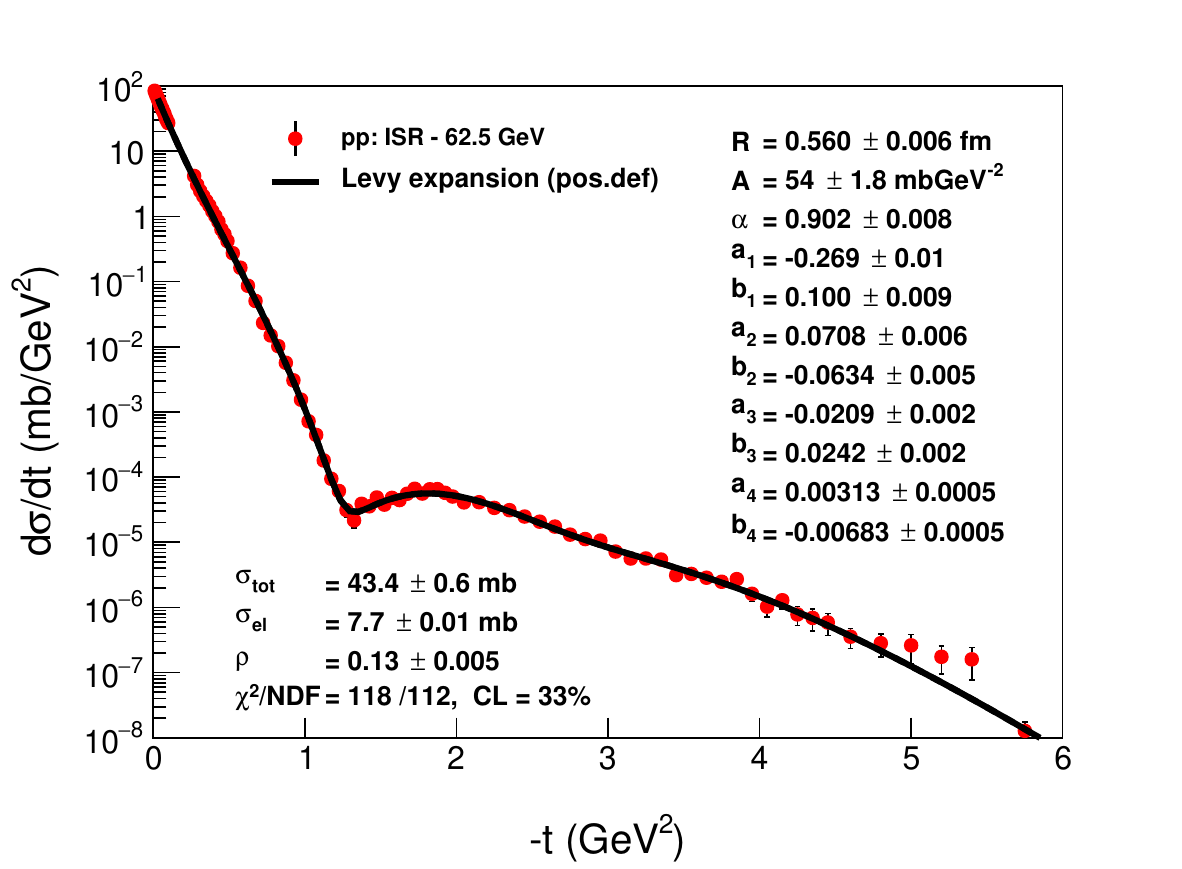}}
\end{minipage}
\end{center}
\caption{Fourth-order L\'evy expansion fits of the differential cross-section data from Ref.~\cite{Amaldi:1979kd} 
on $pp$ elastic scattering at 
$\sqrt{s} = 62.5 $ GeV.
}
\label{f:pp5}
\end{figure}

In this Appendix, we describe the results of the fourth-order L\'evy expansion fits to the elastic scattering data of $pp$ collisions for
five different data sets, measured in the ISR energy range of $\sqrt{s}$ $=$ 23.5, 30.7, 44.7, 52.8 and 62.5 GeV, shown in Figs.~\ref{f:pp1}, \ref{f:pp2}, \ref{f:pp3}, \ref{f:pp4}, and \ref{f:pp5}, respectively.
The parameters of the fourth-order L\'evy expansion are shown on the corresponding plots, together with the extracted value 
of the total cross-section, the value of the $\rho$ parameter and the measures of the fit quality.

Many of the features of these fit results are common for all of these plots, but let us start with noting that the fits, although look
quite reasonable to the naked eye, differ in their quality. The confidence level of fits to the $\sqrt{s}$ $=$ 23.5, 30.7
and 62.5 GeV data sets has a statistically acceptable value, with CL $> 0.1$ \%, while the quality of the fits to the 
data sets at $\sqrt{s} = $ 44.7 and 52.8 GeV is not acceptable from the point of the mathematical statistics, 
corresponding to CL $\ll $ 0.1 \% . This implies that we are not allowed to interpret the results of the fits to the 44.7 and 
52.8 GeV data sets, as the fitted curve, although looks reasonable, does not represent the data well enough 
(from the mathematical statistics point of view).

\section{L\'evy expansion fits to elastic $p\bar p$ collisions: full acceptance region}
\label{s:AppendixB}

\begin{figure}[!h]
\begin{minipage}{0.7\textwidth}
 \centerline{\includegraphics[width=1.0\textwidth]{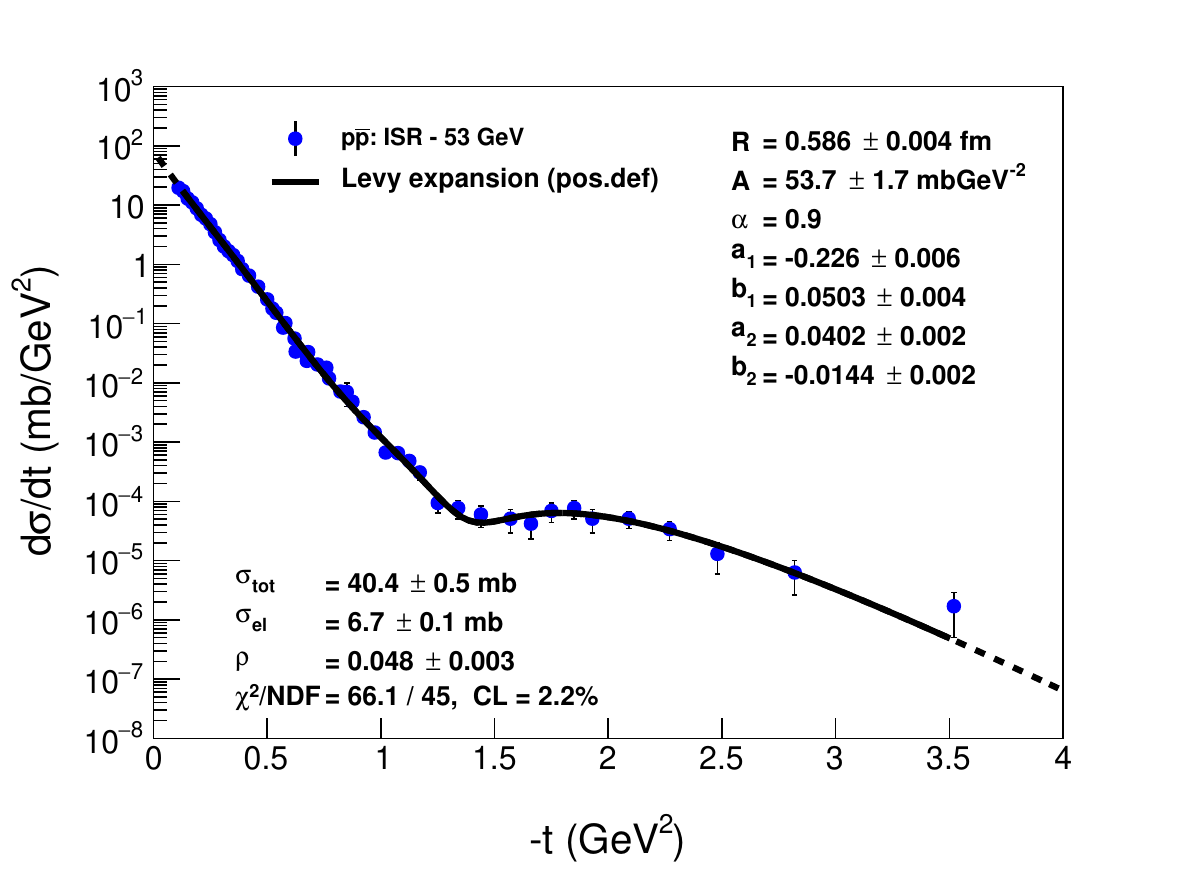}}
\end{minipage}
\caption{
Second-order L\'evy expansion fits of the differential cross-section data on $p\bar p$ elastic scattering 
at $\sqrt{s} = 53 $ GeV \cite{Breakstone:1984te}, 
with fixed $\alpha = 0.9$.
}
\label{f:ppbar1}
\end{figure}

\begin{figure}[!h]
\begin{minipage}{0.7\textwidth}
 \centerline{\includegraphics[width=1.0\textwidth]{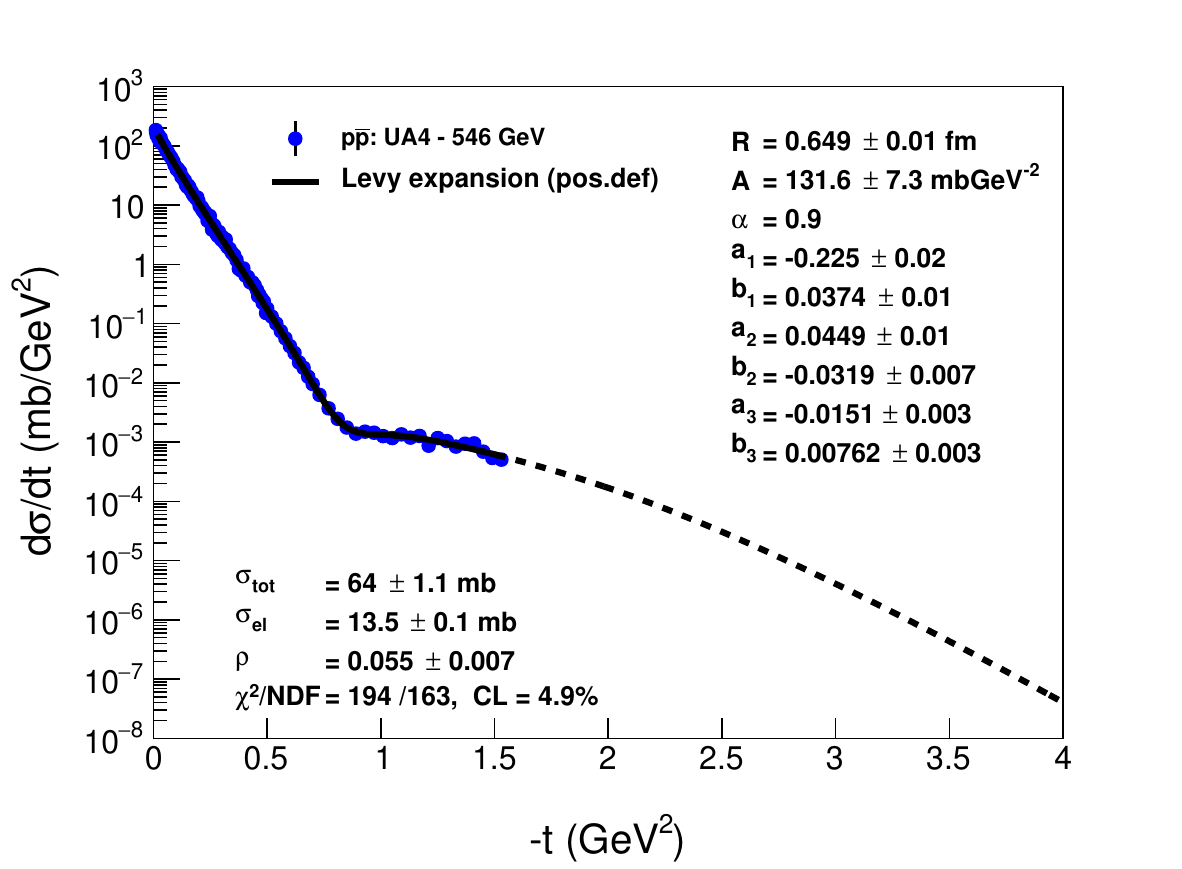}}
\end{minipage}
\caption{
Third-order L\'evy expansion fits of the differential cross-section data on $p\bar p$ elastic scattering 
at $\sqrt{s} = 546 $ GeV \cite{Bernard:1987vq}, 
with fixed $\alpha = 0.9$.
}
\label{f:ppbar2}
\end{figure}

\begin{figure}[!h]
\begin{minipage}{0.7\textwidth}
 \centerline{\includegraphics[width=1.0\textwidth]{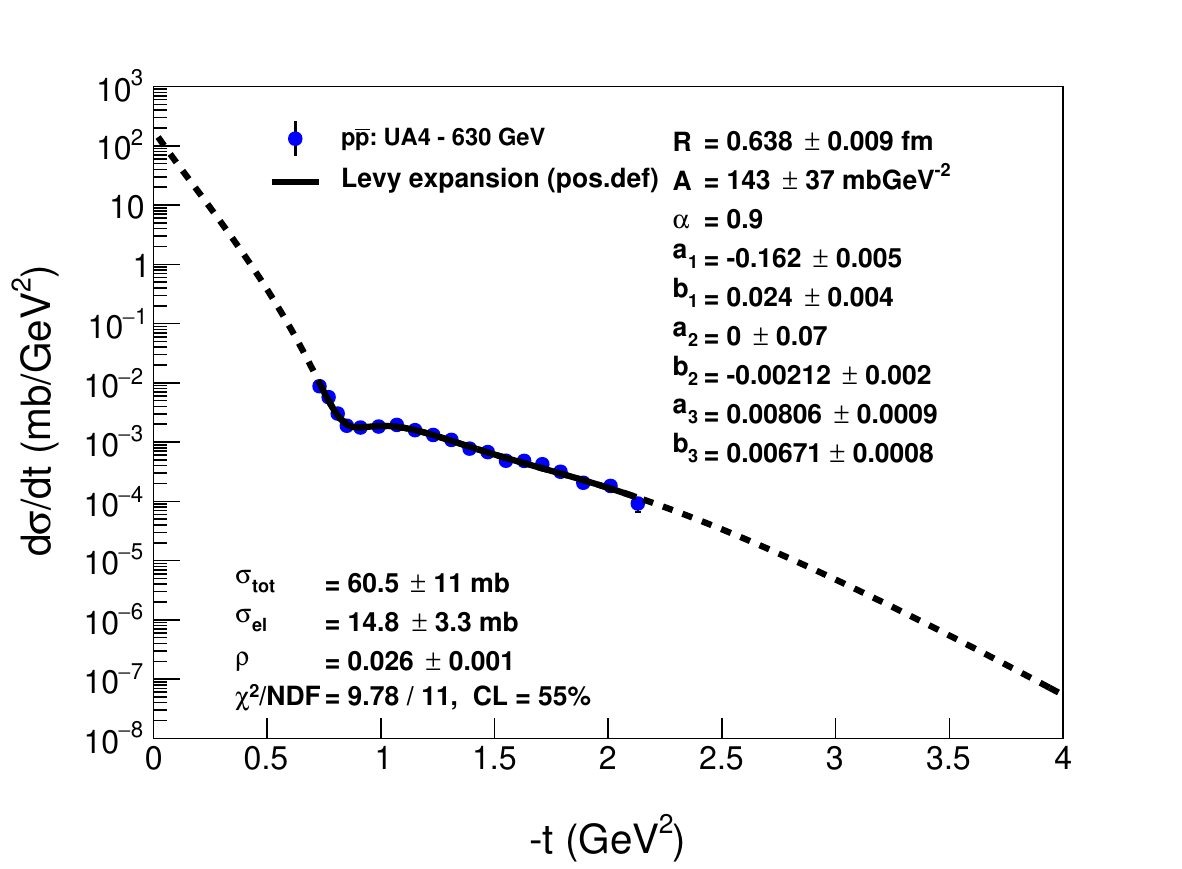}}
\end{minipage}
\caption{
Third-order L\'evy expansion fits of the differential cross-section data on $p\bar p$ elastic scattering 
at $\sqrt{s} = 630 $ GeV \cite{Bernard:1986ye}, 
with fixed $\alpha = 0.9$.
}
\label{f:ppbar3}
\end{figure}

\begin{figure}[!h]
\begin{minipage}{0.7\textwidth}
 \centerline{\includegraphics[width=1.0\textwidth]{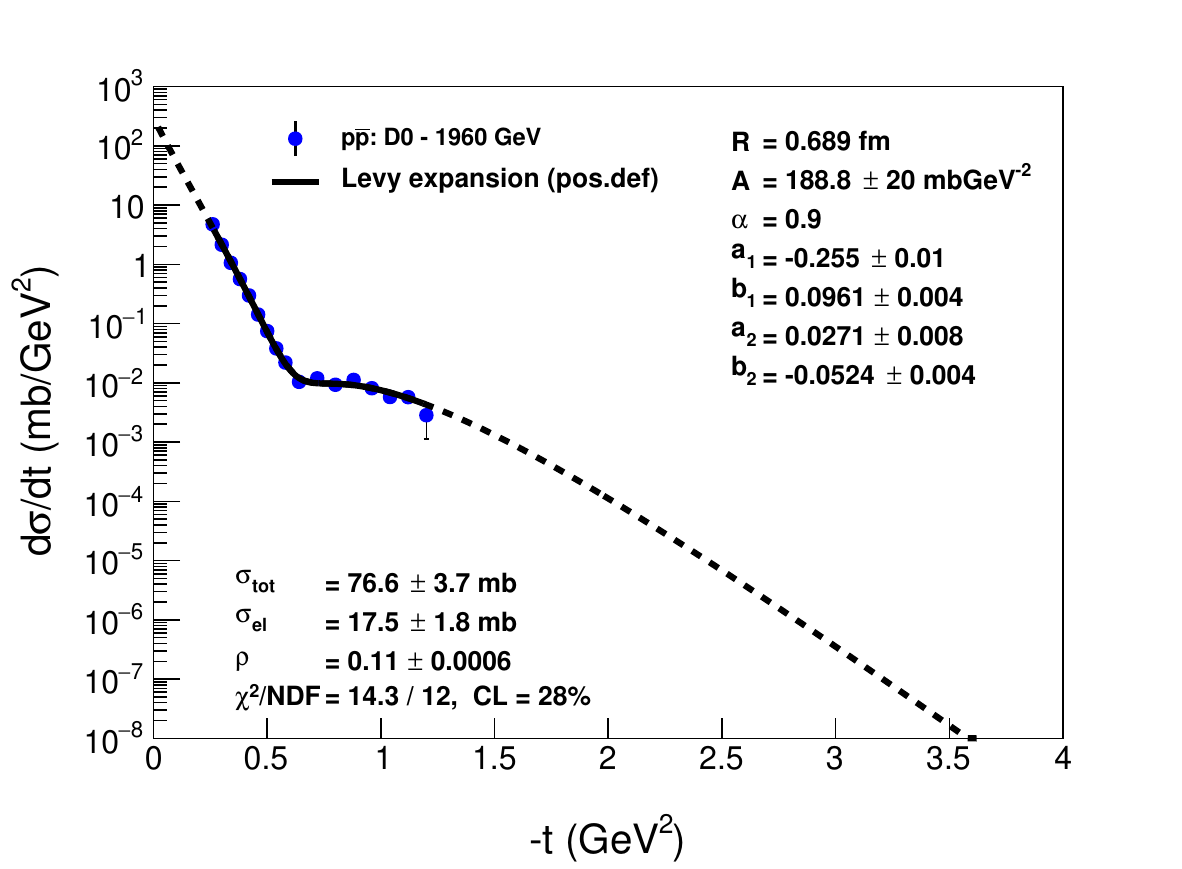}}
\end{minipage}
\caption{
Second-order L\'evy expansion fits of the differential cross-section data on $p\bar p$ elastic scattering 
at $\sqrt{s} = 1960 $ GeV \cite{Abazov:2012qb}, 
with fixed $\alpha = 0.9$ and $R=0.689$ fm.
}
\label{f:ppbar4}
\end{figure}

In this Appendix, we describe the L\'evy expansion fits to the elastic scattering data of $p\bar p$ collisions for
four different data-sets, at $\sqrt{s}$ $=$ 53, 546, 630 and 1960 GeV, illustrated in Figs.~\ref{f:ppbar1}, 
\ref{f:ppbar2}, \ref{f:ppbar3}, and \ref{f:ppbar4}, respectively.

Some of the features are common for all of these plots.

1) The parameter $\alpha$ was found to be within the errors independent of
$\sqrt{s}$, so it was fixed to a collision energy independent constant value
$\alpha = 0.9$ in these cases. It is an interesting and open problem to search
for a physics interpretation of the universality of the $\alpha$ value of
elastic $p\bar p$ collisions, but this topic goes beyond the scope of the
model-independent approach followed in the current manuscript. As an extra
bonus, fixing $\alpha = 0.9 $ has resulted in a reasonably good reproduction of
the $\sqrt{s}$ dependence of the total $p\bar p$ cross-sections, achieved
without an additional tuning. 

2) These $p\bar p$ elastic scattering data were less detailed as compared to
the corresponding $pp$ data, presented in Appendix A.  The ``dip'' region was
covered in all the cases, however, at very low $|t|$ as well as at large $|t|$
values, the acceptance and thus the $t$ range were rather limited. This
prevented us from a reliable analysis of the $\rho$ parameter of elastic
scattering for these collisions even at $t=0$.

3) We have tested if the third-order and fourth-order L\'evy expansions give
similar results for these data sets or not. Within errors, the fourth-order
expansion parameters were found to be consistent with zero, so we have fixed
them to zero and investigated the third-order L\'evy expansion results. In some
cases, e.g. for $\sqrt{s} = 53$ and $1960 $ GeV, the second-order L\'evy fits were
employed as providing better confidence levels for these rather scarce data
sets.

4) We have checked that $\rho(t=0)$ changes well outside the Minuit indicated
errors, if we change the order of the expansion from a third-order to a
fourth-order L\'evy expansion. Usually high precision data at low values of
$|t|$ are needed to reconstruct $\rho$ reliably, so this limitation is not
unexpected. However, it indicates the sensitivity of our method. In contrast to
$\rho(t)$, the $B(t)$ slope parameters and the $P(b)$ shadow profiles were
found to be stable with respect to increasing the order of the L\'evy expansion
from the third to the fourth order.

5) We have found that a third-order L\'evy expansion provides a reasonable
overall description for all of these four data sets: Minuit has converged,
error matrix accurate, and the fit quality is not unacceptable. The confidence
level is sufficiently large, CL $> 0.1$ \% for all the four data sets.  We
concluded, that with the exception of the overly sensitive parameter $\rho$, we
can interpret the physical meaning of the fit parameters, as they represent the
data. Note that the $|t|$ ranges of the various data sets are rather different,
but it is clear that the L\'evy expansion reproduces the elastic $p\bar p$
differential cross-section in the respective $s$ and $|t|$ ranges of these four
data-sets.

The parameters of the L\'evy expansion are shown on the corresponding figures,
together with the extracted values of the total cross-section, and the measures
of the fit quality. The value of the $\rho$ parameter is also shown, but the
fluctuation of these values as a function of $\,\sqrt[]{s}$ and their correlation
with the L\'evy exponent $\alpha$ also indicates that
even the value of $\rho(t =0)$ cannot be reliably determined from the these
$p\bar p$ elastic scattering data, due to lack of data points in the sufficiently 
low $|t|$ region.

\section{L\'evy fits to elastic $pp$ collisions at large $|t|$}
\label{s:AppendixC}

In this Appendix, we detail the L\'evy  fits, $d\sigma/dt = 
A \exp\left(-(R^2 t)^{\alpha} \right)$ to the elastic scattering data of $pp$
collisions for seven different data sets, at $\sqrt{s}$ $=$ 23.5, 30.7, 44.7,
52.8, 62.5 GeV as well as at 7 and 13 TeV. These fits correspond to the zeroth
order, leading terms of the L\'evy expansions, with all the expansion
coefficients set to zero or $c_i = 0$, as detailed in
Subsection~\ref{ss:cross-sections}. Some of the features are common 
for all of these plots.

In the ISR energy domain fits were performed in the 2.5 $ < |t| < $ 5 GeV$^2$
region. In each case, Minuit has converged, error matrix was accurate, and the
confidence level of the fits was acceptable, with CL $> 0.1$ \%. In each case,
the value of the parameter $\alpha$ was, within errors, consistent with $0.9$ so
we have fixed its value to $0.9$ in each case. Let us note, however, that the
fixed value of this  parameter $\alpha$ can be varied between 1.0 and 0.8
without changing the acceptability of the fit, so the criteria of CL $> 0.1 $
\% was satisfied.

The good quality of these fit results indicates, that a substructure is present
inside the protons: the data after the dip and bump region have the same
structure and quality, as a usual low-$|t|$ elastic scattering data below the
dip region, used to be parametrized by a nearly exponential shape. However, the
related cross-sections and length-scales are smaller than that of the protons.
This conclusion is rather obvious after one compares the results of the L\'evy
fits to the tail or large-$|t|$ regions, as presented in Appendix C with the
results of L\'evy fit results in the cone or low-$|t|$ region, as detailed in
Appendix D below. These comparisons were summarized clearly in
Fig.~\ref{f:Summary-tails} for the tail and in Fig.~\ref{f:Summary-cone} for
the cone region. It is quite remarkable, that the proton substructure in the
ISR region has a size that is apparently independent of the change of the
energy in the region of a few tens of GeV, but this substructure appears to be
different (in size) from a substructure that is emerged in $pp$ collisions in
the 7 -- 13 TeV energy range, as evidenced from the parallel dashed lines in
Fig.~\ref{f:Summary-tails}. This observation and the lack of parallel behaviour between
the ISR and the LHC data fits suggests that two substructures, characterized by different
L\'evy source radii but similar L\'evy index of stability 
$\alpha$ are identified  at the ISR and at the LHC energies.

\begin{figure}[!h]
\begin{minipage}{0.7\textwidth}
 \centerline{\includegraphics[width=1.0\textwidth]{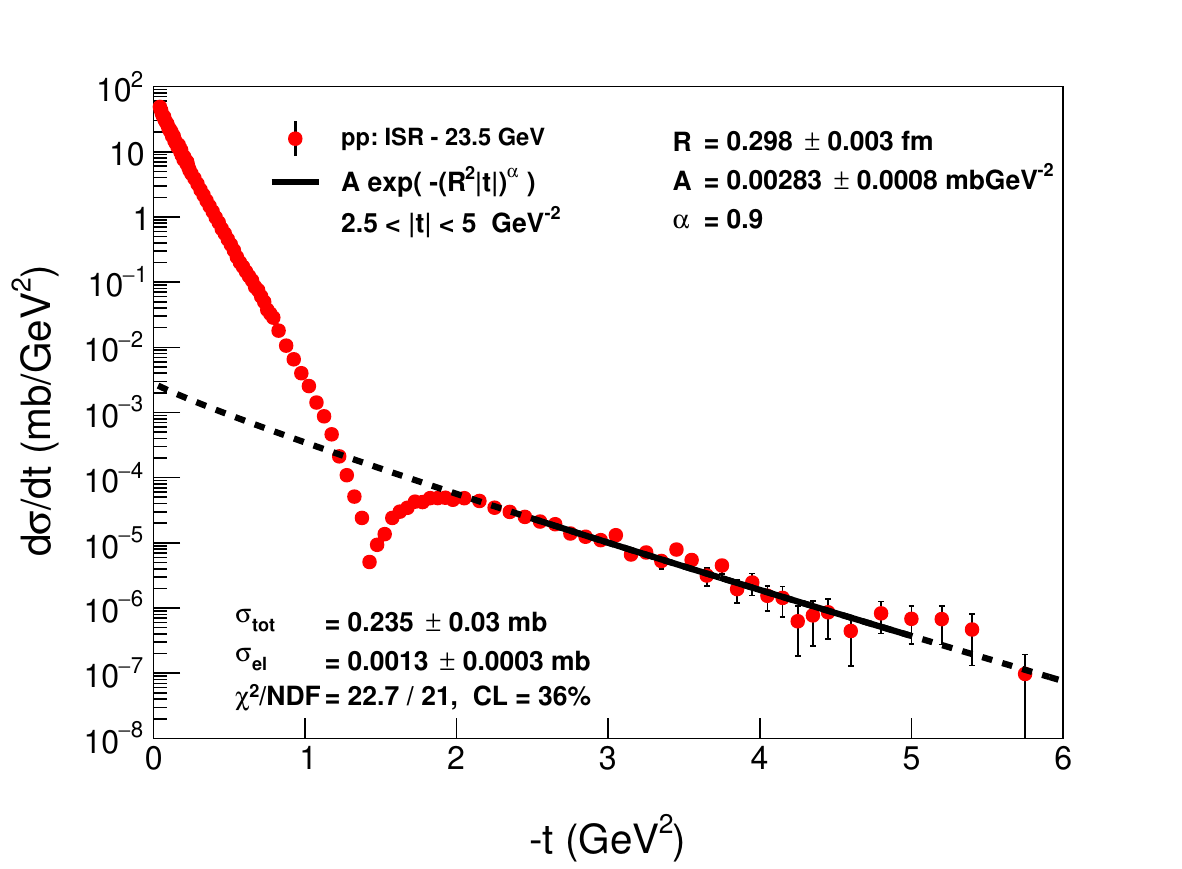}}
\end{minipage}
\caption{Zeroth-order L\'evy fits to the tail of the differential cross-section data on $pp$ elastic scattering at $\sqrt{s} = 23.5 $ GeV \cite{Amaldi:1979kd}, 
with fixed $\alpha = 0.9$, $A$ and $R$ as free fit parameters.
}
\label{f:pp-zeroth-tail1}
\end{figure}

\begin{figure}[!h]
\begin{minipage}{0.7\textwidth}
 \centerline{\includegraphics[width=1.0\textwidth]{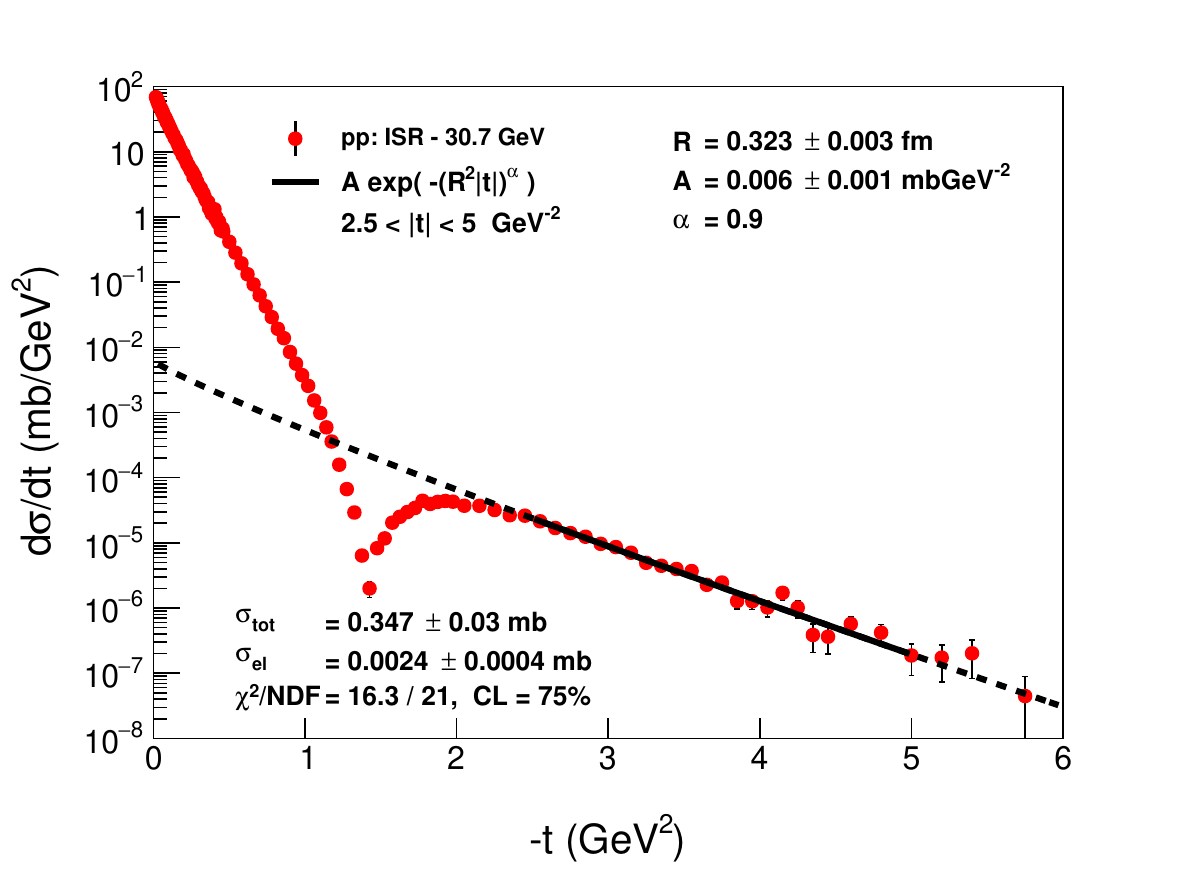}}
\end{minipage}
\caption{Zeroth-order L\'evy fits to the tail of the differential cross-section data on $pp$ elastic scattering at 
$\sqrt{s} = 30.7 $ GeV \cite{Amaldi:1979kd}, 
with fixed $\alpha = 0.9$, $A$ and $R$ as free fit parameters.
}
\label{f:pp-zeroth-tail2}
\end{figure}

\begin{figure}[!h]
\begin{minipage}{0.7\textwidth}
 \centerline{\includegraphics[width=1.0\textwidth]{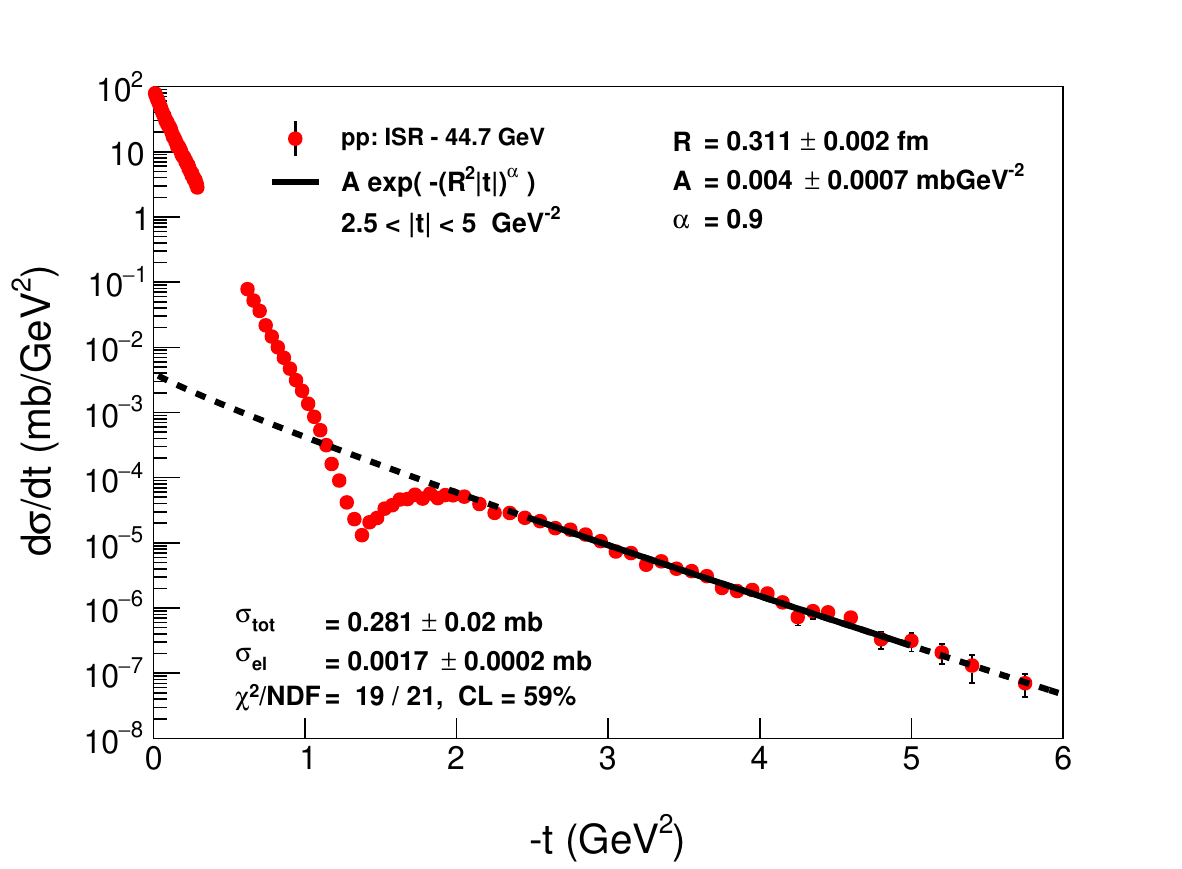}}
\end{minipage}
\caption{Zeroth-order L\'evy fits to the tail of the differential cross-section data on $pp$ elastic scattering at $\sqrt{s} = 44.7$ GeV \cite{Amaldi:1979kd}, 
with fixed $\alpha = 0.9$, $A$ and $R$ as free fit parameters.
}
\label{f:pp-zeroth-tail3}
\end{figure}

\begin{figure}[!h]
\begin{minipage}{0.7\textwidth}
 \centerline{\includegraphics[width=1.0\textwidth]{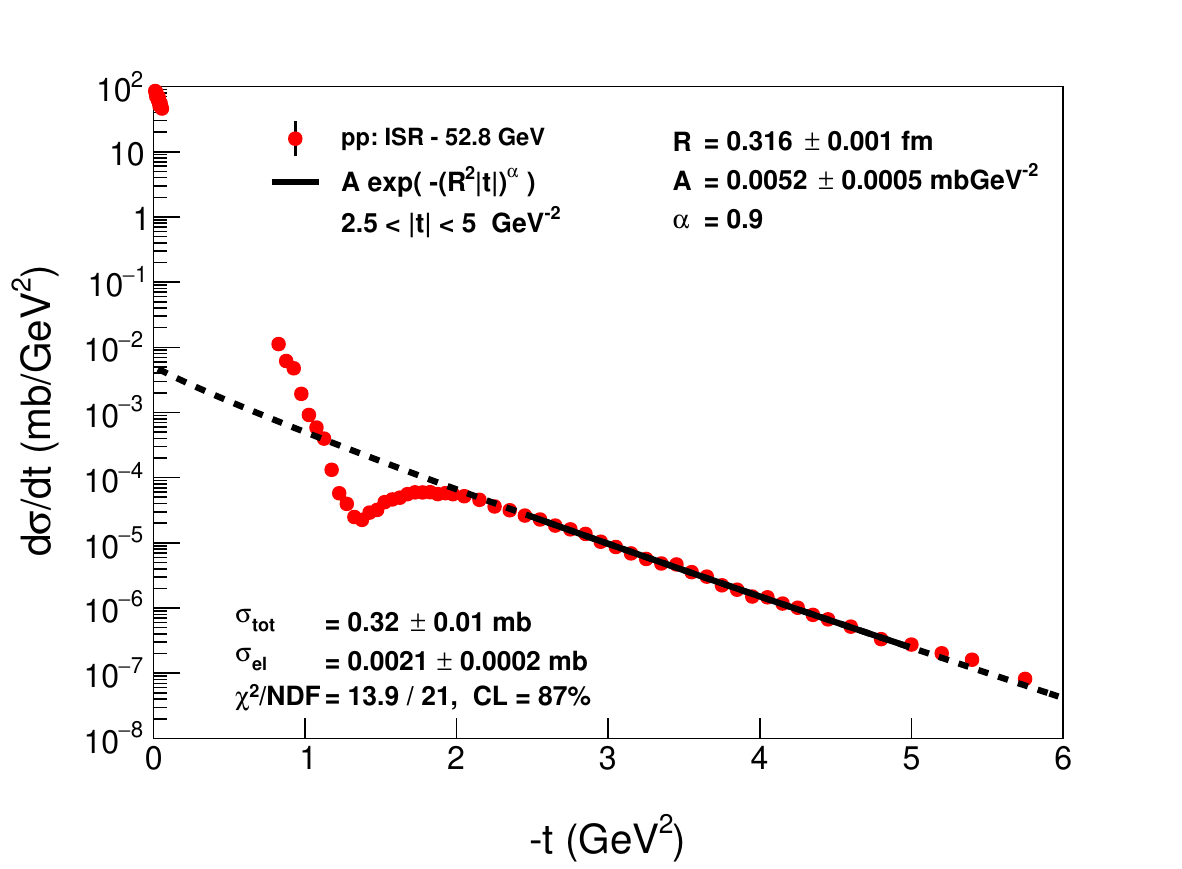}}
\end{minipage}
\caption{Zeroth-order L\'evy fits to the tail of the differential cross-section data on $pp$ elastic scattering at $\sqrt{s} = 52.8 $ GeV \cite{Amaldi:1979kd}, 
with fixed $\alpha = 0.9$, $A$ and $R$ as free fit parameters.
}
\label{f:pp-zeroth-tail4}
\end{figure}

\begin{figure}[!h]
\begin{minipage}{0.7\textwidth}
 \centerline{\includegraphics[width=1.0\textwidth]{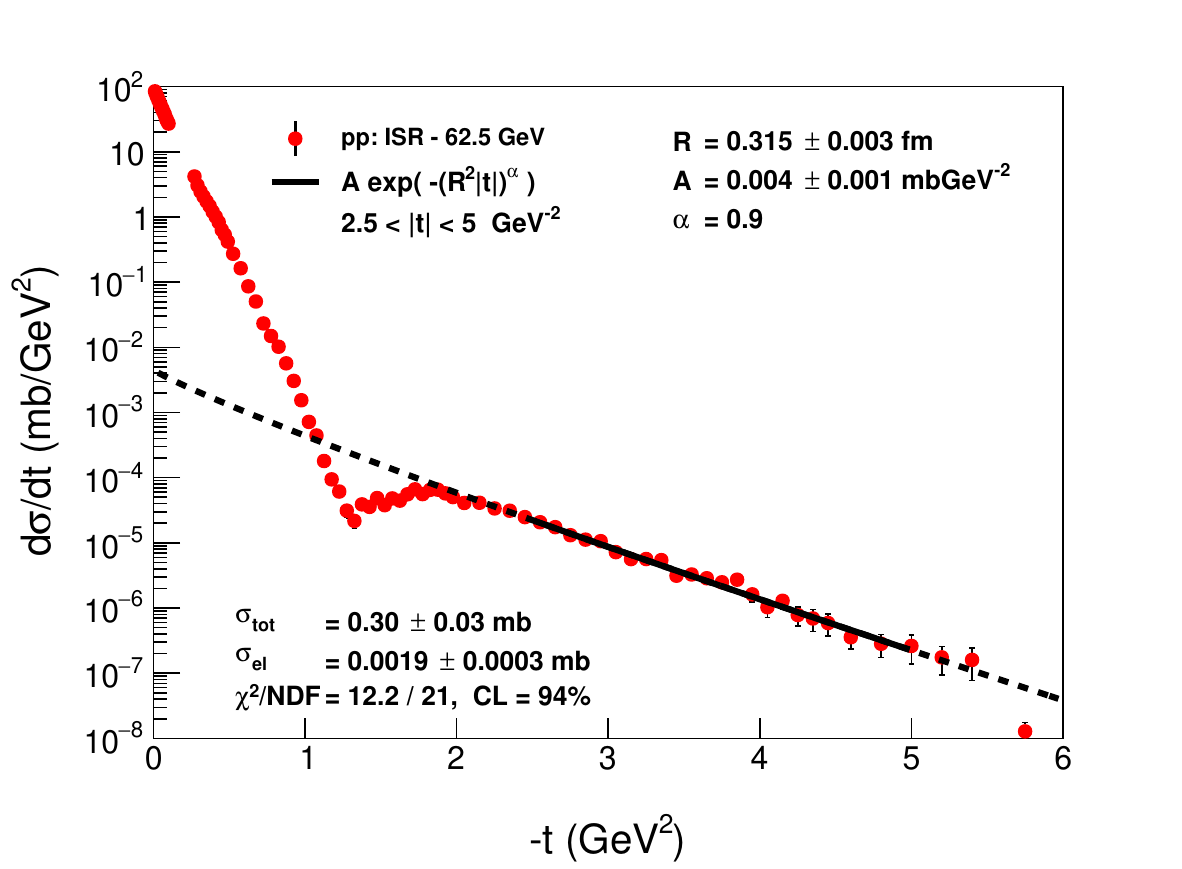}}
\end{minipage}
\caption{Zeroth-order L\'evy fits to the tail of the differential cross-section data on $pp$ elastic scattering at $\sqrt{s} = 62.5 $ GeV \cite{Amaldi:1979kd}, 
with fixed $\alpha = 0.9$, $A$ and $R$ as free fit parameters.
}
\label{f:pp-zeroth-tail5}
\end{figure}

We have also studied the stability of these fits. Increasing the value of
$\alpha$ decreased the L\'evy scale $R$ and the contribution to the total
cross-section. Taking into account a co-variation of the fit parameters with
the fixed value of $\alpha$ we find that in each of these data sets, we are
allowed to vary the value of $\alpha$ in a reasonable range of 0.8 to 1.0.  We
found that in each case the same-size substructure of the proton was present,
with a characteristic L\'evy length scale of $R_{\rm ISR} = 0.3 \pm 0.1 $ fm,
and a contribution to the total cross-section with $\sigma_{\rm ISR} = 0.3
^{+0.3}_{-0.1}$ mb, where the quoted errors take into account the errors coming
from a variation of the value of $\alpha$ as well.  Apparently, the same
structure is seen in these reactions, within the systematic error of the
analysis, as indicated in Figs.~\ref{f:pp-zeroth-tail1},
\ref{f:pp-zeroth-tail2}, \ref{f:pp-zeroth-tail3}, \ref{f:pp-zeroth-tail4}, and
\ref{f:pp-zeroth-tail5}. 


\begin{figure}[!h]
\begin{minipage}{0.7\textwidth}
 \centerline{\includegraphics[width=1.0\textwidth]{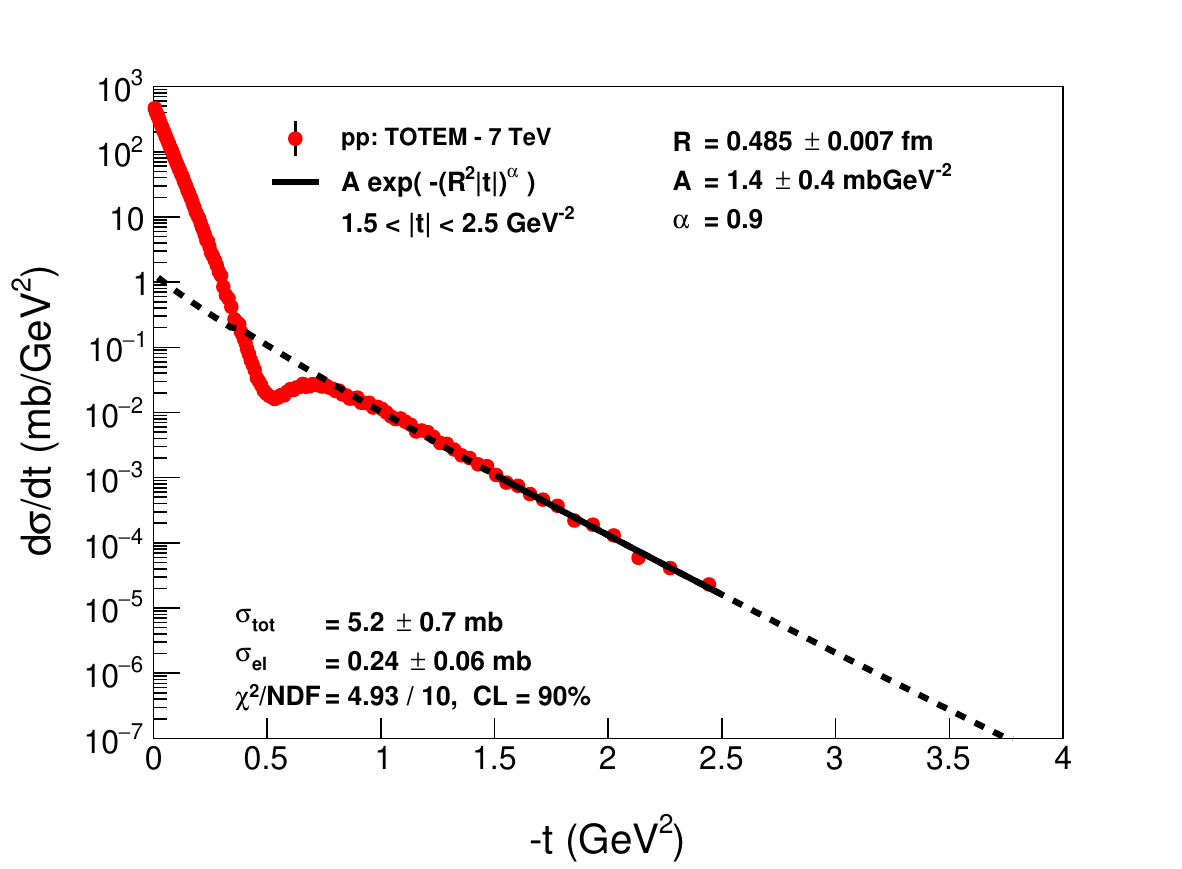}}
\end{minipage}
\caption{Zeroth-order L\'evy fits to the tail of the differential cross-section data on $pp$ elastic scattering at $\sqrt{s} = 7 $ TeV, 
with fixed $\alpha = 0.9$, $A$ and $R$ as free fit parameters.
}
\label{f:pp-zeroth-tail6}
\end{figure}

\begin{figure}[!h]
\begin{center}
\begin{minipage}{0.7\textwidth}
 \centerline{\includegraphics[width=1.0\textwidth]{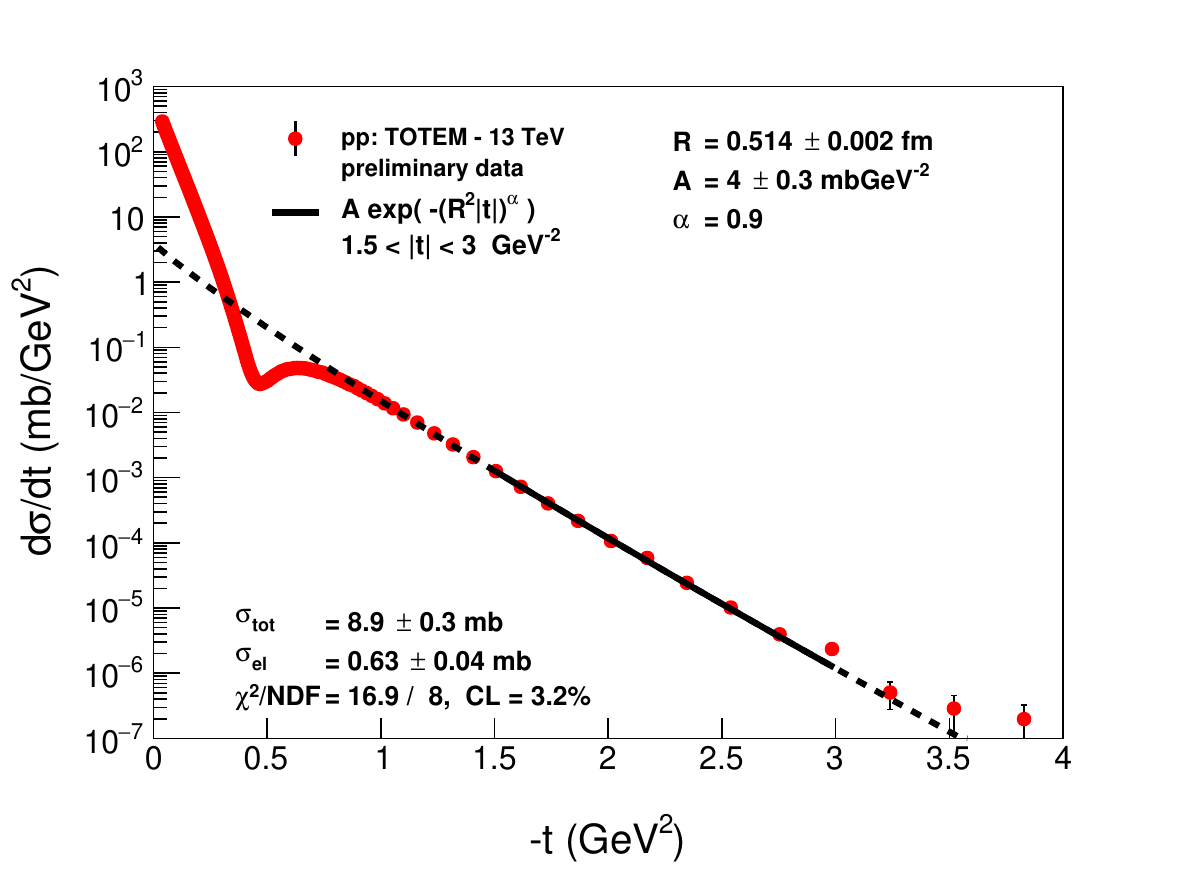}}
\end{minipage}
\end{center}
\vspace{-0.7cm}
\caption{Zeroth-order L\'evy fits to the tail of the differential cross-section data on $pp$ elastic scattering at $\sqrt{s} = 13 $ TeV (TOTEM preliminary), 
with fixed $\alpha = 0.9$, $A$ and $R$ as free fit parameters.
}
\label{f:pp-zeroth-tail7}
\end{figure}

However, the same type of analysis reveals a different-size proton
substructure, when one performs a similar analysis in the TeV energy range.
Fits to the differential cross-sections of elastic $pp$ scattering data at
$\sqrt{s} = 7$ TeV and preliminary TOTEM data at $\sqrt{s} = 13 $ TeV are shown
in Figs.~\ref{f:pp-zeroth-tail6}, and \ref{f:pp-zeroth-tail7}, respectively.
This substructure corresponds to a L\'evy scale of $R_{\rm LHC} = 0.5 \pm 0.1$
fm, and to a contribution to the total cross-section with $\sigma_{\rm
LHC}(7\;{\rm TeV}) = 6.1^{+3.3}_{-2.6}$ mb at 7 TeV, which is within the
errors the same as the preliminary value of $\sigma_{\rm LHC}(13\;{\rm TeV}) =
10.2^{+5.9}_{-4.7}$ mb at 13 TeV. It is clear that these scales of $R$ and the
contributions to the total cross-section at the level of $\sigma_{\rm LHC}
\approx 8.2^{+7.9}_{-4.7} $ mb, that emerge at the TeV scale colliding
energies, are significantly larger than the corresponding cross-sections and
scales in the ISR energy range of 23.5 -- 62.5 GeV. The errors of $R_{\rm LHC}$
and $\sigma_{\rm LHC}$ include the estimated systematic uncertainties as well,
and those are larger than the statistical errors shown on the figures. Indeed,
they also take into account the fact, that similar results were obtained for
$\alpha = 0.8 $ and $\alpha = 1.0$ fixed values as well, where the increasing
or decreasing of $\alpha$ value has resulted in a decrease or increase of $R$
(and $\sigma$), respectively.


The emergence of the proton substructure with two different sizes in elastic $pp$ collisions at the ISR energy range of $\sqrt{s} =$ 23.5 -- 62.5 GeV and at the LHC energy
range of $\,\sqrt[]{s} =$ 7 -- 13 TeV, respectively, is clearly demonstrated on the summary plot that shows the data and the L\'evy fit results to the tails of the differential cross-sections 
of elastic $pp$ collisions on the same plot, also indicating with dashed lines the extrapolation of the fit results outside the fitted domain, see Fig.~\ref{f:Summary-tails}.
\begin{figure}[!h]
\begin{minipage}{0.7\textwidth}
 \centerline{\includegraphics[width=1.0\textwidth]{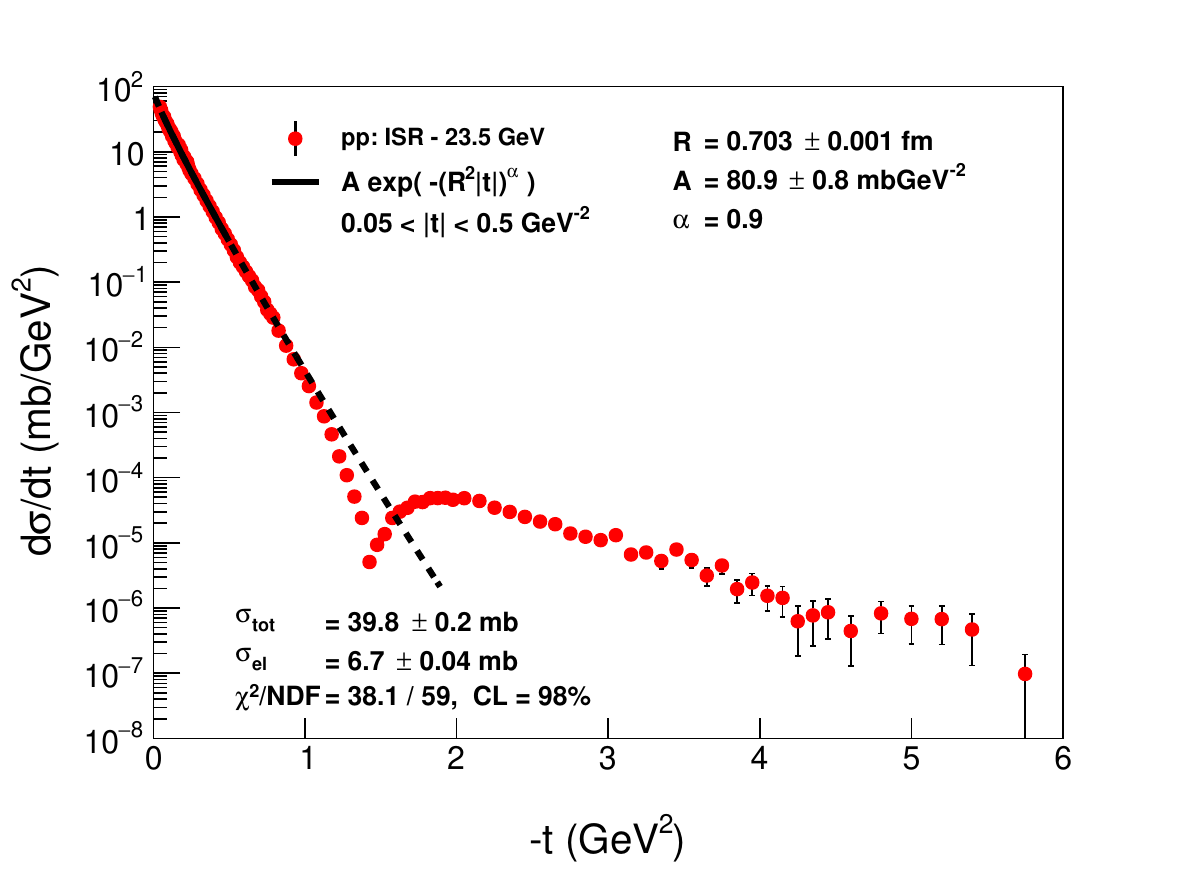}}
\end{minipage}
\caption{Zeroth-order L\'evy fits to the cone region of the differential cross-section data on $pp$ elastic scattering at $\sqrt{s} = 23.5 $ GeV \cite{Amaldi:1979kd}, with fixed $\alpha = 0.9$, $A$ 
and $R$ being free fit parameters.
}
\label{f:pp-zeroth-cone1}
\end{figure}

The change of the slope of these lines from ISR to the LHC energies is so much obvious and striking to the naked eye, that it is rather surprising that such a change was not reported in the literature 
before, at least, to the best of our knowledge. Perhaps, the reason for this effect is the expectation, that perturbative QCD effects should start to be visible soon after the dip-bump structure which could 
result in a power-law decrease of the differential elastic cross-section. Indeed, such a power-law fit was performed and published by the TOTEM Collaboration in Ref.~\cite{Antchev:2011zz}
when analysing the first set of 7 TeV elastic scattering data. In particular, the TOTEM Collaboration has reported that for $|t|$-values larger than $\sim$1.5 GeV$^2$,  the differential cross-section 
of elastic $pp$ collisions exhibits a power law behaviour, with an exponent of $-7.8\, \pm 0.3$ (stat) $\pm\, 0.1$ (syst).
\begin{figure}[!h]
\begin{minipage}{0.7\textwidth}
 \centerline{\includegraphics[width=1.0\textwidth]{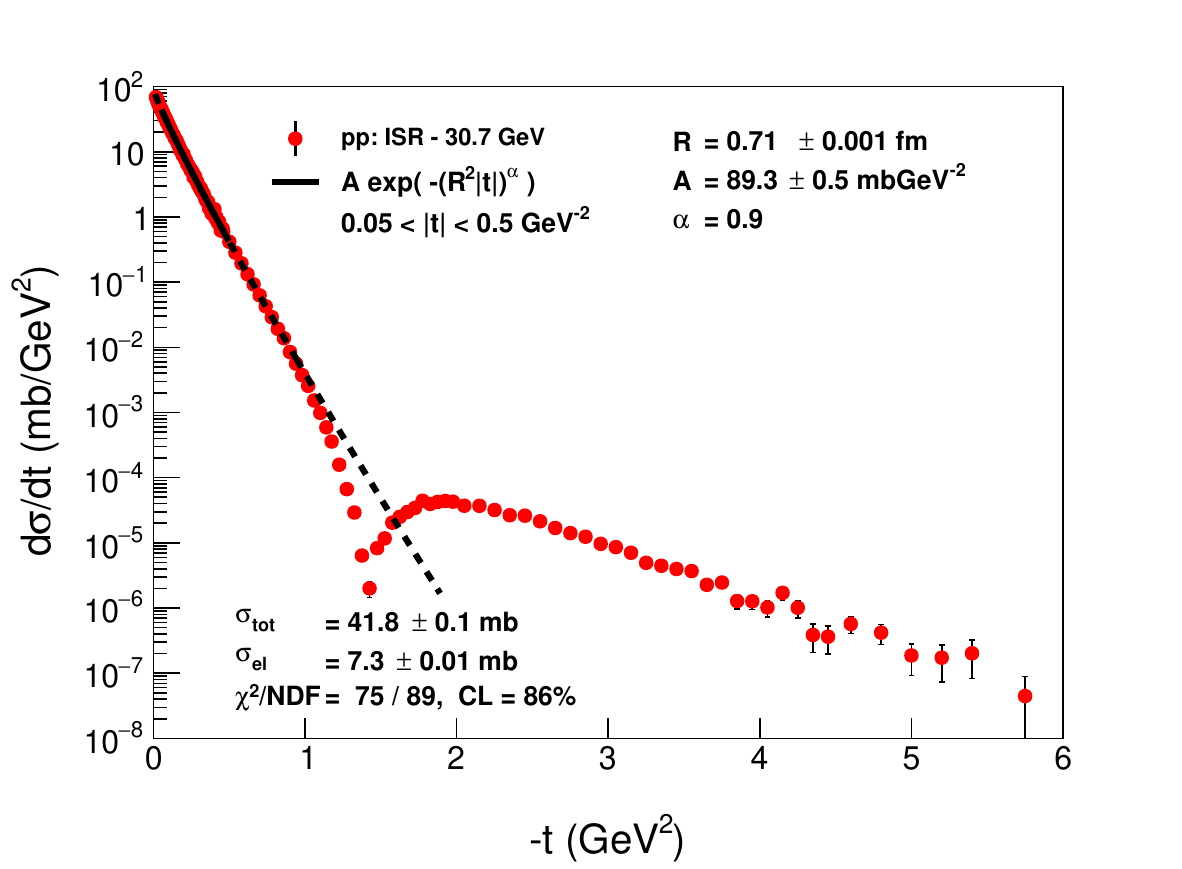}}
\end{minipage}
\caption{Zeroth-order L\'evy fits to the cone region of the differential cross-section data on $pp$ elastic scattering at 
$\sqrt{s} = 30.7 $ GeV \cite{Amaldi:1979kd}, with fixed $\alpha = 0.9$, $A$ 
and $R$ being free fit parameters.
}
\label{f:pp-zeroth-cone2}
\end{figure}
\begin{figure}[!h]
\begin{minipage}{0.7\textwidth}
 \centerline{\includegraphics[width=1.0\textwidth]{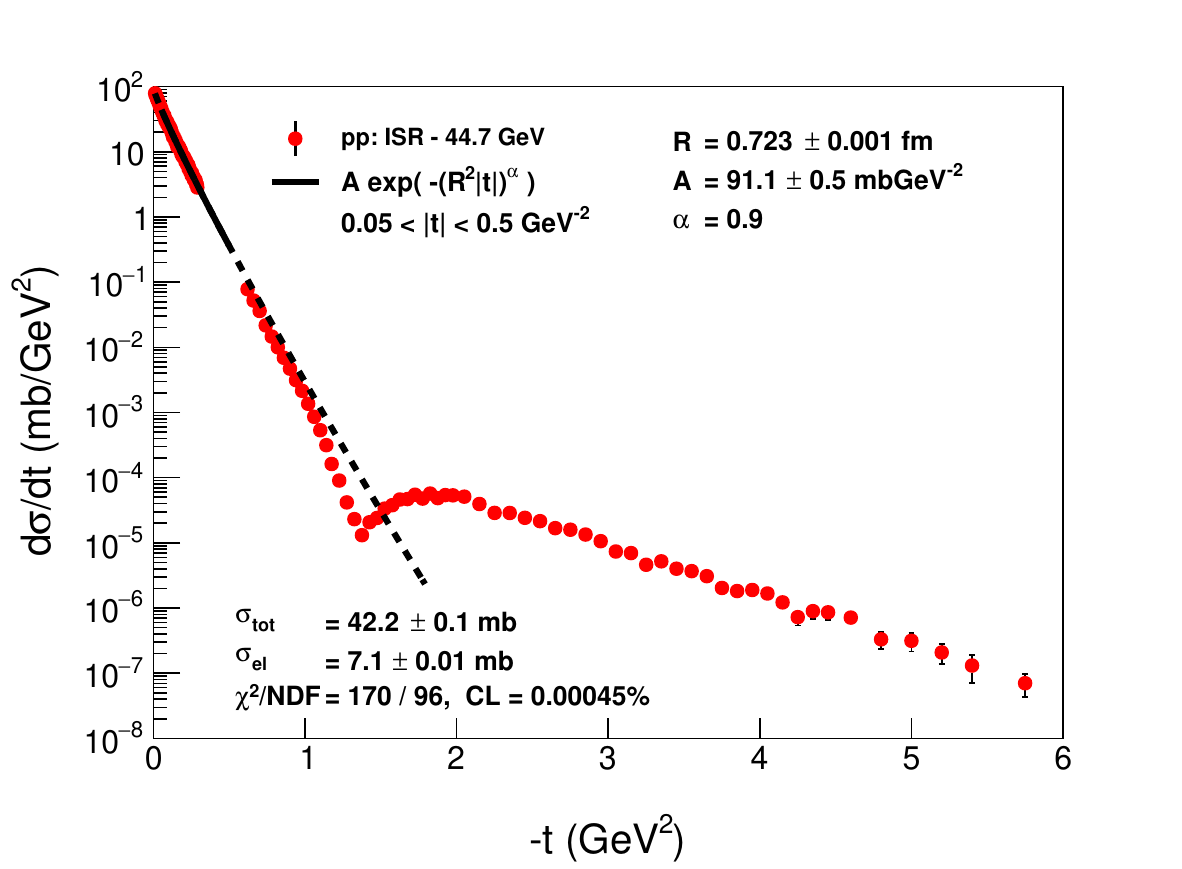}}
\end{minipage}
\caption{Zeroth-order L\'evy fits to the cone region of the differential cross-section data on $pp$ elastic scattering at $\sqrt{s} = 44.7$ GeV \cite{Amaldi:1979kd}, with fixed $\alpha = 0.9$, $A$ 
and $R$ being free fit parameters.
}
\label{f:pp-zeroth-cone3}
\end{figure}
\begin{figure}[!h]
\begin{minipage}{0.7\textwidth}
 \centerline{\includegraphics[width=1.0\textwidth]{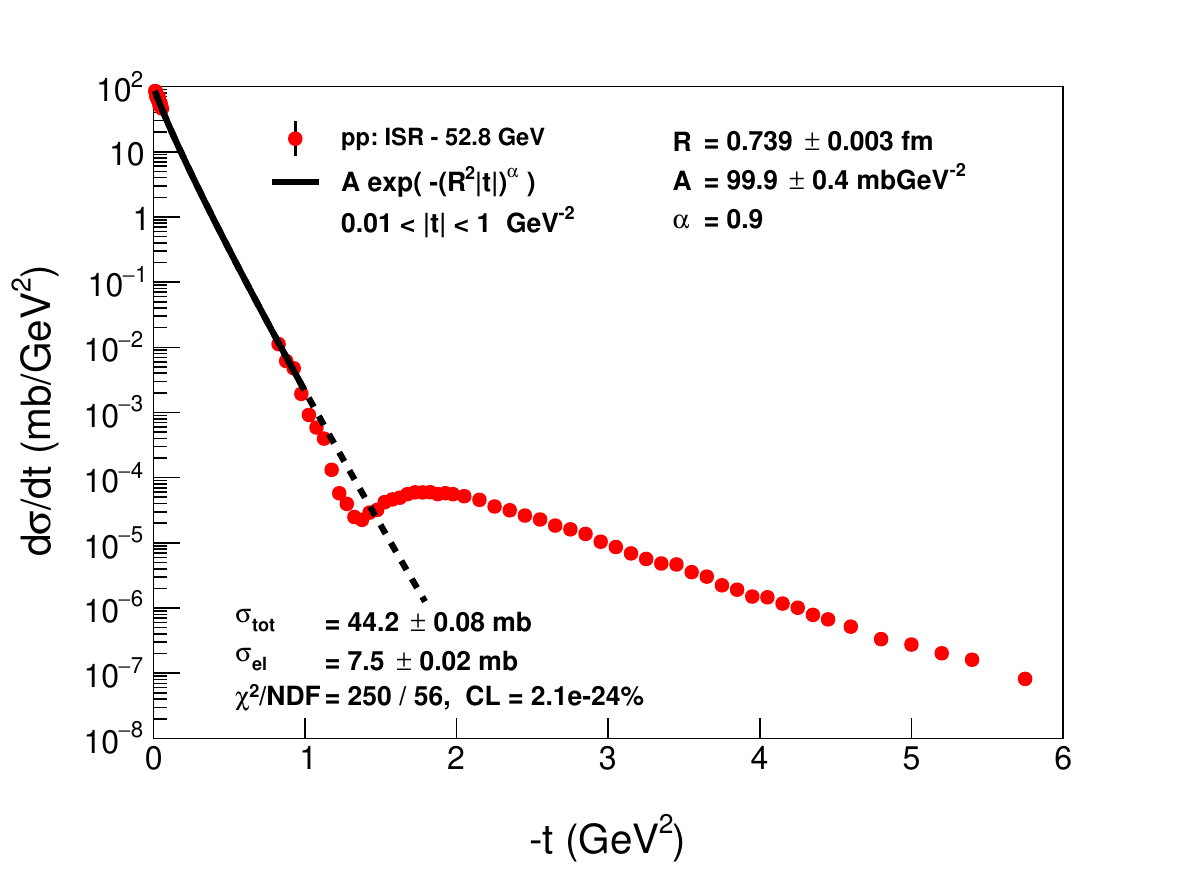}}
\end{minipage}
\caption{Zeroth-order L\'evy fits to the cone region of the differential cross-section data on $pp$ elastic scattering at $\sqrt{s} = 52.8 $ GeV \cite{Amaldi:1979kd}, with fixed $\alpha = 0.9$, $A$ 
and $R$ being free fit parameters.
}
\label{f:pp-zeroth-cone4}
\end{figure}

Given that the new (but still preliminary) TOTEM data at 13 TeV are much more detailed and cover an extended $|t|$ range after the dip and bump structure, we can easily test the presence (or not) 
of the perturbative QCD predicted power-law tails in these elastic scattering data: if $d\sigma/dt \propto |t|^n$ then the power-law tail emerges as a straight line on a log-log plot.
In order to illustrate this point, we have prepared a log-log plot that
includes the TOTEM preliminary $d\sigma/dt$ data at $\sqrt{s} = 13$ TeV, and
the L\'evy expansion fit results.  As clearly seen in this 
Fig.~\ref{f:Appendix-C-13TeV-log-log}, the tail is not a straight line on this
plot, so the $|t|$-range is apparently not yet in the domain of perturbative
QCD, with a possible exception for data points close to the end of the
acceptance region at large $|t|$. These data points, however, have rather big
error bars.

\section{L\'evy fits to elastic $pp$ collisions at small $|t|$}
\label{s:AppendixD}

In this Appendix, we describe the zeroth-order L\'evy fits, $d\sigma/dt = A \exp\left(-(R^2 t)^{\alpha} \right)$ to the low-$|t|$ or cone region of elastic $pp$ scattering data for seven different 
data sets, at $\sqrt{s}$ $=$ 23.5, 30.7, 44.7, 52.8, 62.5 GeV as well as at 7 and 13 TeV. 
\begin{figure}[!h]
\begin{minipage}{0.7\textwidth}
 \centerline{\includegraphics[width=1.0\textwidth]{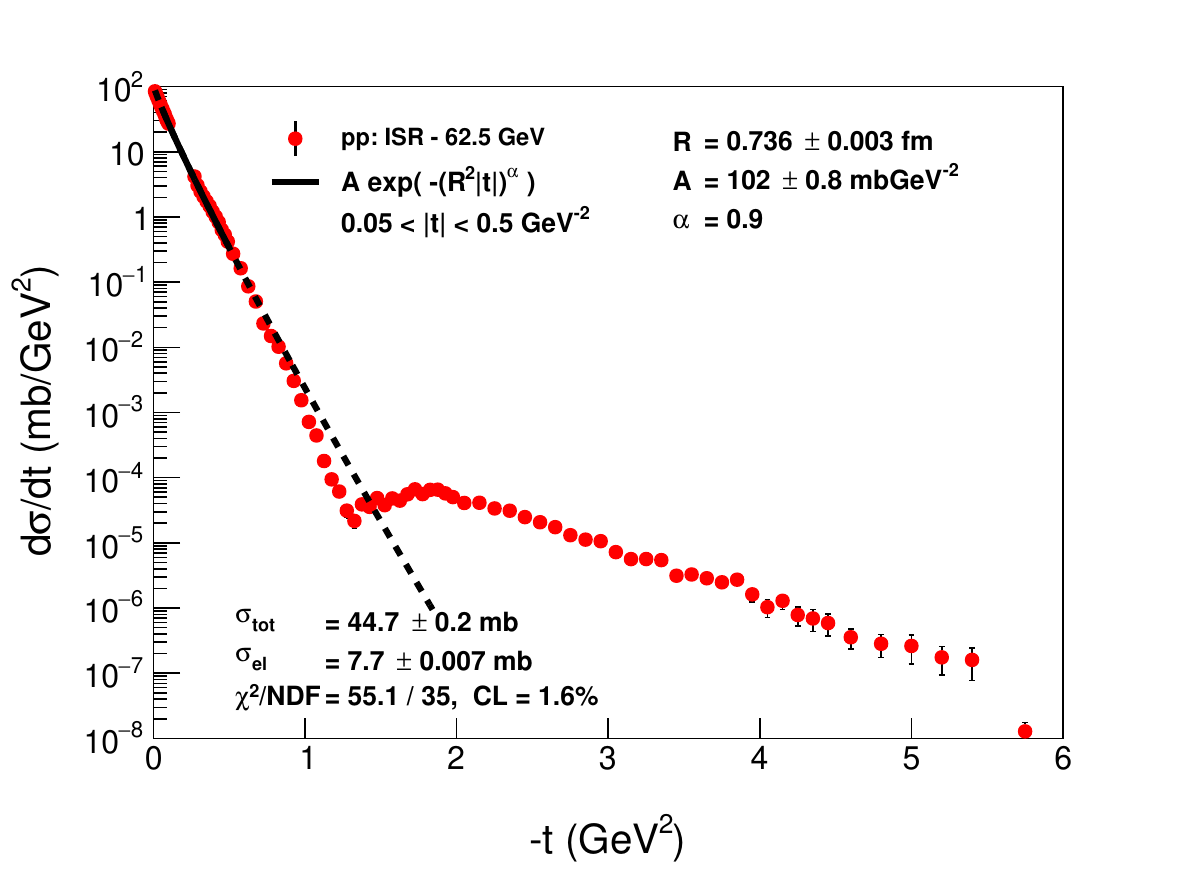}}
\end{minipage}
\caption{Zeroth-order L\'evy fits to the cone region of the differential cross-section data on $pp$ elastic scattering at
$\sqrt{s} = 62.5 $ GeV \cite{Amaldi:1979kd}, with fixed $\alpha = 0.9$, $A$ 
and $R$ being free fit parameters.
}
\label{f:pp-zeroth-cone5}
\end{figure}

Traditionally, an exponential behaviour is assumed for this kinematic region, with $\alpha = 1$ fixed fits. However, recently the TOTEM collaboration demonstrated a non-exponen\-tial behaviour in 
$\sqrt{s} = 8$ TeV $pp$ collisions, that was found to be significant, corresponding to a more than 7$\sigma$ effect~\cite{Antchev:2015zza}. These results were highlighted and some of their 
implications and ramifications were detailed also in Ref.~\cite{Csorgo:2016qyr}. In this Appendix, we re-examine the other, already published data sets using the zeroth-order L\'evy fits,
where the selected low-$|t|$ regions are very similar to the 8 TeV analysis of TOTEM, namely we fit a $|t|$ region that is similar to the TOTEM analysis at 8 TeV, when $|t|$ is measured in units of 
$t_{\rm dip}$. Our fit region is thus $0.04 \, |t_{\rm dip}| \le |t| \le 0.4 \, |t_{\rm dip}|$. We have performed a detailed investigation of this region and found that all the data sets were described 
with a good confidence level with the zeroth-order L\'evy fit, with only three free parameters, $A$, $R$  and $\alpha$. Within the errors, in the ISR region, the values of the parameter $\alpha$ 
were in the region of 0.90 $\pm$ 0.02 in these fits. Hence, we have fixed the value of $\alpha $ to $0.9$, that allowed us to demonstrate the trends more clearly and also to compare the results 
with the similar analysis in the large $|t|$ region.
\begin{figure}[!h]
\begin{minipage}{0.7\textwidth}
 \centerline{\includegraphics[width=1.0\textwidth]{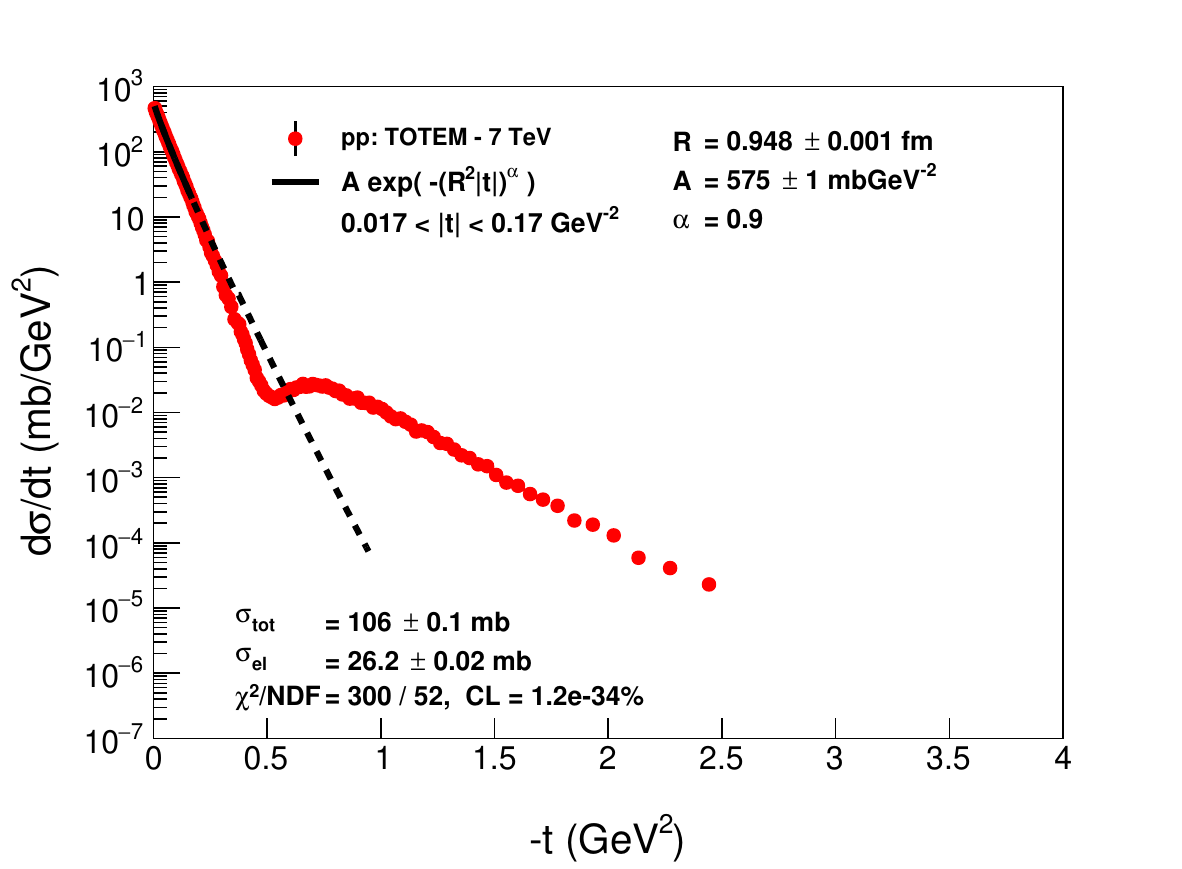}}
\end{minipage}
\caption{Zeroth-order L\'evy fits to the cone region of the differential cross-section data on $pp$ elastic scattering at $\sqrt{s} = 7 $ TeV, with fixed $\alpha = 0.9$, $A$ 
and $R$ being free fit parameters.
}
\label{f:pp-zeroth-cone6}
\end{figure}

All the data in the ISR energy range were fitted successfully with a CL $\geq $
0.1 \% for the $\alpha = 0.9$ fixed case as well. The L\'evy radii $R$ kept on
increasing monotonically with increasing $\sqrt{s}$. The best fit parameters as
well as the fit quality measures are indicated on
Figs.~\ref{f:pp-zeroth-cone1}, \ref{f:pp-zeroth-cone2},
\ref{f:pp-zeroth-cone3}, \ref{f:pp-zeroth-cone4}, \ref{f:pp-zeroth-cone5}.

\begin{figure}[!h]
\begin{center}
 \centerline{\includegraphics[width=0.7\textwidth]{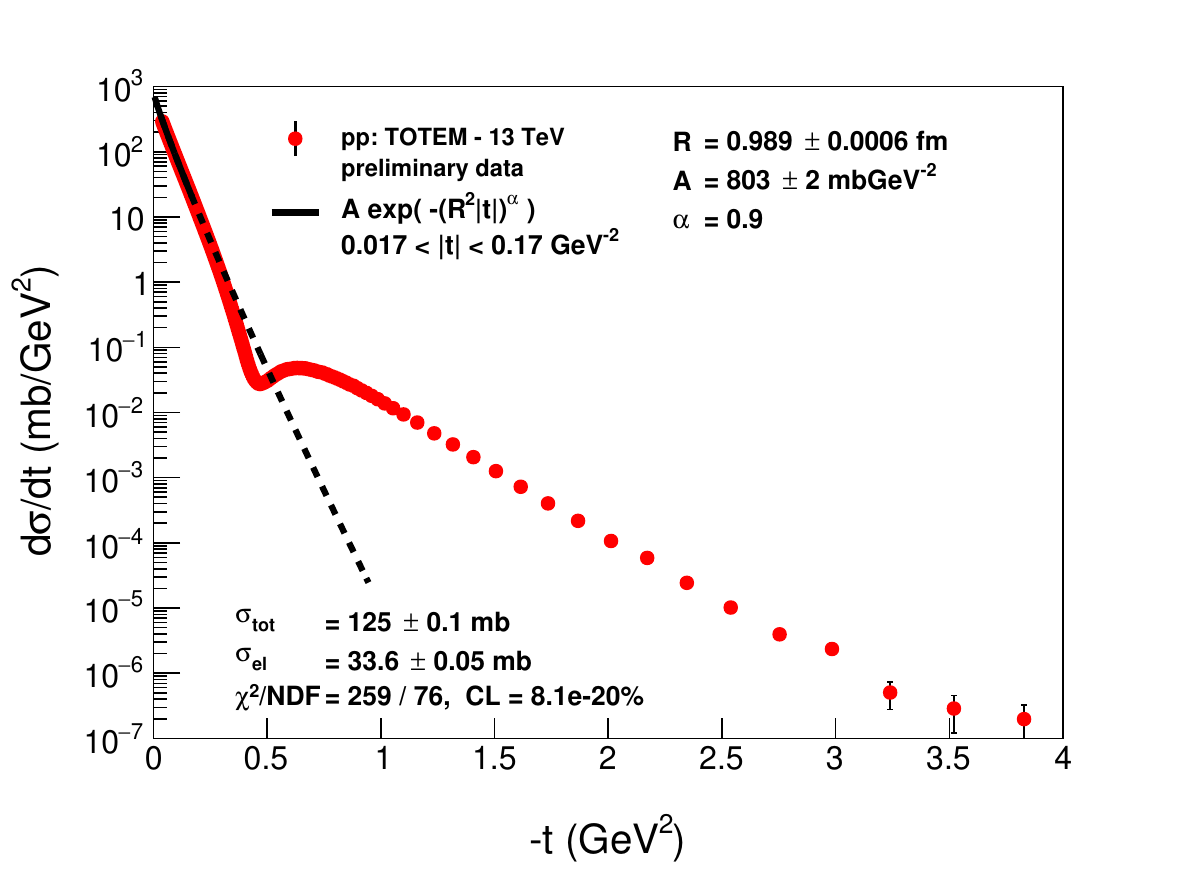}}
\end{center}
\caption{Zeroth-order L\'evy fits to the cone region of the differential cross-section data on $pp$ elastic scattering at $\sqrt{s} = 13 $ TeV (TOTEM preliminary), with fixed $\alpha = 0.9$, $A$ 
and $R$ being free fit parameters.
}
\label{f:pp-zeroth-cone7}
\end{figure}

However, the fixed $\alpha = 0.9$ fits failed both on the $\sqrt{s} = 7 $ TeV
final and on the $\sqrt{s} = 13$ TeV preliminary data, as indicated on 
Figs.~\ref{f:pp-zeroth-cone6} and~\ref{f:pp-zeroth-cone7}.
These fit results, although on a qualitative level apparently may be called as ``reasonable'',
are  in fact statistically not acceptable. This implies that in the TeV energy region,
some new mechanism starts to work, that changes not only the L\'evy scale but
also the shape of the proton. Actually, from the analysis of the shadow profile
function, we know that such a new effect in the TeV energy range corresponds to
the saturation of the shadow profile functions $P(b)$ in the region of $b \le
0.4 - 0.5 $ fm. This result, detailed in Subsection~\ref{ss:shadow-results} is
supported strongly and independently by the fits detailed above.



\begin{thebibliography}{99}

\bibitem{Antchev:2017dia} 
  G.~Antchev {\it et al.} [TOTEM Collaboration],
  arXiv:1712.06153 [hep-ex], Preprint CERN-EP-2017-321,
  \url{http://cds.cern.ch/record/2298154}.
 
%
\bibitem{Antchev:2017yns} 
  G.~Antchev {\it et al.} [TOTEM Collaboration],
  Preprint CERN-EP-2017-335, \url{http://cds.cern.ch/record/2298154}.

%
\bibitem{Lukaszuk:1973nt} 
  L.~Lukaszuk and B.~Nicolescu,
  Lett.\ Nuovo Cim.\  {\bf 8}, 405 (1973).

%
\bibitem{Ster:2015esa} 
  A.~Ster, L.~Jenkovszky and T.~Cs\"org\H{o},
  Phys.\ Rev.\ D {\bf 91}, no. 7, 074018 (2015);
  [arXiv:1501.03860 [hep-ph]].
  
%
\bibitem{Samokhin:2017kde} 
  A.~P.~Samokhin and V.~A.~Petrov,
		Nucl.\ Phys.\ A {\bf 974}, 45 (2018);
  [arXiv:1708.02879 [hep-ph]].

\bibitem{Khoze:2017swe} 
  V.~A.~Khoze, A.~D.~Martin and M.~G.~Ryskin,
  Phys.\ Rev.\ D {\bf 97}, no. 3, 034019 (2018);
  [arXiv:1712.00325 [hep-ph]].
  
%
\bibitem{Petrov:2018xma} 
  V.~A.~Petrov,
  Eur.\ Phys.\ J.\ C {\bf 78}, no. 3, 221 (2018);
  Erratum: [Eur.\ Phys.\ J.\ C {\bf 78}, no. 5, 414 (2018)];
  [arXiv:1801.01815 [hep-ph]].

%
\bibitem{Khoze:2018bus} 
  V.~A.~Khoze, A.~D.~Martin and M.~G.~Ryskin,
  Phys.\ Lett.\ B {\bf 780}, 352 (2018);
  [arXiv:1801.07065 [hep-ph]].
  
%
\bibitem{Goncalves:2018yxc} 
  V.~P.~Gon\c{c}alves and B.~D.~Moreira,
  Phys.\ Rev.\ D {\bf 97}, no. 9, 094009 (2018);
  [arXiv:1801.10501 [hep-ph]].

%
\bibitem{Shabelski:2018jfq} 
  Y.~M.~Shabelski and A.~G.~Shuvaev,
  Eur.\ Phys.\ J.\ C {\bf 78}, no. 6, 497 (2018);
  [arXiv:1802.02812 [hep-ph]].
 
\bibitem{Broilo:2018brv} 
  M.~Broilo, E.~G.~S.~Luna and M.~J.~Menon,
  arXiv:1803.06560 [hep-ph].

%
\bibitem{Broilo:2018els} 
  M.~Broilo, E.~G.~S.~Luna and M.~J.~Menon,
  Phys.\ Lett.\ B {\bf 781}, 616 (2018);
  [arXiv:1803.07167 [hep-ph]].
  
%
\bibitem{Lebiedowicz:2018eui} 
  P.~Lebiedowicz, O.~Nachtmann and A.~Szczurek,
  arXiv:1804.04706 [hep-ph].
  
%
\bibitem{Martynov:2018nyb} 
  E.~Martynov and B.~Nicolescu,
  arXiv:1804.10139 [hep-ph].
  
%
\bibitem{Troshin:2018ihb} 
  S.~M.~Troshin and N.~E.~Tyurin,
  arXiv:1805.05161 [hep-ph].

%
\bibitem{Dremin:2018uwt} 
  I.~M.~Dremin,
  Universe {\bf 4}, no. 5, 65 (2018).

%
\bibitem{Broniowski:2018xbg} 
  W.~Broniowski, L.~Jenkovszky, E.~Ruiz Arriola and I.~Szanyi,
  arXiv:1806.04756 [hep-ph].

%
\bibitem{Khoze:2018kna} 
  V.~A.~Khoze, A.~D.~Martin and M.~G.~Ryskin,
  arXiv:1806.05970 [hep-ph].
  
\bibitem{Frici-la-Biodola:2018vvv}
F. Nemes for the TOTEM Collaboration,
Proc. of the 4th Elba workshop on Forward Physics @ LHC energy, 
24-26 May 2018, Elba, Italy, in preparation, 
to be published in a special issue of Instruments;
\url{https://indico.cern.ch/event/705748/}\\
Fabio Ravera for the TOTEM collaboration, 
134th LHCC meeting - open session, 30 May 2018;
\url{https://indico.cern.ch/event/726320/} .

%
\bibitem{Csorgo:2000pf}
  T.~Cs\"org\H{o} and S.~Hegyi, %
  Phys.\ Lett.\ B \textbf{489} (2000) 15.

%
\bibitem{DeKock:2012gp}
  M.~B.~De Kock, H.~C.~Eggers and T.~Cs\"org\H{o},
  PoS WPCF {\bf 2011} (2011) 033.

%
\bibitem{Adare:2017vig} 
  A.~Adare {\it et al.} [PHENIX Collaboration],
  Phys.\ Rev.\ C {\bf 97}, no. 6, 064911 (2018);
  [arXiv:1709.05649 [nucl-ex]].

%
\bibitem{Novak:2016cyc} 
  T.~Nov\'ak, T.~Cs\"org\H{o}, H.~C.~Eggers and M.~de Kock,
  Acta Phys.\ Polon.\ Supp.\  {\bf 9}, 289 (2016);
  [arXiv:1604.05513 [physics.data-an]].

%
\bibitem{Block:2006hy} 
  M.~M.~Block,
  Phys.\ Rept.\  {\bf 436}, 71 (2006);
  [hep-ph/0606215].

  \bibitem{Antchev:2015zza} 
  G.~Antchev {\it et al.} [TOTEM Collaboration],
  Nucl.\ Phys.\ B {\bf 899}, 527 (2015);
  [arXiv:1503.08111 [hep-ex]].

%
\bibitem{Tsallis:1995zz} 
  C.~Tsallis, S.~V.~F.~L\'evy, A.~M.~C.~Souza and R.~Maynard,
  Phys.\ Rev.\ Lett.\  {\bf 75}, 3589 (1995).

%
\bibitem{Weisstein:Stable}
E. W. Weisstein,
"Stable Distribution." From MathWorld--A Wolfram Web Resource. 
\url{http://mathworld.wolfram.com/StableDistribution.html}

%
\bibitem{Nolan:2016st}
  J.~P.~Nolan,  {\it Stable distributions: Models for Heavy Tailed Data},
  Springer-Verlag, Imprint Birkhauser,  ISBN10 0817641599, (2016), pp.~1-352.

%
\bibitem{Ster-BGL17}
A. Ster et al, 
{\it Model independent analysis of data with nearly Gaussian or L\'evy shape},\\
proceedings of the 10th Bolyai-Gauss-Lobachevsky conference BGL17,
Gy\"ongy\"os, Hungary, August 20-26, 2017,
\url{https://indico.cern.ch/event/586799/contributions/2695964/}

%
\bibitem{Novak-WPCF18}
T. Nov\'ak et al,
{\it Model independent analysis of nearly L\'evy sources},\\
proceedings of the 12th Workshop on Particle Correlations and Femtoscopy,
Cracow, Poland, May 22-26, 2018,
\url{https://indico.ifj.edu.pl/event/199/contributions/1166/}

%
\bibitem{Csorgo:2003uv}
  T.~Cs\"org\H{o}, S.~Hegyi and W.~A.~Zajc,
  Eur.\ Phys.\ J.\ C {\bf 36} (2004) 67;
[nucl-th/0310042].
  
\bibitem{Antchev:2013gaa} 
  G.~Antchev {\it et al.} [TOTEM Collaboration],
  EPL {\bf 101}, no. 2, 21002 (2013).
  
 %
\bibitem{Jenkovszky:2011hu} 
  L.~L.~Jenkovszky, A.~I.~Lengyel and D.~I.~Lontkovskyi,
  Int.\ J.\ Mod.\ Phys.\ A {\bf 26}, 4755 (2011);
  [arXiv:1105.1202 [hep-ph]].
  
%
\bibitem{Phillips:1974vt} 
  R.~J.~N.~Phillips and V.~D.~Barger,
  Phys.\ Lett.\  {\bf 46B}, 412 (1973).

%
\bibitem{Nemes:2015iia} 
  F.~Nemes, T.~Cs\"org\H{o} and M.~Csan\'ad,
  Int.\ J.\ Mod.\ Phys.\ A {\bf 30}, no. 14, 1550076 (2015);
  [arXiv:1505.01415 [hep-ph]].

%
\bibitem{Kohara:2014waa} 
  A.~K.~Kohara, E.~Ferreira and T.~Kodama,
  Eur.\ Phys.\ J.\ C {\bf 74}, 3175 (2014);
  [arXiv:1408.1599 [hep-ph]].

%
\bibitem{Lipari:2013kta} 
  P.~Lipari and M.~Lusignoli,
  Eur.\ Phys.\ J.\ C {\bf 73}, no. 11, 2630 (2013);
  [arXiv:1305.7216 [hep-ph]].

%
\bibitem{Dremin:2017vtf} 
  I.~M.~Dremin,
  EPJ Web Conf.\  {\bf 145}, 10003 (2017).

\bibitem{Csanad:2016add} 
  M.~Csan\'ad, T.~Cs\"org\H{o}, Z.~F.~Jiang and C.~B.~Yang,
  Universe {\bf 3}, no. 1, 9 (2017);
  [arXiv:1609.07176 [hep-ph]].

%
\bibitem{Kopeliovich:2000ef} 
  B.~Z.~Kopeliovich, I.~K.~Potashnikova, B.~Povh and E.~Predazzi,
  Phys.\ Rev.\ Lett.\  {\bf 85}, 507 (2000);
[hep-ph/0002241].

%
\bibitem{Kopeliovich:2000pc} 
  B.~Z.~Kopeliovich, I.~K.~Potashnikova, B.~Povh and E.~Predazzi,
  Phys.\ Rev.\ D {\bf 63}, 054001 (2001);
[hep-ph/0009008].

%
\bibitem{Kopeliovich:2012yy} 
  B.~Z.~Kopeliovich, I.~K.~Potashnikova and B.~Povh,
  Phys.\ Rev.\ D {\bf 86}, 051502 (2012);
[arXiv:1208.5446 [hep-ph]].

%
\bibitem{Amos:1991bp} 
  N.~A.~Amos {\it et al.} [E710 Collaboration],
  Phys.\ Rev.\ Lett.\  {\bf 68}, 2433 (1992).

%
\bibitem{Brodsky:2017qno} 
  S.~J.~Brodsky,
  Adv.\ High Energy Phys.\  {\bf 2018}, 7236382 (2018);
[arXiv:1709.01191 [hep-ph]].

%
\bibitem{Amaldi:1979kd} 
  U.~Amaldi and K.~R.~Schubert,
  Nucl.\ Phys.\ B {\bf 166}, 301 (1980).

%
\bibitem{Breakstone:1984te} 
  A.~Breakstone {\it et al.} [AMES-BOLOGNA-CERN-DORTMUND-HEIDELBERG-WARSAW Collaboration],
  Nucl.\ Phys.\ B {\bf 248}, 253 (1984).

%
\bibitem{Bernard:1987vq} 
  D.~Bernard {\it et al.} [UA4 Collaboration],
  Phys.\ Lett.\ B {\bf 198}, 583 (1987).

%
\bibitem{Bernard:1986ye} 
  D.~Bernard {\it et al.} [UA4 Collaboration],
  Phys.\ Lett.\ B {\bf 171}, 142 (1986).

%
\bibitem{Antchev:2011zz} 
  G.~Antchev {\it et al.} [TOTEM Collaboration],
  EPL {\bf 95}, no. 4, 41001 (2011);
[arXiv:1110.1385 [hep-ex]].

%
\bibitem{Abazov:2012qb} 
  V.~M.~Abazov {\it et al.} [D0 Collaboration],
  Phys.\ Rev.\ D {\bf 86}, 012009 (2012);
[arXiv:1206.0687 [hep-ex]].

\bibitem{Csorgo:2016qyr} 
  T.~Cs\"org\H{o} [TOTEM Collaboration],
  EPJ Web Conf.\  {\bf 120}, 02004 (2016);
[arXiv:1602.00219 [hep-ex]].
  
\end{thebibliography}
\end{document}